\documentclass[structabstract]{aa}

\usepackage{array}
\usepackage{graphicx}
\usepackage{booktabs}
\usepackage[round]{natbib}
\usepackage[version=3]{mhchem}   %chemische Formeln
\usepackage{aalongtable}         %long table
\usepackage{multirow}            %multirow
\usepackage{txfonts}

\newcolumntype{C}[1]{>{\centering}m{#1}}
\newcolumntype{L}[1]{>{\raggedright}m{#1}}

\begin{document}

   \title{Dust and molecular shells in asymptotic giant branch stars\thanks{Based on observations made with the Very Large Telescope Interferometer (VLTI) at the Paranal Observatory under program IDs 079.D-0140, 080.D-0005, 081.D-0198, 082.D-0641 and 083.D-0294.}$^{,}$\thanks{Color versions of the figures and Tables~\ref{App_VisFit} to \ref{TableApp_log} are only available in electronic form via http://www.aanda.org. FITS files of the calibrated visibilities are only available at the CDS via anonymous ftp to cdsarc.u-strasbg.fr (130.79.128.5) or via http://cdsarc.u-strasbg.fr/viz-bin/qcat?J/A+A/vol/page.}}
   
   %\footnote{Color versions of the figures are available in electronic form via http://www.aanda.org}

   \subtitle{Mid-infrared interferometric observations of\\ \object{R~Aql}, \object{R~Aqr}, \object{R~Hya}, \object{W~Hya} and \object{V~Hya}}
 
   \author{R. Zhao-Geisler\inst{1,2}\fnmsep\thanks{Fellow of the International Max Planck Research School (IMPRS)}
          \and A. Quirrenbach\inst{1}
          \and R. K\"ohler\inst{1,3}
          \and B. Lopez\inst{4}
          }

   \institute{Zentrum f\"ur Astronomie der Universit\"at Heidelberg (ZAH),
              Landessternwarte, K\"onigstuhl 12, 69120 Heidelberg, Germany, \email{rgeisler@ntnu.edu.tw}
    \and
              National Taiwan Normal University, Department of Earth Sciences, 88 Sec.~4, Ting-Chou Rd, Wenshan District, Taipei, 11677 Taiwan, ROC
    \and
              Max-Planck-Institut f\"ur Astronomie, K\"onigstuhl 17, D-69120 Heidelberg, Germany
    \and
              Laboratoire J.-L. Lagrange, Universit\'e de Nice Sophia-Antipolis and Observatoire de la C$\hat{\mathrm{o}}$te d'Azur BP 4229, 06304 Nice Cedex 4, France
              }

   \date{Received September 26, 2011; accepted June 21, 2012}

%###########################################################################################
%###########################################################################################

  \abstract % 5 {} token are mandatory
  % context heading (optional)
   {Asymptotic giant branch (AGB) stars are one of the largest distributors of dust into the interstellar medium. However, the wind formation mechanism and dust condensation sequence leading to the observed high mass-loss rates have not yet been constrained well observationally, in particular for oxygen-rich AGB stars.}
  % aims heading (mandatory)
   {The immediate objective in this work is to identify molecules and dust species which are present in the layers above the photosphere, and which have emission and absorption features in the mid-infrared (IR), causing the diameter to vary across the N-band, and are potentially relevant for the wind formation.}
  % methods heading (mandatory)
   {Mid-IR ($8-13$~$\mu$m) interferometric data of four oxygen-rich AGB stars (R~Aql, R~Aqr, R~Hya, and W~Hya) and one carbon-rich AGB star (V~Hya) were obtained with MIDI/VLTI between April~2007 and September~2009. The spectrally dispersed visibility data are analyzed by fitting a circular fully limb-darkened disk (FDD).}
  % results heading (mandatory)
   {The FDD diameter as function of wavelength is similar for all oxygen-rich stars. The apparent size is almost constant between 8 and 10~$\mu$m and gradually increases at wavelengths longer than 10~$\mu$m. The apparent FDD diameter in the carbon-rich star V~Hya essentially decreases from 8 to 12~$\mu$m. The FDD diameters are about 2.2 times larger than the photospheric diameters estimated from K-band observations found in the literature. The silicate dust shells of R~Aql, R~Hya and W~Hya are located fairly far away from the star, while the silicate dust shell of R~Aqr and the amorphous carbon (AMC) and SiC dust shell of V Hya are found to be closer to the star at around 8~photospheric radii. Phase-to-phase variations of the diameters of the oxygen-rich stars could be measured and are on the order of 15\% but with large uncertainties.}
  % conclusions heading (optional)
   {From a comparison of the diameter trend with the trends in RR~Sco and S~Ori it can be concluded that in oxygen-rich stars the overall larger diameter originates from a warm molecular layer of \ce{H2O}, and the gradual increase longward of 10~$\mu$m can be most likely attributed to the contribution of a close \ce{Al2O3} dust shell. The chromatic trend of the Gaussian FWHM in V~Hya can be explained with the presence of AMC and SiC dust. The observations suggest that the formation of amorphous \ce{Al2O3} in oxygen-rich stars occurs mainly around or after visual minimum. However, no firm conclusions can be drawn concerning the mass-loss mechanism. Future modeling with hydrostatic and self-consistent dynamical stellar atmospheric models will be required for a more certain understanding.}

% A small increase of the diameter shortward of 10~$\mu$m can be assigned to the presence of SiO.

             
   \titlerunning{Dust and molecular shells in asymptotic giant branch stars}

   \maketitle

%###########################################################################################
%###########################################################################################

%\renewcommand{\topfraction}{1.0}
%\renewcommand{\bottomfraction}{1.0}

\section{Introduction}\label{secIntro}
 
   Asymptotic giant branch (AGB) stars are among the most important distributors of dust into the interstellar medium due to their high mass-loss rates in combination with an effective dust condensation. Especially, the dust plays a crucial role for the formation and acceleration of the dense wind. By providing the seed particles for interstellar grains, AGB stars contribute to the chemical evolution of the interstellar medium (ISM) and facilitate further star and planet formation. Progress in the theoretical understanding has been made \citep{Hoefner2003,Woitke2006a,Hoefner08,Norris2012}, but the wind formation mechanism and dust condensation sequence in oxygen-rich AGB stars need further observational constraints. Also for low mass-loss carbon-rich objects the wind formation has not been fully understood \citep{Mattsson2011,Sacuto2011}.

   The slow wind of AGB stars is driven by stellar pulsation in combination with radiation pressure on dust. In order to support this current understanding, the location and composition of newly formed dust as function of the pulsation cycle, mass-loss rate and underlying chemistry are investigated. The size of the inner region free of dust is one of the parameters best constrained from mid-IR interferometric observations. This quantity is fundamental for understanding the condition of dust formation.

   The low surface gravity and the stellar pulsation lead to an increased scale height of the atmosphere and make it possible to provide the conditions for dust grain formation. The consequence of a very extended atmosphere is a not clearly definable and measurable photospheric radius $R_{\mathrm{phot}}$\footnote{The definition of $R_{\mathrm{phot}} = \theta_{\mathrm{phot}}$/2 used in this work is given in Sect.~\ref{secResSWaveDep}.} \citep[e.g.][]{Mennesson2002,Tej2003,Ohnaka2004,Woodruff2004,Fedele2005} and observations at different wavelengths probe different atmospheric layers \citep{Baschek1991,Scholz2001}. The measured apparent diameters are correlated to the absorption and emission features of the most abundant and radiatively important molecular species \citep[e.g.][and references therein]{Hofmann1998,Jacob2000} as well as first dust grain species with high sublimation temperatures \citep[e.g.][]{LorenzM_Pompeia2000,Verhoelst09}.
     
   The immediate objective in this work is to identify molecules and dust which are present in close layers above the photosphere (at around 2~$R_{\mathrm{phot}}$), and which have absorption features in the mid-IR, causing the diameter to vary across the N-band. Candidates in O-rich environments are \ce{H2O}, SiO, CO, and TiO molecules, and dust grains composed of \ce{Mg2SiO4}, \ce{MgSiO3}, \ce{SiO2}, \ce{Al2O3}, and \ce{TiO2} \citep[e.g.][]{Woitke2006a}. In C-rich stars, molecular layers of \ce{C2H2} and HCN, and dust composed of SiC and amorphous carbon (AMC) can be expected.
      
   The size of the photosphere, close molecular and dust layers are constrained by a broad sampling of the visibility in the first lobe, i.e.~by carrying out observations over a large projected baseline range. Visibility points at the shortest baselines are used to measure the contribution of surrounding dust in the more extended circumstellar environment. Observations were conducted at several orientations to establish whether the sources are elongated or not.

   From infrared spectroscopy, it is already known that there is no simple relation between the spatial distribution of different molecules and pulsation phase \citep[e.g.][and references therein]{Woodruff2009}. By monitoring the stars regularly over a few pulsation cycles the dynamic behavior is studied, in particular the variation in the distribution of the close warm layers of dust and molecules.

   One star in this study (W~Hya) was already extensively described in \citet{ZhaoGeisler2011} (thereafter paper~I) and is included here only for completeness. In addition, some content refers to paper~I. Section~\ref{secObsSubTar} gives an overview of the target stars, and Sect.~\ref{secObsData} describes the observation and data reduction. This is followed by Sect.~\ref{secLCSpecVis} showing the light curves, spectra and results of the visibility modeling. The data are interpreted and discussed in Sect.~\ref{secIntDis} including an investigation of the dynamic properties and morphology. A summary is given in Sect.~\ref{secConc}.

%###########################################################################################
%###########################################################################################
\section{Target properties}\label{secObsSubTar}

   The observations concentrated on a relative small number of stars to obtain a fair number of visibility points to sample the uv-plane as well as the pulsation phase. In total five stars, namely R~Aql, R~Aqr, R~Hya, W~Hya, and V~Hya, were observed in the framework of a Guaranteed Time Observation (GTO) program (cf.~Sect.~\ref{secObsSubInt}). The targets were chosen to cover different chemistries, evolution stages and mass-loss rates. R~Aql, R~Aqr, R~Hya, and W~Hya are oxygen-rich AGB stars, with R~Aqr being a symbiotic system, while V~Hya is an evolved carbon-rich AGB star with a fast collimated wind.

   In the following, important theoretical and observational work from previous publications are summarized for each star giving an overview of their characteristic properties. Table~\ref{Table_Phenomenology} lists some of the relevant phenomenological features for each star. The given K band diameter, but also other optical and IR interferometric angular diameter measurements, are discussed in Sect.~\ref{secResSWaveDep}. The distances given in Table~\ref{Table_Phenomenology} are assumed throughout this paper.

%---------------------------------------------------------------
\begin{table*}
  \caption{Target properties and phenomenology.}
  \label{Table_Phenomenology}
  \centering
      \begin{tabular}{lccccc}
%        \noalign{\medskip}
        \hline
%        \noalign{\medskip}
        \noalign{\smallskip}
      Target name:                                  & \textbf{R~Aql} & \textbf{R~Aqr} & \textbf{R~Hya}   & \textbf{W~Hya} & \textbf{V~Hya} \\
                                                    &   HIP 93820    &  HIP 117054    &  HIP 65835       &  HIP 67419     &  HIP 53085     \\
        \noalign{\smallskip}
      Associated calibrator$^{\mathrm{a}}$:         & $\eta$ Sgr, $\epsilon$ Peg & 30 Psc & 2 Cen        & 2 Cen          &  $\alpha$ Hya  \\
        \noalign{\smallskip}
        \hline
%        \noalign{\smallskip}
        \noalign{\smallskip}           
      Type:                                         & Mira           & Mira           & Mira             & Mira/SRa       & SRa/L       \\
        \noalign{\smallskip}
      Chemistry:                                    & O-rich         & O-rich         & O-rich           & O-rich         & C-rich         \\
        \noalign{\smallskip}
      Mass$^{\mathrm{b}}$ ($M_{\odot}$):            & 1.0            & 1.5            & 2.0              & 1.0            & 1.0            \\
        \noalign{\smallskip}
      Distance$^{\mathrm{c}}$ (pc):                 & 214$^{+45}_{-32}$ & 250 $\pm$ 50 & 130 $\pm$ 25    & 98$^{+30}_{-18}$ & 360 $\pm$ 70 \\
        \noalign{\smallskip}
      12 $\mu$m flux$^{\mathrm{d}}$ (Jy):           & 402 $\pm$ 36   & 1580 $\pm$ 60  & 1590 $\pm$ 80    & 4200 $\pm$ 210 & 1110 $\pm$ 60  \\
        \noalign{\smallskip}
      Spectral classification$^{\mathrm{e}}$:       & M7 III e var   & M7 III pev     & M7 III e g       & M7 III e g     & N; C7,5e       \\
        \noalign{\smallskip}
      Effective temperature$^{\mathrm{b}}$ (K):     & 3000 $\pm$ 300 & 2800 $\pm$ 300 &  2700 $\pm$ 300  & 2500 $\pm$ 300 & 2650 $\pm$ 300 \\
        \noalign{\smallskip}
      Luminosity$^{\mathrm{b}}$ ($L_{\odot}$):      & 3470 $\pm$ 500 & 5000 $\pm$ 500 & 11600 $\pm$ 1000 & 5400 $\pm$ 500 & 7850 $\pm$ 500 \\
        \noalign{\smallskip}
      K-band diameter$^{\mathrm{f}}$ (mas):         & 11.5 $\pm$ 1.4 & 16.6 $\pm$ 3.0 &  24.5 $\pm$ 1.2  & 42.8 $\pm$ 3.2 & 14.5 $\pm$ 0.3 \\
        \noalign{\smallskip}
      Mass-loss rate$^{\mathrm{g}}$ ($10^{-7}$~$M_\odot$yr$^{-1}$):  & 6 -- 35  & 0.1 -- 9  & 0.1 -- 2  & 0.8 -- 5  & 10 -- 600   \\
        \noalign{\smallskip}
      Multiplicity$^{\mathrm{b}}$:                  & triple system  & SB             & wide binary      & wide binary    & wide binary    \\
        \noalign{\smallskip}
      Evolutionary stage$^{\mathrm{b}}$:            & TP-AGB         & TP-AGB         & TP-AGB           & early-AGB      & post-AGB       \\
        \noalign{\smallskip}
      Jets/fast wind$^{\mathrm{b}}$:                & no             & yes            & no               & no             & yes            \\
        \noalign{\smallskip}
      Asymmetry$^{\mathrm{b}}$:                     & maybe          & yes            & no               & yes            & yes            \\
        \noalign{\smallskip}
      Maser$^{\mathrm{b}}$:                         & SiO, \ce{H2O}, OH & SiO, \ce{H2O}, OH & SiO, \ce{H2O}, OH & SiO, \ce{H2O}, OH & CO ? \\
        \noalign{\smallskip}
      Period$^{\mathrm{b}}$:                        & decreasing     & stable         & decreasing       & stable         & two periods    \\
        \noalign{\smallskip}
      Others$^{\mathrm{b}}$:                        & recent         & symbiotic      & detached         & --             & common         \\
                                                    & He flash       & system         & shell, recent    &                & envelope,      \\
                                                    &                & (Mira+WD)      & He flash         &                & rapid rotator  \\
        \noalign{\smallskip}
        \hline
%        \noalign{\smallskip}
      \end{tabular}
  \newline
  \begin{flushleft}
    \textbf{Notes. }
    $^{\mathrm{a}}$~For calibrator properties see Sect.~\ref{secObsSubInt} and Table~\ref{Table_Calibrators}. 
    $^{\mathrm{b}}$~See text for references, errors are approximations derived from the values found in the literature. 
    $^{\mathrm{c}}$~For R~Aqr, R~Hya and V~Hya derived from the period-luminosity relation \citep[cf.~][]{Whitelock2008}, and for R~Aql and W~Hya obtained from maser measurements \citep[][respectively]{Kamohara2010,Vlemmings2003}. 
    $^{\mathrm{d}}$~IRAS flux \citep{Neugebauer1984}; 
    $^{\mathrm{e}}$~SIMBAD; 
    $^{\mathrm{f}}$~Values are averages from the literature (cf.~Sect.~\ref{secResSWaveDep}); 
    $^{\mathrm{g}}$~Range of current mass-loss rates as given in Sect.~\ref{secObsSubTar}. For W~Hya see e.g.~\citet{DeBeck2010}, \citet{Muller2008} and \citet{Justtanont2005}.
  \end{flushleft}
\end{table*}
%---------------------------------------------------------------

\subsection{R~Aquilae}\label{secPropSSubRAql}

   The visual period of the Mira R~Aql has declined quite dramatically with an average rate of approximately 0.4~days per cycle since 1900 \citep{Greaves98,GreavesHowarth2000}. While in 1915 the period was about 320~days, it declined to about 264~days in 2010. This has been accompanied by a decrease in amplitude of about 1~mag \citep{Bedding2000}. The shrinking of the period is attributed to a recent thermal pulse \citep{WoodZarro81}. R~Aql is listed in the Washington Visual Double Star Catalog \citep[WDS][]{Mason2001} as triple system. However, the two companion stars have dissimilar proper motions and are probably only optical companions \citep{GreavesHowarth2000}.

   The distance to R~Aql is relatively well known. The Hipparcos catalog \citep{Perryman1997b} gives a value of 211$^{+71}_{-42}$~pc, while \citet{Kamohara2010} estimated the distance to 214$^{+45}_{-32}$~pc via maser observations. The mass and luminosity, obtained from modeling and observations, are 1~$M_\odot$ and 3470~$L_\odot$, respectively \citep{Hofmann2000}. Pulsation phase dependent effective temperature determinations range from (2550~$\pm$~150)~K \citep{Haniff1995}, over 3072~$\pm$~161~K \citep{Hofmann2000} to 3198~$\pm$~147~K \citep{vanBelle1996}. From the relations between pulsation period, mass and linear radius, \citet{Hofmann2000} came to the result that R~Aql is a fundamental mode pulsator in agreement with recent considerations.

   SiO, \ce{H2O} and OH maser emissions have been monitored for this O-rich star. SiO maser emission was reported by \citet{Benson1990}, \citet{Pardo2004} and \citet{Cotton2010}. The latter author derived for the 43.1 and 42.8~GHz emission average ring diameters of (26.1~$\pm$~1.5)~mas and (22.3~$\pm$~1.9)~mas, respectively, in agreement with the supposed extended atmosphere. At a larger diameter of around 330~mas, \citet{Lane1987} and \citet{Brand1994} detected \ce{H2O} maser emission. Studies of OH masers were performed by \citet{Bowers1989}, \citet{EtokaLeSqueren2000} and \citet{He2005}, showing that this maser emission originates in a region with a diameter between 2000 and 3000~mas. 

   \citet{Matsuura2002} modeled the extended molecular spheres of R~Aql with two layers of water vapor of different temperatures. The radius of the hot layer changes from 1 to 2~photospheric radii between visual minima and maxima. The model could explain the emission and absorption features seen in the near infrared ($2.5-4.0$~$\mu$m). While \citet{Cotton2010} could not identify strong asymmetries in the SiO maser distribution, \citet{Lane1987} and \citet{Bowers1989} saw highly elongated and complex structures in the \ce{H2O} (NW-SE oriented) and OH masers (NE-SW oriented), respectively. In addition, the interferometric measurements by \citet{Ragland2006} gave a significantly non-zero closure phase, indicating an asymmetry as well.

   Total mass-loss rates were estimated to 8~$\times$~$10^{-7}$~$M_{\odot}$yr$^{-1}$ and 5.6~$\times$~$10^{-7}$~$M_{\odot}$yr$^{-1}$ by \citet{Gehrz1971} and \citet{Hagen1982}, respectively, whereas \citet{KnappMorris1985} found a much higher value of 3.5~$\times$~$10^{-6}$~$M_{\odot}$yr$^{-1}$. The dust mass-loss rate was derived by \citet{Hagen1982} with a value of 6~$\times$~$10^{-8}$~$M_{\odot}$yr$^{-1}$. Wind velocities are measured in a range from 7 to 10~km/s \citep[and references therein]{Bowers1989}.

\subsection{R~Aquarii}\label{secPropSSubRAqr}

   R~Aqr is the closest known symbiotic binary at a distance of about 250~pc \citep[derived from the period-luminosity relation by][]{Whitelock2008}. This D-type (dusty) symbiotic system consists of a $1.0-2.0$~$M_\odot$ Mira variable and a $0.6-1.0$~$M_\odot$ white dwarf~(WD) \citep{Hollis1997,Boboltz1997,Tatebe2006,GromadzkiMikolajewska2009} that accretes matter through a disk \citep[e.g.][]{Hollis2000}. The R~Aqr binary system was for the first time resolved in the continuum radio emission at 7~mm by \citet{Hollis1997}. \citet{Hollis1997} and \citet{GromadzkiMikolajewska2009} derived orbital solutions with a period of about 44~years, while \citet{McIntoshRustan2007} estimated the orbit period to be 34.6~years. The fitted semi-major axis is in all cases on the order of 15~AU (60~mas), meaning that even during periastron passage, the Mira variable remains relatively far from filling the Roche lobe \citep{GromadzkiMikolajewska2009}.

   The binary system hosts a compact H\textsc{ii} region within a filamentary oval nebula of 30~arcsec size \citep[e.g.][]{Kafatos1986}. It is surrounded by a large and expanding hour-glass shaped nebulosity with an extension of at least 120~arcsec \citep{WallersteinGreenstein1980,Hollis1985,HenneyDyson1992,Corradi1999}. The accretion disk, formed around the compact component, gives rise to prominent jets, detected across all spectral domains \citep[cf.~e.g.][]{Kellogg2007,NicholsSlavin2009,Hollis1991,ParesceHack1994,Hollis1985}. The symmetric jets, oriented along a NE-SW axis, extend up to 2500~AU (10~arcsec) and have expansion velocities of 90 to 200~km/s.

   The current mass-loss rate of this star is rather low, but uncertainties are very high due to difficulties in obtaining reliable gas-to-dust ratios. Rate estimates range from 1.3~$\times$~$10^{-8}$~$M_{\odot}$yr$^{-1}$ \citep{HenneyDyson1992} and 6~$\times$~$10^{-8}$~$M_{\odot}$yr$^{-1}$ \citep{MatthewsReid2007} to 3~$\times$~$10^{-7}$~$M_{\odot}$yr$^{-1}$ \citep{Mennesson2002} and 8.9~$\times$~$10^{-7}$~$M_{\odot}$yr$^{-1}$ \citep{Danchi1994}. In most calculations an effective temperature of 2800~K \citep{Burgarella1992,MatthewsReid2007} and a luminosity of 5000~$L_\odot$ \citep{Ragland2008} were assumed.

   The O-rich Mira variable R~Aqr is one of only three among 48 symbiotic Miras that exhibits \ce{H2O} masers \citep{Ivison1994,Ivison1998,Whitelock2003}. A complementary study of SiO masers showed strong emission at 42.8 and 43.1~GHz. SiO maser ring diameters were obtained by \citet{Boboltz1997}, \citet{Hollis2001} and \citet{Cotton2004,Cotton2006}, with relatively consistent values between 30 and 33~mas and errors on the order of 1.5~mas. This corresponds to a radius of 1.9~photospheric radii. OH maser and CO thermal lines are very weak and only tentative detections have been published by \citet{Ivison1998} and \citet{Groenewegen1999}, respectively. In the most popular model, UV radiation and a fast wind from the companion remove the outer envelope of dusty molecular gas, where an OH maser or a thermal CO line could originate \citep{Ivison1998}.

   \citet{Ragland2008} modeled the star with a three-component model, consisting of a symmetric central star surrounded by a water vapor shell with a radius of about 2.25~photospheric radii, and an off-axis compact feature at about 2~photospheric radii at a position angle of 131$^\circ$. They concluded that the observations are best explained with a clumpy, extended \ce{H2O} circumstellar envelope. In this model the SiO masers appear at the outer edge of the molecular envelope, as reported for other Mira stars.

   Possibly caused by interactions with the close companion, asymmetries could be identified in the close environment around R~Aqr. \citet{Ragland2006,Ragland2008} found non-zero closure phases, and \citet{Hollis2001} and \citet{Cotton2004,Cotton2006} detected an asymmetric distribution of the SiO maser emission with position angles between 150$^\circ$ and 180$^\circ$. Terminal wind velocity measurements are rare. \citet{KotnikKaruza2007} obtained a value of $6-7$~km/s, while \citet{Dougherty1995} assumed $10-30$~km/s.

\subsection{R~Hydrae}\label{secPropSSubRHya}

   The O-rich variable star R~Hya is well known for its declining period \citep{WoodZarro81,Zijlstra2002} and the presence of a detached shell. In the past few centuries, the period of R~Hya has declined by over a hundred days and has remained constant since 1950 \citep{Zijlstra2002}. The steady decrease in the period can be possibly attributed to a recent thermal pulse \citep{WoodZarro81,Zijlstra2002}. In the post thermal-pulse evolution the decline in luminosity causes a reduction in stellar radius, which in turn, causes the period to become shorter.

   The detached shell observed around R~Hya \citep{Young1993,Hashimoto1998} indicates a change in the mass-loss rate some 220 years ago. The shell is located about 1.9~arcsec from the star ($\approx$~250~AU). While the mass-loss rate before 1770 is estimated to be between 1.5~$\times$~$10^{-7}$~$M_{\odot}$yr$^{-1}$ and 3~$\times$~$10^{-7}$~$M_{\odot}$yr$^{-1}$ \citep{Hashimoto1998,Zijlstra2002}, the present day mass-loss rate is a factor of $\approx$~20 lower and between 1~$\times$~$10^{-8}$~$M_{\odot}$yr$^{-1}$ and 4~$\times$~$10^{-8}$~$M_{\odot}$yr$^{-1}$ \citep[][respectively]{Decin2008,Teyssier2006}. \citet{DeBeck2010} gives an average mass-loss rate of 1.6~$\times$~$10^{-7}$~$M_{\odot}$yr$^{-1}$.

   This derived mass-loss history nicely agrees with the period decline analyzed by \citet{Zijlstra2002}. Even the stellar evolution tracks calculated by \citet{VassiliadisWood1993} confirmed that mass-loss fluctuations during the thermal pulse cycle can lead to detached circumstellar shells. An apparent large detached shell may also arise from the interaction of the AGB wind with the ISM. \citet{Ueta2006} and \citet{Wareing2006} detected a far-IR nebula at a distance of about 100~arcsec to the west. This is explained by a slowing-down of the stellar wind by surrounding matter. Therefore, no extra mass-loss modulation at around 100~arcsec, i.e.~around 10,000 years ago, needs to be invoked \citep{Decin2008}.

   Distance estimations for R~Hya range from (110~$\pm$~21)~pc \citep{JuraKleinmann1992} to 165~pc \citep{Zijlstra2002}, while the period-luminosity relation gives an intermediate value of 130~pc \citep{Whitelock2008}. The mass and luminosity, inferred from modeling and observations, are 2~$M_\odot$ and 11,600~$L_\odot$, respectively \citep{Zijlstra2002}. Effective temperature determinations range from 2600~K \citep{Teyssier2006}, over 2680~$\pm$~70~K \citep{Haniff1995} to 2830~K \citep{Zijlstra2002}. R~Hya is believed to be a wide binary system with a very long orbital period. The angular separation is 21~arcsec (WDS).

   The wind in the inner CSE starts with a velocity of 1.5~km/s and accelerates to 6.5~km/s farther out, as modeled by \citet{Teyssier2006}, before reaching a terminal velocity of 7.5~km/s \citep{Justtanont1998} to 10.0~km/s \citep{Hashimoto1998,Knapp1998}. \citet{Justtanont1998} found that a 13~$\mu$m dust feature and the appearance of strong emission lines of \ce{CO2} originate in a warm layer close to the star. R~Hya exhibits maser emission of SiO \citep{Humphreys1997}, \ce{H2O} \citep{Takaba2001} and OH \citep{Lewis1995}. The resolution of these observations were too low to derive any spatial information. Thus, asymmetries were not studied with masers. A morphology study in the visual \citep[$0.7-1.0$~$\mu$m,][]{Ireland2004a} and K-band \citep{Monnier2004} gave no indications of a departure from symmetry.

\subsection{W~Hydrae}\label{secPropSSubWHya}

   The main characteristics of the O-rich star W~Hya were extensively described in paper~I, but are summarized in Table~\ref{Table_Phenomenology}.

\subsection{V~Hydrae}\label{secPropSSubVHya}

   V~Hya is a classical (N~type) carbon star \citep{Zuckerman1977} and is believed to be in a short-lived but critical stage in the evolution of a mass-losing AGB star into a bipolar PN \citep{Tsuji1988,Kahane1996}. This dust-enshrouded star has a C/O~ratio of 1.05 and an effective temperature of about 2650~K \citep{Lambert1986}. The luminosity is estimated to be 7850~$L_\odot$ \citep{Knapp1997}. Distance calculations range from 340~pc \citep{Barnbaum1995} and 380~pc \citep{Knapp1997} to 440~pc \citep{Olivier2001} and 550~pc \citep{Bergeat1998}, while the period-luminosity relation \citep{Whitelock2008} gives a value at the lower end with 360~pc.

   V~Hya has two variability periods, (529.4~$\pm$~30)$\,$d with a peak-to-peak variation of 1.5~mag and (6160~$\pm$~400)$\,$d ($\approx$~17~years) with a peak-to-peak variation\footnote{The long secondary pulsation amplitudes in JHKL are 2.4, 2.1, 1.7, and 0.7 mag, respectively \citep{Olivier2001}.} of 3.5~mag, and is classified as semi-regular (SRa) by \citet{Mayall1965} and Mira variable by \citet{Knapp1999}. While the 529.4~d period is typical of a luminous AGB star, \citet{Knapp1999} suggests that the regular long period dimming of V~Hya is due to a thick dust cloud orbiting the star.

   The rotation velocity of an AGB star, evolving in isolation, is not likely to be greater than 2~kms$^{-1}$, even if its main-sequence progenitor rotated at breakup velocity. In contrast, the rotation velocity, $v$~sin$\,i$, derived for V~Hya from a high resolution optical spectral broadening analysis by \citet{Barnbaum1995} is on the order of 11 to 14~kms$^{-1}$. \citet{Barnbaum1995} and \citet{Kahane1996} concluded that this rapid rotation is due to the spin-up by a companion in a common envelope configuration, but alternative explanations have also been proposed \citep{LuttermoserBrown1992,OlivierWood2003}. The secondary star could be an early G or F~star, implying a mass of $1.0-1.5$~$M_\odot$, or possibly a WD \citep{Barnbaum1995}. V~Hya is in contrast inferred to have a mass of about 1~$M_\odot$ \citep{Kahane1996}. In addition, a wide companion, with a separation of 46~arcsec, exists (WDS).

   A consequence of the fast rotation is an enhanced equatorial mass loss producing a disk and a jet-like structure \citep[e.g.][]{Soker1992,Morris1987}. High angular resolution CO maps, millimeter, infrared and optical spectra suggest that the circumstellar structure of V Hya consists of three kinematic components: a low-velocity disk with a radial velocity of $\Delta v$~$\approx$~$\pm$~(8~to~16)~km/s, a intermediate-velocity wind with $\Delta v$~$\approx$~$\pm$~60~km/s, and a high-velocity jet with $\Delta v$~$\approx$~$\pm$~(60~to~200)~km/s \citep{ZuckermanDyck1986,SahaiWannier1988,Lloyd1991,Knapp1997,Olivier2001,Hirano2004,Sahai2009}.

   The low-velocity circumstellar environment component is flattened or has a disk-like shape, and is elongated along the north-south direction \citep{Tsuji1988,Kahane1988,Kahane1996,Sahai2003}. This may enable or enhance the formation of an accretion disk and supports a model in which the jet is driven by an accretion disk around an unseen, compact companion. Images obtained at 9.8 and 11.7~$\mu$m by \citet{Lagadec2005} show an additional slightly elongated structure in the south-west direction, tracing the dust emission from material blown away, while the overall structure is roughly spherically symmetric.

   The star has a large infrared excess and strong millimeter molecular line emission, showing that it is losing mass at a fairly high rate. Total mass-loss rates were estimated to 6.1~$\times$~$10^{-5}$~$M_{\odot}$yr$^{-1}$ \citep{DeBeck2010}, 2.5~$\times$~$10^{-5}$~$M_{\odot}$yr$^{-1}$ \citep{Knapp1997},  1.5~$\times$~$10^{-6}$~$M_{\odot}$yr$^{-1}$ \citep{Kahane1996,Knapp2000}, and 1.0~$\times$~$10^{-6}$~$M_{\odot}$yr$^{-1}$ \citep{Barnbaum1995} with dust mass-loss rates of 2.0~to~5.7~$\times$~$10^{-8}$~$M_{\odot}$yr$^{-1}$ \citep[][respectively]{Knapp1997,Hirano2004}. \citet{Knapp1997} modeled the dust envelope of V~Hya by assuming grains consisting of amorphous carbon with dimensions of 0.2~$\mu$m. \citet{ZuckermanDyck1986} discovered the presence of a narrow CO emission feature superposed on a standard broad stellar CO profile, which probably represents the first example of a CO maser ever seen in any interstellar or circumstellar source.

%###########################################################################################
%###########################################################################################

\section{Observations and data reduction}\label{secObsData}

%%%%%%%%%%%%%%%%%%%%%%%%%%%%%%%%%%%%%%%%%%%%%%%%%%%%%%%%%%%%%%%%%%%%%%%%%%%%%%%%%%%%%%%%%%%%
\subsection{Interferometric observations with MIDI/VLTI}\label{secObsSubInt}

   The data presented here were obtained with the mid-IR ($8-13$~$\mu$m) interferometer MIDI \citep{Leinert2003,Leinert2004} at the Very Large Telescope Interferometer (VLTI) in service mode using the Auxiliary Telescopes (ATs). All five stars were monitored from P79 to P83 (April~2007 to September~2009) under the program IDs 079.D-0140, 080.D-0005, 081.D-0198, 082.D-0641 and 083.D-0294 in GTO time. An overview of the course of observations is given in Table~\ref{Table_ObsOverview}. It should be noted that observations were still ongoing at the time of analyzing the data, and that observations not executed in a specific semester are shifted to the next semester. A complete observation log is given in Table~\ref{TableApp_log} in the appendix. Projected baselines range from 11 to 71~m and the position angles (PA, $\vartheta$; East of North) are differently distributed for each star. The uv-coverages are shown later in the left hand panels of Fig.~\ref{FigUVdia}.

%---------------------------------------------------------------
\begin{table*}
\caption{Number of visibility measurements used in this study (see Appendix~\ref{App_ObsLog} for a detailed observation log and paper~I in the case of W~Hya).}
    \label{Table_ObsOverview}
    \centering
    \begin{tabular}{cccccccc}
%            \noalign{\smallskip}
%            \noalign{\smallskip}
            \hline
            \noalign{\smallskip}
%            \noalign{\smallskip}           
            Semester & Program ID & Disperser & R Aql & R Aqr & R Hya & W Hya & V Hya\\
            \noalign{\smallskip}
            \hline
            \noalign{\smallskip}
%            \noalign{\smallskip}           
            79$^{\mathrm{ }}$ (1$^{\mathrm{st}}$~Apr, 2007 -- 30$^{\mathrm{th}}$~Sep, 2007) & 079.D-0140 & GRISM & 11 & 10 & 18 & 22 &  8 \\
            80$^{\mathrm{ }}$ (1$^{\mathrm{st}}$~Oct, 2007 -- 31$^{\mathrm{st}}$~Mar, 2008) & 080.D-0005 & PRISM &  6 & 10 & 16 & 17 & 26 \\
            81$^{\mathrm{ }}$ (1$^{\mathrm{st}}$~Apr, 2008 -- 30$^{\mathrm{th}}$~Sep, 2008) & 081.D-0198 & PRISM & 15 &  7 & 16 & 18 &  8 \\
            82$^{\mathrm{a}}$ (1$^{\mathrm{st}}$~Oct, 2008 -- 31$^{\mathrm{st}}$~Mar, 2009) & 082.D-0641 & PRISM &  0 &  4 & 12 & 14 & 20 \\
            83$^{\mathrm{a}}$ (1$^{\mathrm{st}}$~Apr, 2009 -- 30$^{\mathrm{th}}$~Sep, 2009) & 083.D-0294 & PRISM &  8 &  8 & 10 & 12 &  1 \\
            \noalign{\smallskip}
            \hline
    \end{tabular}
  \newline
  \begin{flushleft}
    \textbf{Notes. }
    $^{\mathrm{a}}$~The observation program was still ongoing at the time of analyzing the data.
  \end{flushleft}
\end{table*}
%---------------------------------------------------------------

   Before or after each target observation a calibrator star is observed with the same setup in order to calibrate the visibilities and fluxes. The properties of the calibrator stars are listed in Table~\ref{Table_Calibrators}. Necessary input parameters for the calibration are the angular diameter \citep[model diameter from][]{Verhoelst2005}\footnote{http://www.ster.kuleuven.ac.be/$\sim$tijl/MIDI\_calibration/mcc.txt;~see also ESO CalVin database: http://www.eso.org/observing/etc/} and the IRAS\footnote{http://irsa.ipac.caltech.edu/Missions/iras.html} 12~$\mu$m flux. The second row lists the associated targets for which the calibrator is used. The angular separation between calibrator and target is given in brackets and shows that they are located relatively far away from each other as a consequence of the low number of available calibrators. However, this is fortunately not a big concern in the mid-infrared. In particular, the visibility calibration is not affected by different airmasses since both, the correlated and uncorrelated flux, changes proportionally with it. Nevertheless, observations need to be carried out under good seeing conditions (typical $\leq$~1.4$''$) and a clear sky. Both are necessary to have a long coherence time to obtain fringes and to reduce the infrared background.

   In order to identify molecules and dust species, the mid-infrared fringes are spectrally dispersed. The high spectral resolution GRISM mode of MIDI, with $R = \lambda/\Delta\lambda = 230$, would be therefore of advantage, but introduces additional problems in the reduction process (e.g.~the photometric channels are unfavorably illuminated). Hence, mainly the PRISM with a spectral resolution of 30 was used. Since the targets are bright enough, observations were executed in \texttt{SCI-PHOT} mode, where the photometric and the interferometric spectra are recorded simultaneously. This has the advantage that the photometric spectrum and the fringe signal are observed under the same atmospheric conditions.

%---------------------------------------------------------------
\begin{table*}
  \caption{Calibrator properties.}
  \label{Table_Calibrators}
  \centering
  %\footnotesize 
      \begin{tabular}{lccccc}
%        \noalign{\smallskip}
        \hline
%        \noalign{\smallskip}
        \noalign{\smallskip}
      Calibrator name:                  &\textbf{$\eta$ Sgr} &\textbf{$\epsilon$ Peg}&\textbf{30 Psc}    &\textbf{2 Cen}     &\textbf{$\alpha$ Hya}\\
                                        &\object{HD 167618}  &\object{HD 206778}     &\object{HD 224935} &\object{HD 120323} &\object{HD 81797}    \\
        \noalign{\smallskip}
      Associated target$^{\mathrm{a}}$: &R~Aql (46.4$^\circ$)&R~Aql (42.8$^\circ$)&R~Aqr (10.3$^\circ$) &W~Hya (6.1$^\circ$) &V~Hya (23.8$^\circ$) \\
                                        &                    &                    &                     &R~Hya (12.0$^\circ$)&                     \\
        \noalign{\smallskip}
        \hline
%        \noalign{\smallskip}           
        \noalign{\smallskip}
      Spectral classification$^{\mathrm{b}}$: & M2 III           & K2 Ib var       & M3 III          & M5 III           & K3 III            \\
%        \noalign{\smallskip}
      Parallax$^{\mathrm{b}}$ (mas):          & 21.87 $\pm$ 0.92 & 4.85 $\pm$ 0.84 & 7.86 $\pm$ 0.94 & 18.39 $\pm$ 0.74 & 18.40 $\pm$ 0.78  \\
%        \noalign{\smallskip}
      12 $\mu$m flux$^{\mathrm{c}}$ (Jy):     & 214 $\pm$ 9      & 104 $\pm$ 5     & 87 $\pm$ 4      & 255 $\pm$ 26     & 158 $\pm$ 10      \\
%        \noalign{\smallskip}
      Angular diameter$^{\mathrm{c}}$ (mas):  & 11.66 $\pm$ 0.04 & 7.59 $\pm$ 0.05 & 7.24 $\pm$ 0.03 & 13.25 $\pm$ 0.06 & 9.14 $\pm$ 0.05   \\
%        \noalign{\smallskip}
      Quality flag$^{\mathrm{d}}$:            & 1                & 1               & 1               & 2                & 1                 \\
        \noalign{\smallskip}
        \hline
%        \noalign{\smallskip}
      \end{tabular}
  \newline
  \begin{flushleft}
    \textbf{Notes. }
    $^{\mathrm{a}}$~For target properties see Table~\ref{Table_Phenomenology}. The angular separation between calibrator and target is given in brackets. 
    $^{\mathrm{b}}$~Hipparcos \citep{Perryman1997b}; 
    $^{\mathrm{c}}$~IRAS flux and model diameter from \citet{Verhoelst2005}, http://www.ster.kuleuven.ac.be/$\sim$tijl/MIDI{\_}calibration/mcc.txt; 
    $^{\mathrm{d}}$~ESO CalVin database, http://www.eso.org/observing/etc/
  \end{flushleft}
\end{table*}
%---------------------------------------------------------------

%%%%%%%%%%%%%%%%%%%%%%%%%%%%%%%%%%%%%%%%%%%%%%%%%%%%%%%%%%%%%%%%%%%%%%%%%%%%%%%%%%%%%%%%%%%%
\subsection{MIDI \texttt{SCI-PHOT} data reduction}\label{secObsSubRed}

    The standard \texttt{MIA+EWS}\footnote{http://www.strw.leidenuniv.nl/$\sim$nevec/MIDI} (version~1.6) data reduction package with additional routines for processing \texttt{SCI-PHOT} data (Walter Jaffe, private communication) was used. Measurements at wavelengths beyond 12.0~$\mu$m were excluded due to too low fluxes of the calibrator stars in that wavelength regime and therefore difficulties to determine the signal in the presence of a high infrared background\footnote{Remaining flux, coming from short term variations of the high sky background and instrumental imperfections, not subtracted by the un-chopping has to be determined and removed. The reduction routines have to distinguish between sky, tunnel and source flux, and have to know the slit position. This changes with AT position, projected baseline and pointing. If the flux of the calibrator is too low, its flux can not be reliably recovered and the calibration process fails.}. The remaining wavelength range from 8 to~12~$\mu$m is then binned into 25 wavelength bins. A detailed description of the reduction process and how the errors are derived are given in paper~I and \citet{ZhaoGeisler2010}. In the end, 32~of~40, 26~of~39, 64~of~72, 75~of~83 and 48~of~63 observations could be adequately reduced for R~Aql, R~Aqr, R~Hya, W~Hya and V~Hya, respectively. Rejected are observations where the reduction process failed or unphysical visibilities arose due to bad environmental conditions.

%###########################################################################################
%###########################################################################################

\section{Light curves, spectra and visibility modeling results}\label{secLCSpecVis}

%%%%%%%%%%%%%%%%%%%%%%%%%%%%%%%%%%%%%%%%%%%%%%%%%%%%%%%%%%%%%%%%%%%%%%%%%%%%%%%%%%%%%%%%%%%%
\subsection{Light curves}\label{secObsSubLC}

%---------------------------------------------------------------
   \begin{figure*} %5stars: 0.47 and 0.43 and \hspace{0.2cm}
    \centering
    \includegraphics[width=0.52\linewidth]{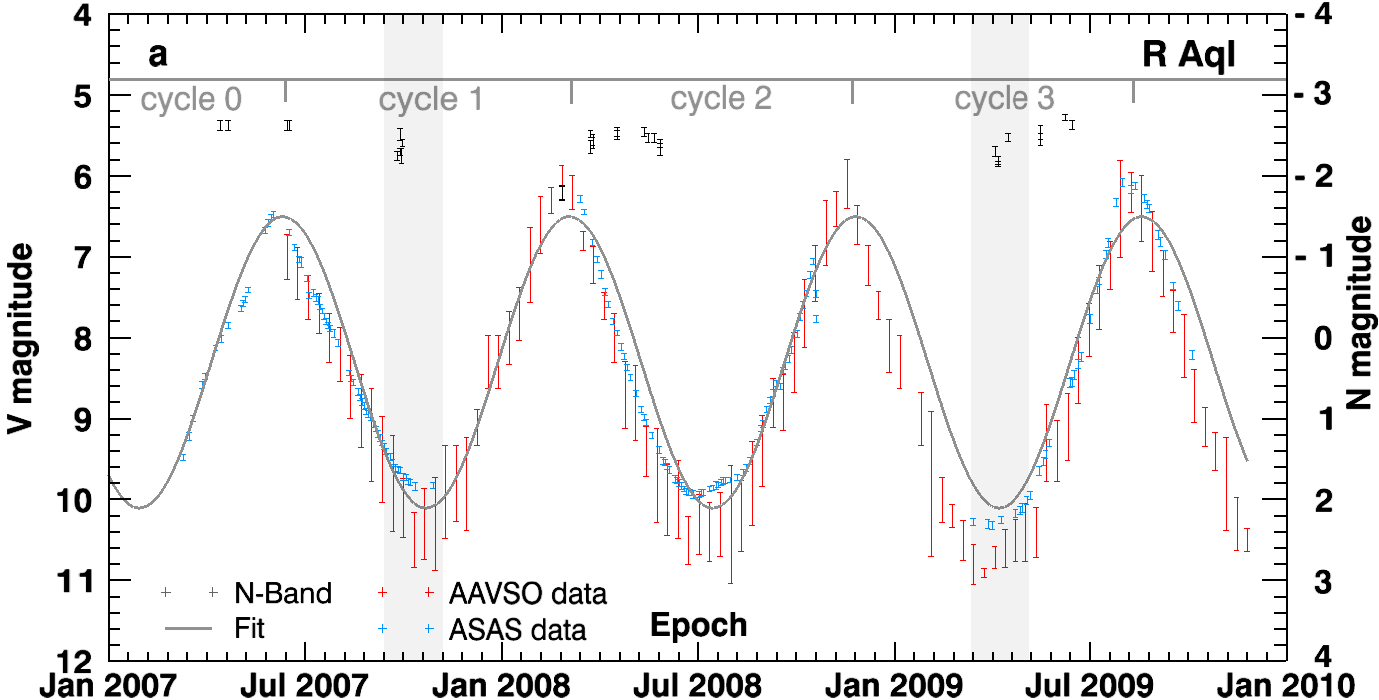}
    %\hspace{0.2cm}
    \includegraphics[width=0.47\linewidth]{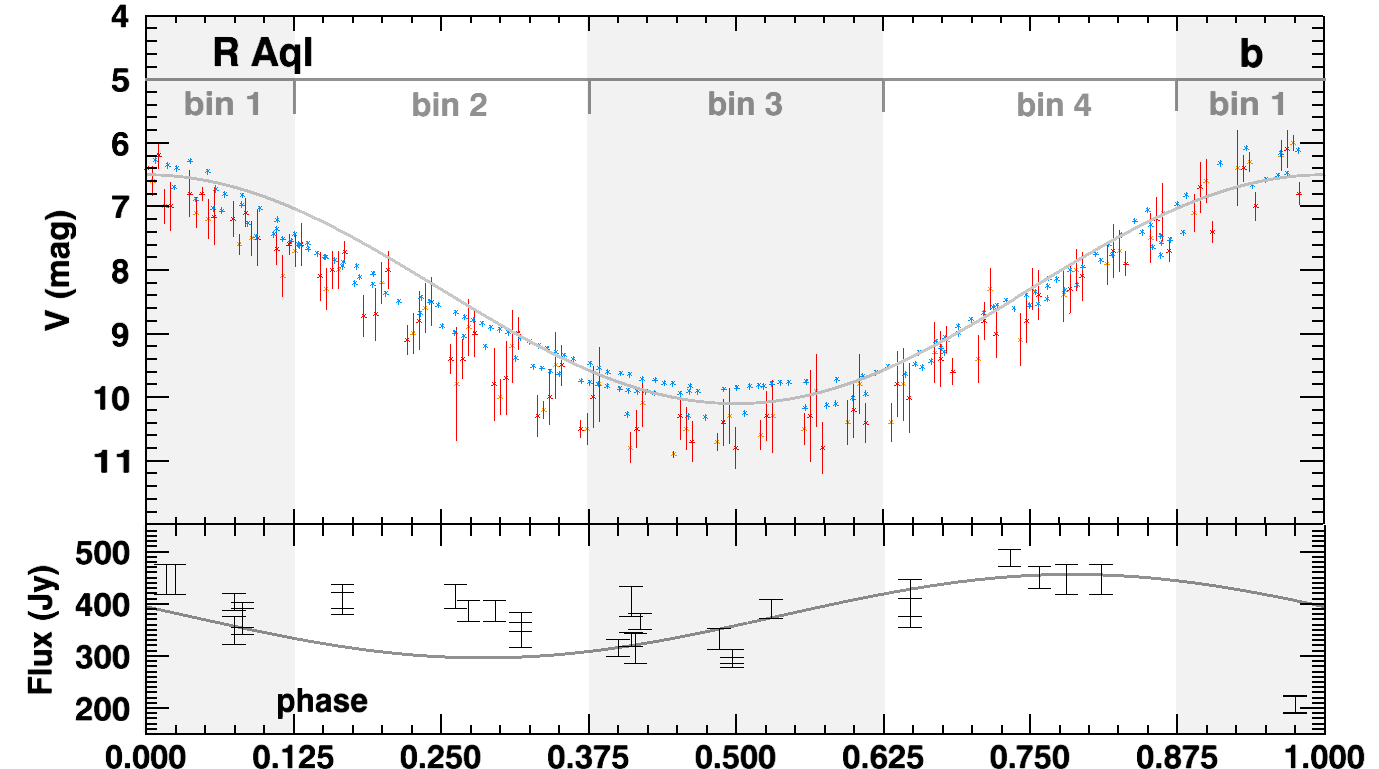}
    \includegraphics[width=0.52\linewidth]{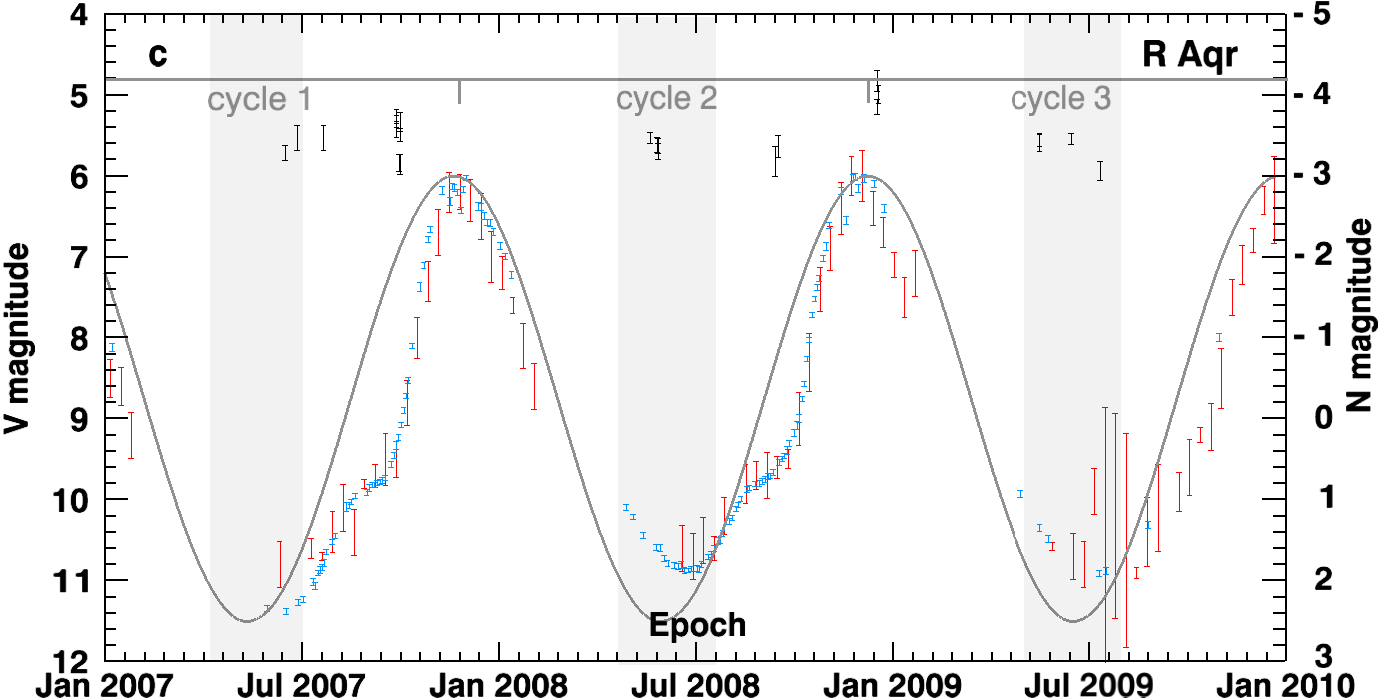}
    %\hspace{0.2cm}
    \includegraphics[width=0.47\linewidth]{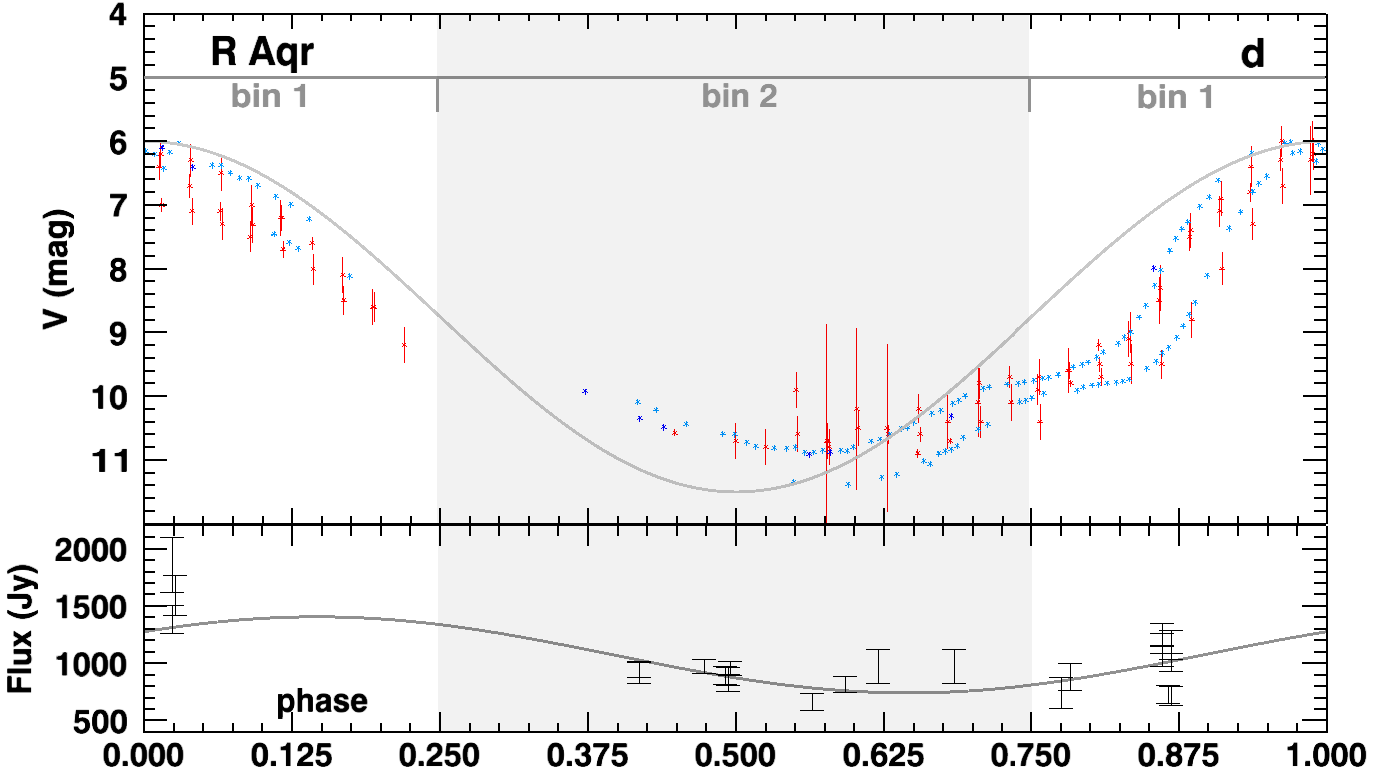}
    \includegraphics[width=0.52\linewidth]{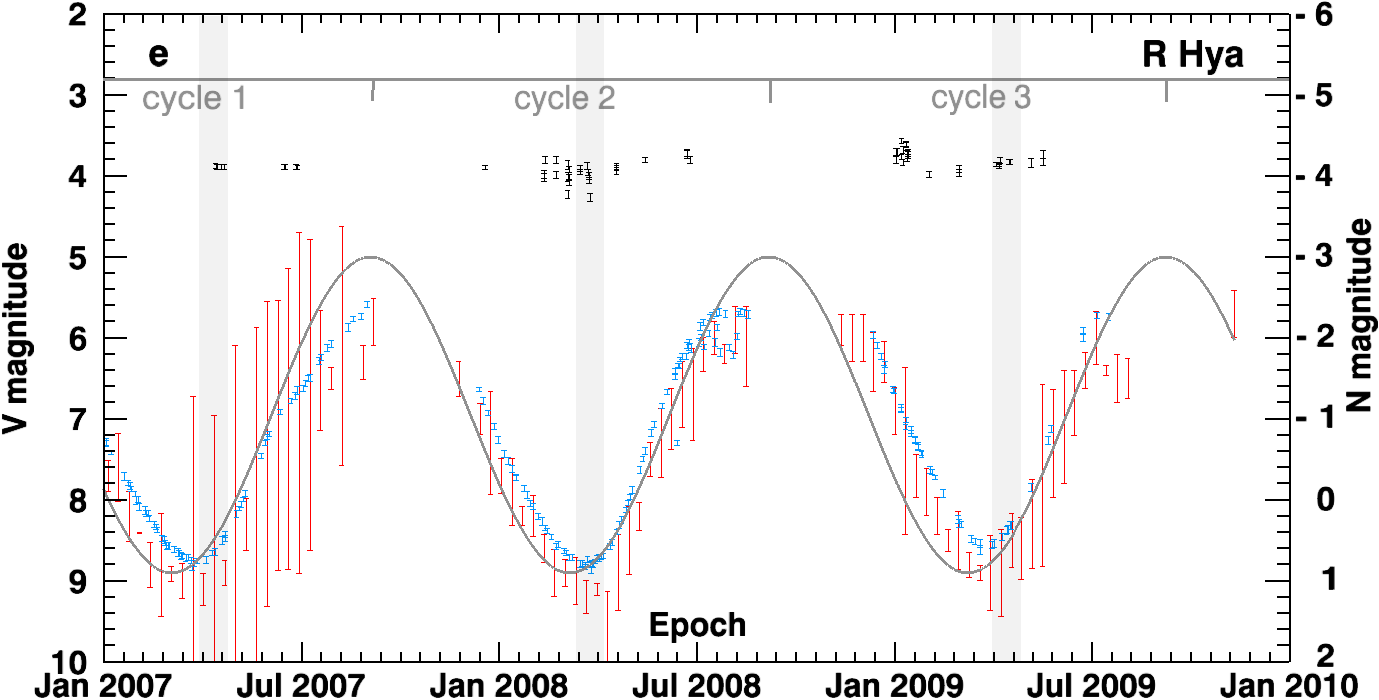} 
    %\hspace{0.2cm}
    \includegraphics[width=0.47\linewidth]{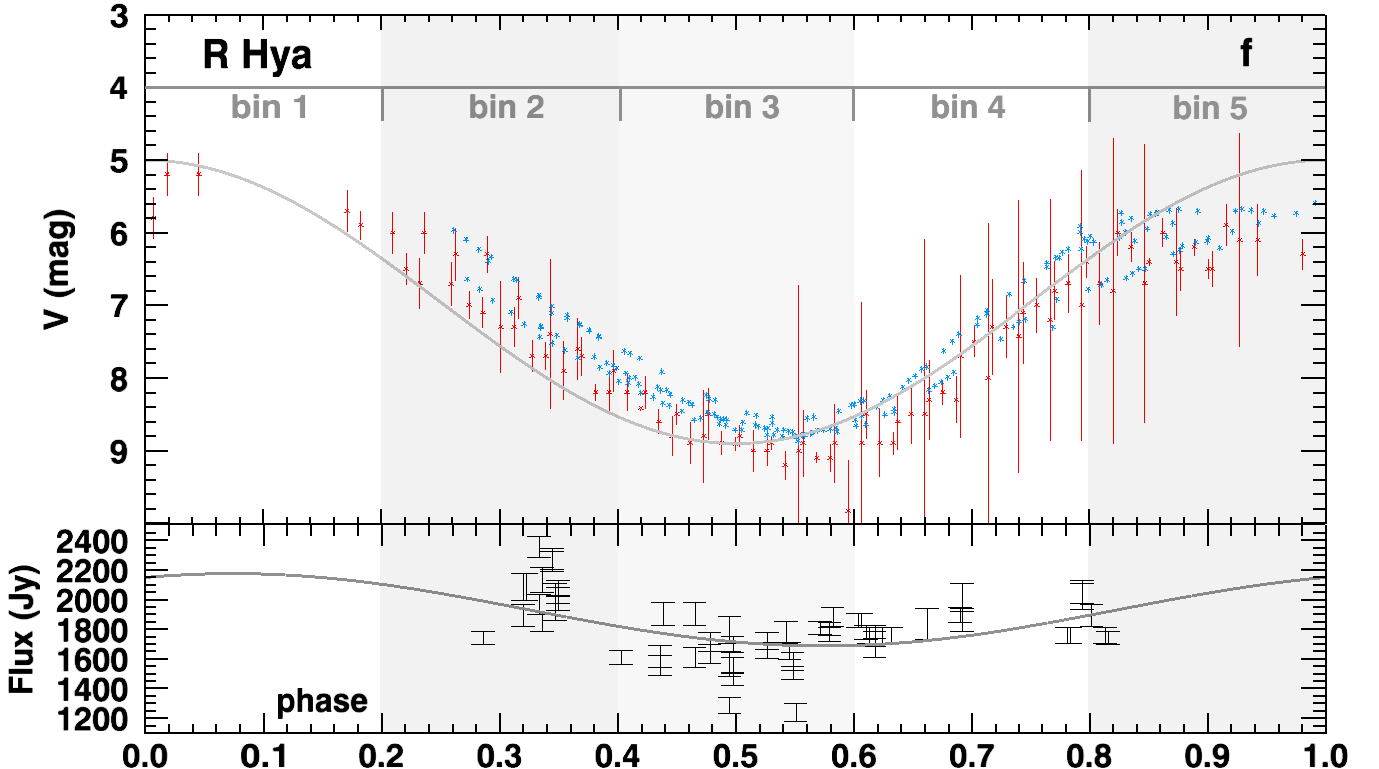}
    \includegraphics[width=0.52\linewidth]{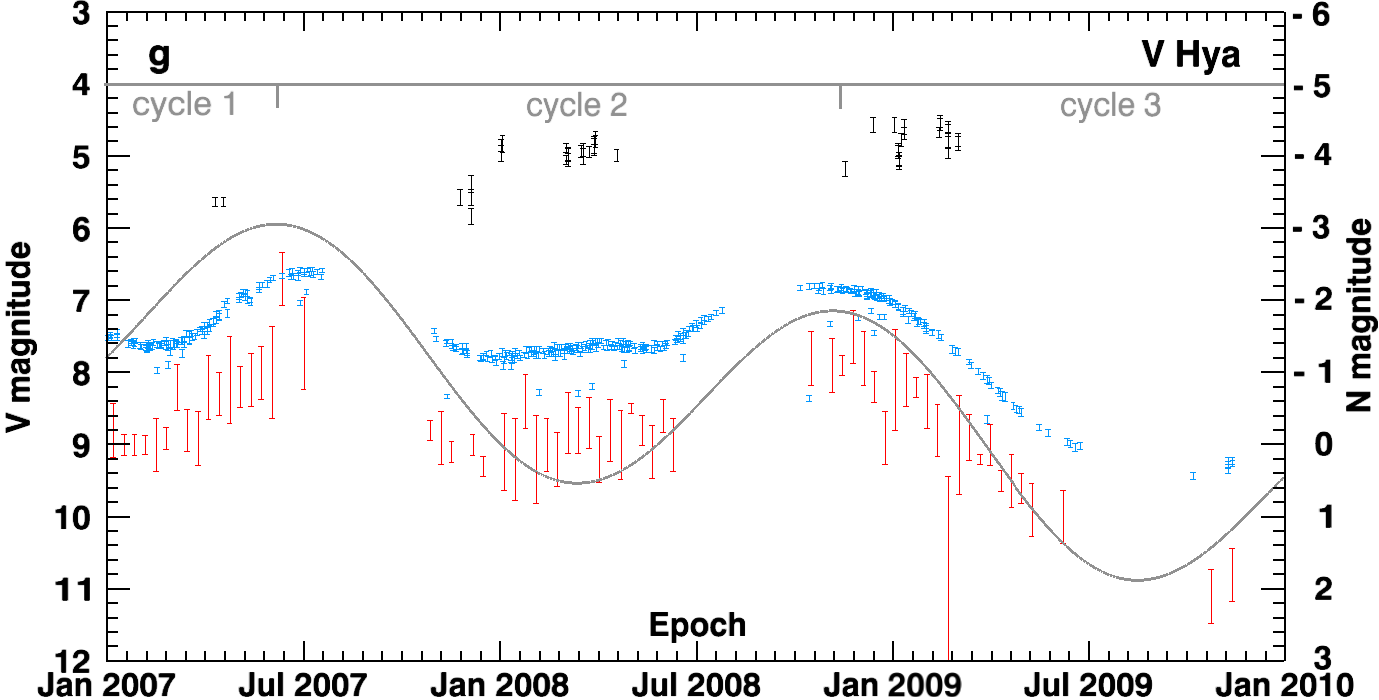}
    %\hspace{0.2cm}
    \includegraphics[width=0.47\linewidth]{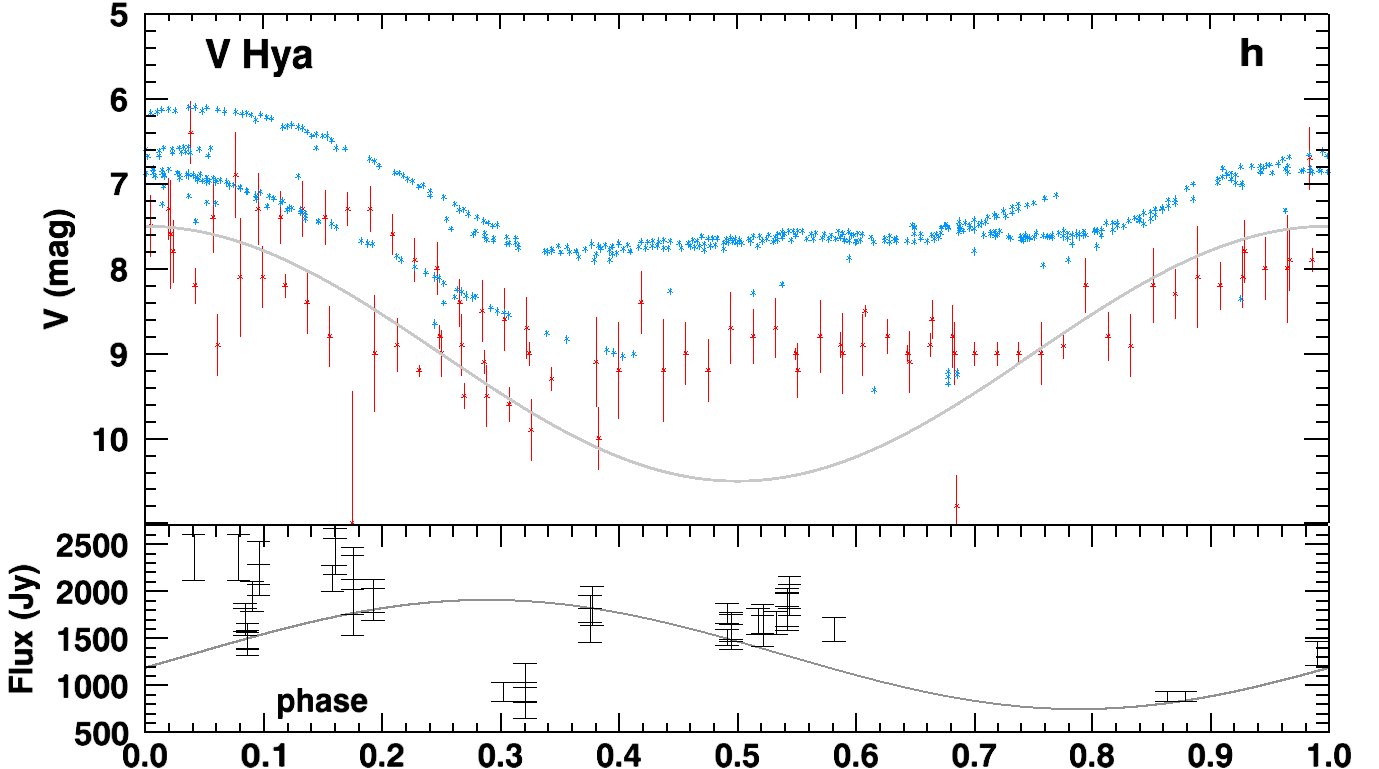}
    \caption{\textit{Left:} Visual light curves covering the period of the MIDI observations. Simple sinusoidal fits are included to determine the pulsation phases used throughout this paper. The MIDI fluxes at around 12~$\mu$m are shown as well with the magnitude scale given on the right. \textit{Right:} Same as left, but plotted versus visual phase. The MIDI fluxes at around 12~$\mu$m are shown in the lower panels of each plot. The data within the shaded regions in both columns are used for size and flux variation studies in Sect.~\ref{secIntDis_Intra}. The corresponding plots for W~Hya can be found in paper~I (Fig.~3 and 4, respectively).}
    \label{FigLightPhase}
   \end{figure*}
%---------------------------------------------------------------
%---------------------------------------------------------------
\begin{table*}
   \caption{Fitted visual light and mid-IR light curves$^{\mathrm{a}}$.}
   \label{Table_Periods}
   \centering
   \begin{tabular}{c|cccc|ccc}
%       \noalign{\smallskip}
       \hline
       \noalign{\smallskip}
                    & \multicolumn{4}{c}{visual (AAVSO+ASAS)}            & \multicolumn{3}{c}{mid-IR (MIDI)}          \\
         Target~~~~ & $P$   & $T_0$ &~~~~$V_0$~~~~ & $V_{\mathrm{ampl}}$ & $\Delta\phi$ & $N_0$ & $N_{\mathrm{ampl}}$ \\
                    & days  & JD    & mag          & mag                 &              & Jy    & Jy                  \\
       \noalign{\smallskip}
       \hline
       \noalign{\smallskip}
         R Aql& 271~$\pm$~4  &2452910&8.30&1.80~$\pm$~0.15 & $-0.21$~$\pm$~0.50 &  377~$\pm$~50  & ~80~$\pm$~50  (21\%) \\
         R Aqr& 390~$\pm$~5  &2452480&8.75&2.75~$\pm$~0.20 & $+0.15$~$\pm$~0.05 & 1080~$\pm$~100 & 330~$\pm$~50  (31\%) \\
         R Hya& 374~$\pm$~5  &2453231&6.95&1.95~$\pm$~0.15 & $+0.08$~$\pm$~0.05 & 1930~$\pm$~200 & 240~$\pm$~50  (12\%) \\
         W Hya& 388~$\pm$~5  &2452922&7.56&1.41~$\pm$~0.10 & $+0.15$~$\pm$~0.05 & 4910~$\pm$~200 & 510~$\pm$~100 (10\%) \\
         V Hya$^{\mathrm{b}}$& 529~$\pm$~10 &2450045&8.50&1.50~$\pm$~0.10 & $+0.29$~$\pm$~0.05 & 1330~$\pm$~200 & 580~$\pm$~200 (44\%) \\
       \noalign{\smallskip}
       \hline
%       \noalign{\smallskip}
   \end{tabular}
  \newline
  \begin{flushleft}
    \textbf{Notes. }
    $^{\mathrm{a}}$~$V(t) = V_{\mathrm{ampl}}\sin{[2\pi(t-T_0)/P] + V_0}$ and $N(t) = N_{\mathrm{ampl}}\sin{[2\pi(t-T_0)/P-\Delta\phi] + N_0}$, respectively. The phase shift $\Delta\phi$ gives the positive offset in respect to the visual phase. 
    $^{\mathrm{b}}$~V~Hya has a second period with $P_2$~=~(6160~$\pm$~400) days, $T_{0,2}$~=~2446937 days, $V_{0,2}$~=~8.50 mag and $V_{\mathrm{ampl},2}~=~$(3.50~$\pm$~0.50) mag \citep{Knapp1999}.
  \end{flushleft}
\end{table*}
%---------------------------------------------------------------

   In order to assign a pulsation phase to the observations, visual data from the American Association of Variable Star Observers (AAVSO)\footnote{http://www.aavso.org/} and the All Sky Automated Survey \citep[ASAS][]{Pojmanski2005}\footnote{http://www.astrouw.edu.pl/asas/} are used. After binning the AAVSO data into 10 day bins a simple sinusoid is fitted to the AAVSO and ASAS data over a period of about 10~years ($2000-2010$). The resulting periods~$P$, Julian Dates of maximum brightness~$T_0$ (defined as phase~0.0), mean visual magnitudes~$V_0$ and visual semi-amplitudes~$V_{\mathrm{ampl}}$ are given in Table~\ref{Table_Periods}. The errors given are estimates derived from the fitting process. The uncertainties for $T_0$ and $V_0$ are of the same order as the uncertainties for the period and the visual semi-amplitude, respectively. Note that V~Hya has two periods and the longer period is taken from \citet{Knapp1999}.

   For each star the photometric data as well as the fit are plotted in Fig.~\ref{FigLightPhase} versus time and versus visual phase, covering the period of the MIDI observations. In addition, the mid-IR fluxes in Jy (averaged between 11.5 and 12.5~$\mu$m) obtained with MIDI are shown in the right hand panels as function of visual phase. As described in paper~I, the MIDI fluxes are strongly error-prone due to difficulties in the reduction removing instrumental imperfections and the large sky background which varies on short time scales. Additional scatter in the N-band flux originates from cycle-to-cycle variations caused by the movement of the dust shell in the case of a dust driven wind \citep[e.g.][]{Nowotny2010}.
   
   However, a clear phase dependence is detectable in each of the phase-folded plots. The results of a sinusoidal fit are summarized in Table~\ref{Table_Periods}. The table gives the relative phase shifts $\Delta\phi$ of the mid-IR maximum with respect to the visual maximum, the mean N-band fluxes~$N_0$ and the semi-amplitudes~$N_{\mathrm{ampl}}$ with the corresponding percentage flux variations given in brackets. R~Aql shows only a very small variation in the mid-IR by eye, suggesting that the amplitude is overestimated by the fit, and it is even consistent with no variation at all. The poor fit could be the reason why a mid~IR maximum before the visual maximum is obtained in comparison to the other objects.
   
   The Mira R~Aqr and the semiregular variable V~Hya show the highest mid-IR flux variations. Both systems are close binary systems and might therefore contain a large amount of circumstellar material. The fit to the mid-IR flux data of R~Aqr is fair and the flux variation of about 30\% seems to be real. Even though the mass-loss rate is low, the orbiting dust in this symbiotic system could be the reason for an increased mid-IR flux variation.

   In contrast, the mid-IR flux of V~Hya varies not only with a high amplitude but also with a large scatter, resulting in a fit which is not very robust. Since V~Hya displays two superimposed visual periods, an assignment of a visual phase is not an adequate description. Even if the short period is Mira-like, the cause and influence of the long secondary period is not well known. Many different causes probably contribute to its large and varying mid-IR flux. In C-rich stars the production of carbonaceous dust is more effective, and hence more dust can radiate. The fact that V~Hya is probably a common envelope system might be related to this. However, V~Hya's evolutionary status as post AGB star, exhibiting the superwind phase with a fast and dense mass loss, might be the main reason for the varying mid-IR flux.

   R~Hya had a very smooth and regular visual light curve over the last decade. MIDI observations were only possible around minimum visual light, since the period is close to one year. As a result, the mid-IR light curve could not be well constrained and the observed flux variation is relatively small, only on the order of 10\%. The mid-IR and visual light curves of W~Hya are very stable with only moderate amplitudes as well. For both stars, the low mid-IR flux variations might be related to the low mass-loss rates. Since W~Hya and R~Hya are fairly nearby, both stars have high absolute mid-IR fluxes.

   Even if the mid-IR flux variations of all five AGB stars are much smaller than the visual ones, they are still on the order of 10\% to 30\%. Similar flux amplitudes in the mid-IR have been reported for other AGB stars as well (cf.~references given in paper~I). Except for R~Aql, the mid-IR maximum occurs always after the visual maximum at an average visual phase of 0.15~$\pm$~0.05. This phase shift is consistent with previous studies of AGB stars \citep[cf.~e.g.][and references therein]{Lattanzio2004,Smith2006,Nowotny2010}. The reason for this phase-lag is most probably related to the dynamic processes of shock front and dust formation.

   Since the mid-IR fits are relatively uncertain, all data are interpreted with respect to the visual light curve in the following. It should also be kept in mind that folding consecutive cycles into one cycle might not always be appropriate, since the pulsation is not strictly regular, in particular for V~Hya (see Fig.~\ref{FigLightPhase}).

%%%%%%%%%%%%%%%%%%%%%%%%%%%%%%%%%%%%%%%%%%%%%%%%%%%%%%%%%%%%%%%%%%%%%%%%%%%%%%%%%%%%%%%%%%%%
\subsection{Spectra}\label{secObsSubSpec}

%---------------------------------------------------------------
   % bei 5 Sternen: width=0.41 and width=0.37
   % bei 4 Sternen: width=0.52 and width=0.47 and \hspace{0.5cm}
   \begin{figure*} %0.42 and 0.38
    \centering 
    \includegraphics[width=0.44\linewidth]{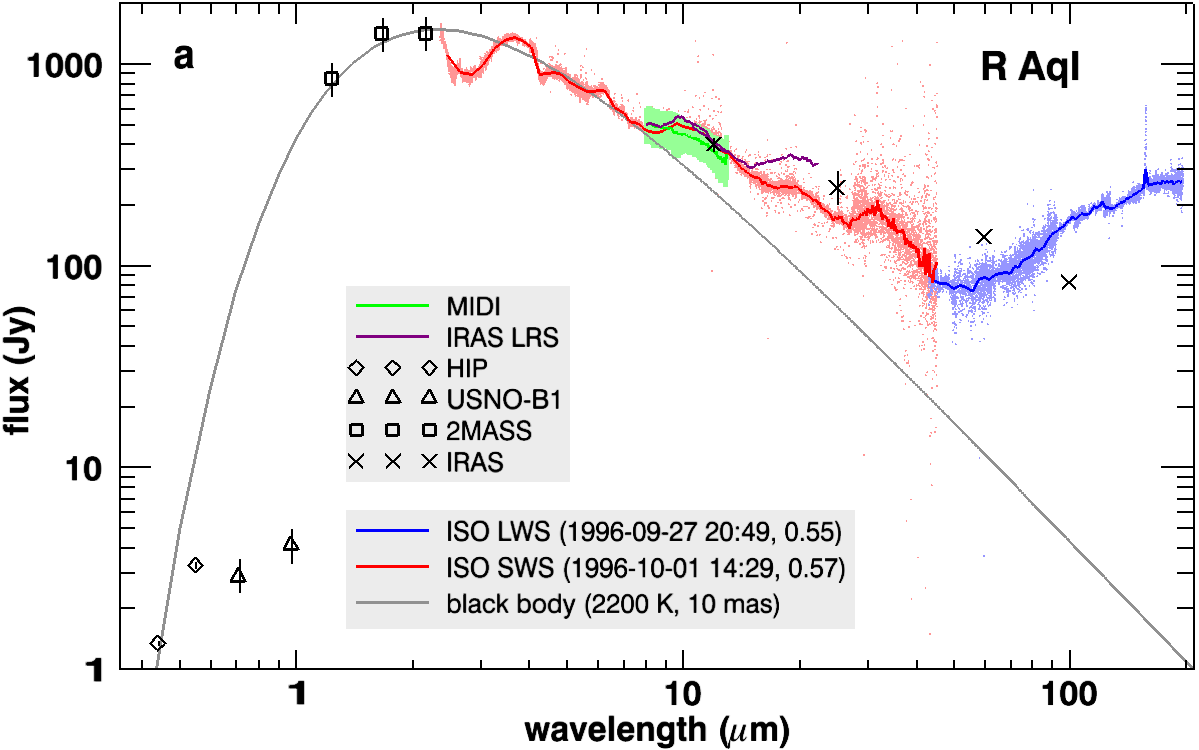}
    \hspace{0.2cm}
    \includegraphics[width=0.44\linewidth]{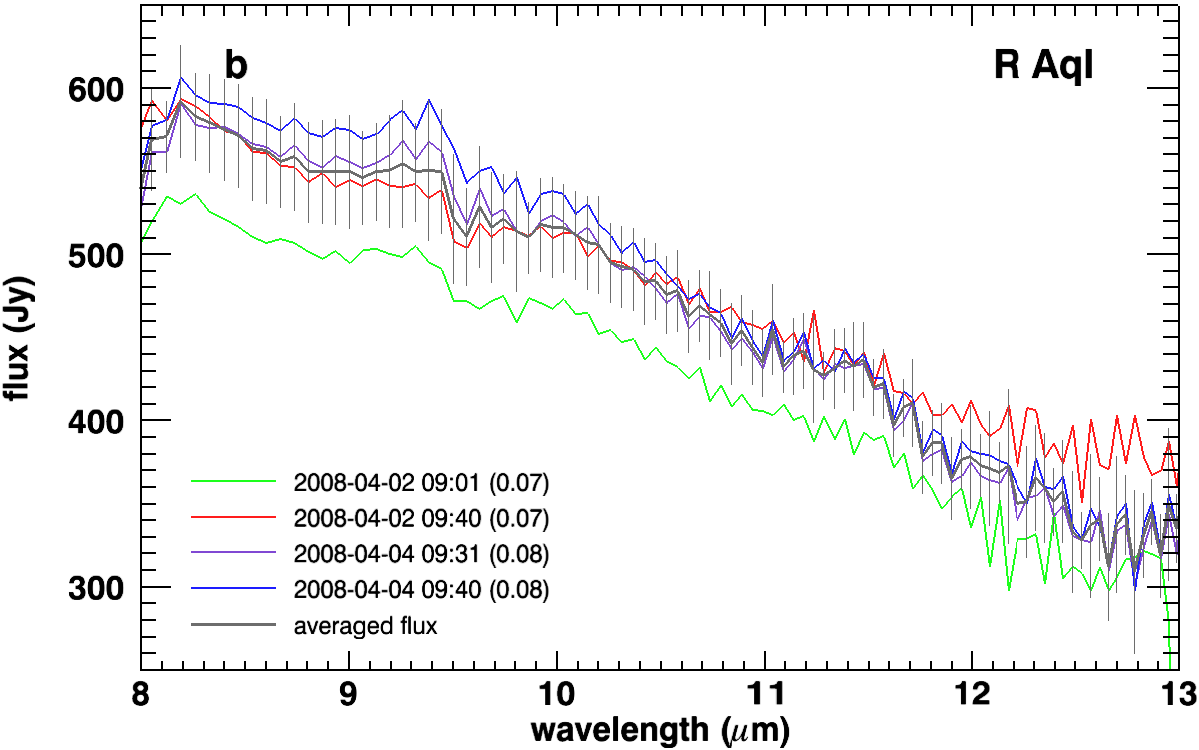}
    \includegraphics[width=0.44\linewidth]{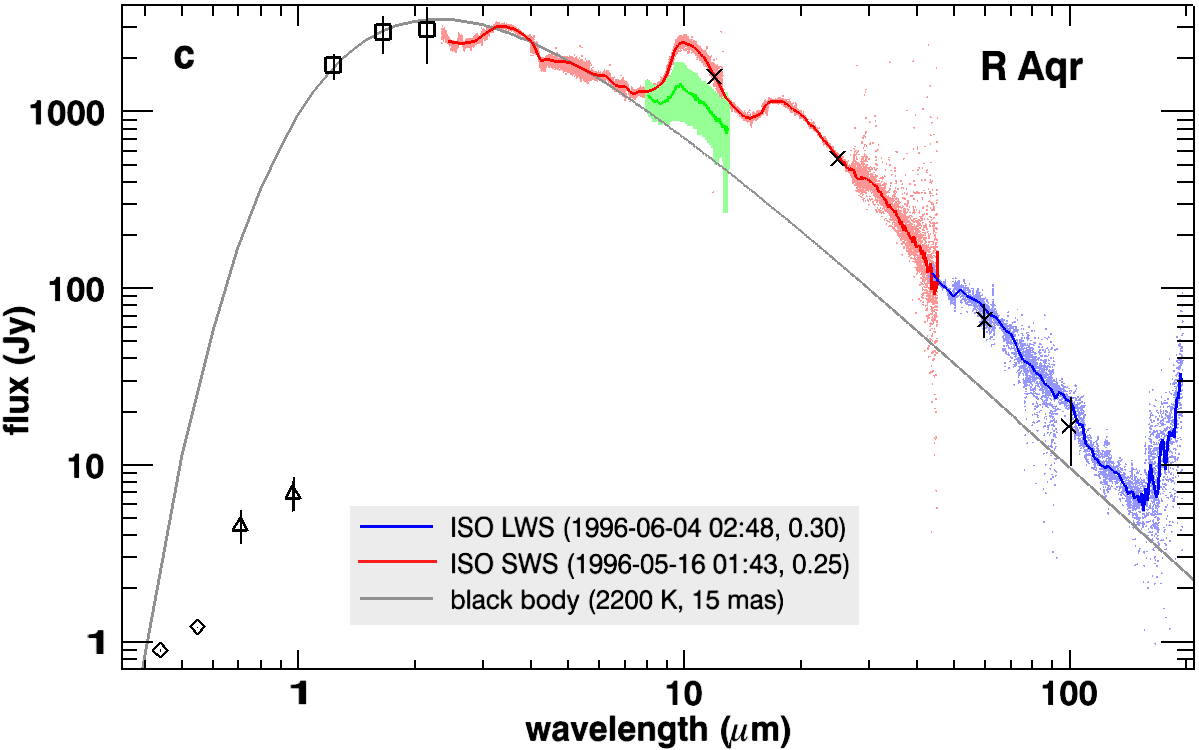}
    \hspace{0.2cm}
    \includegraphics[width=0.44\linewidth]{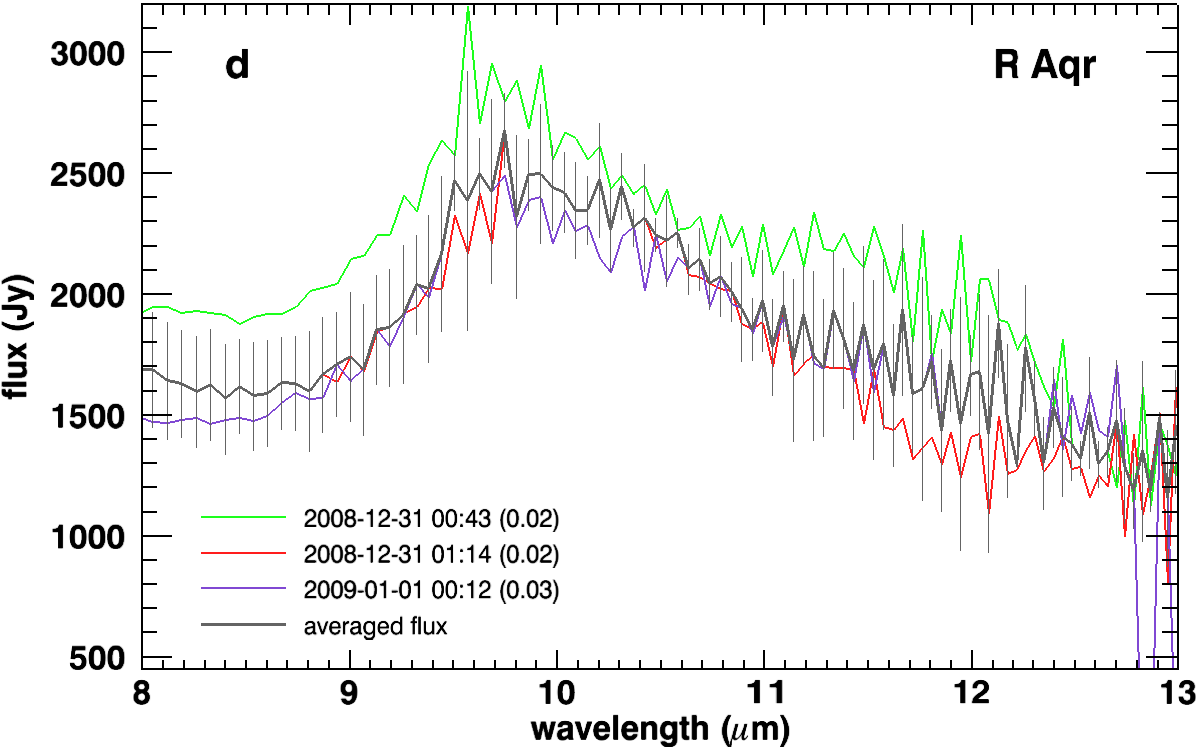}
    \includegraphics[width=0.44\linewidth]{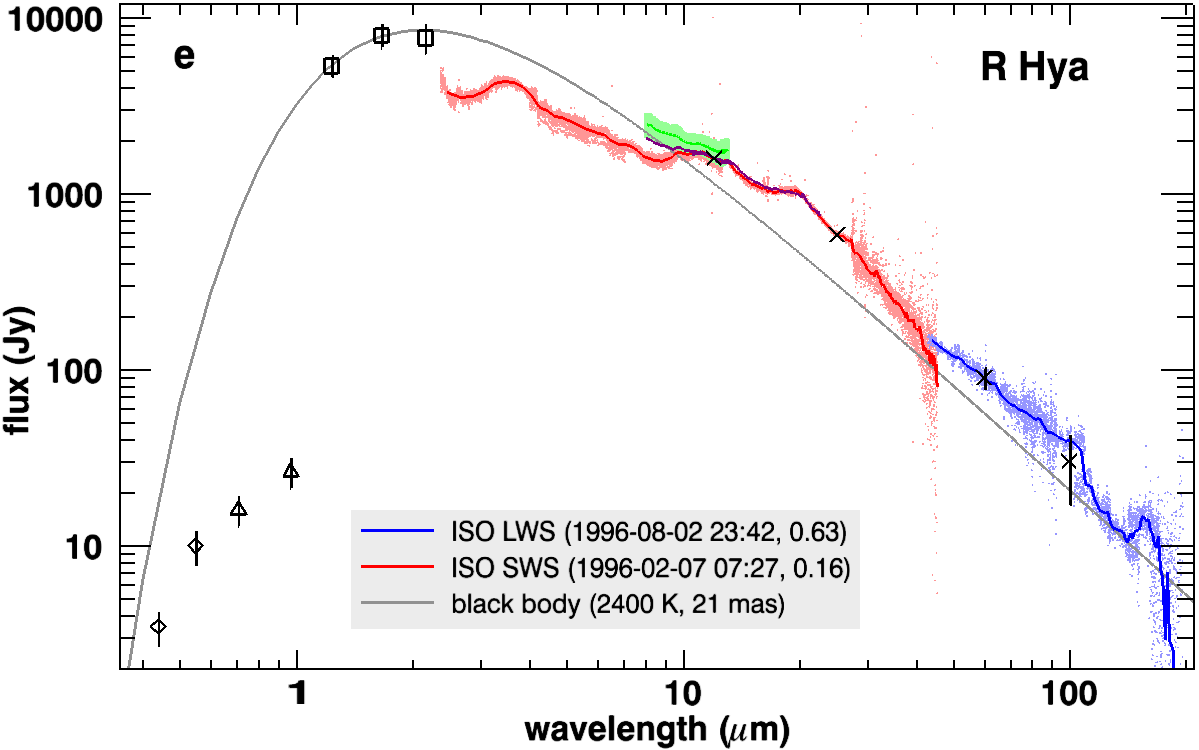}
    \hspace{0.2cm}
    \includegraphics[width=0.44\linewidth]{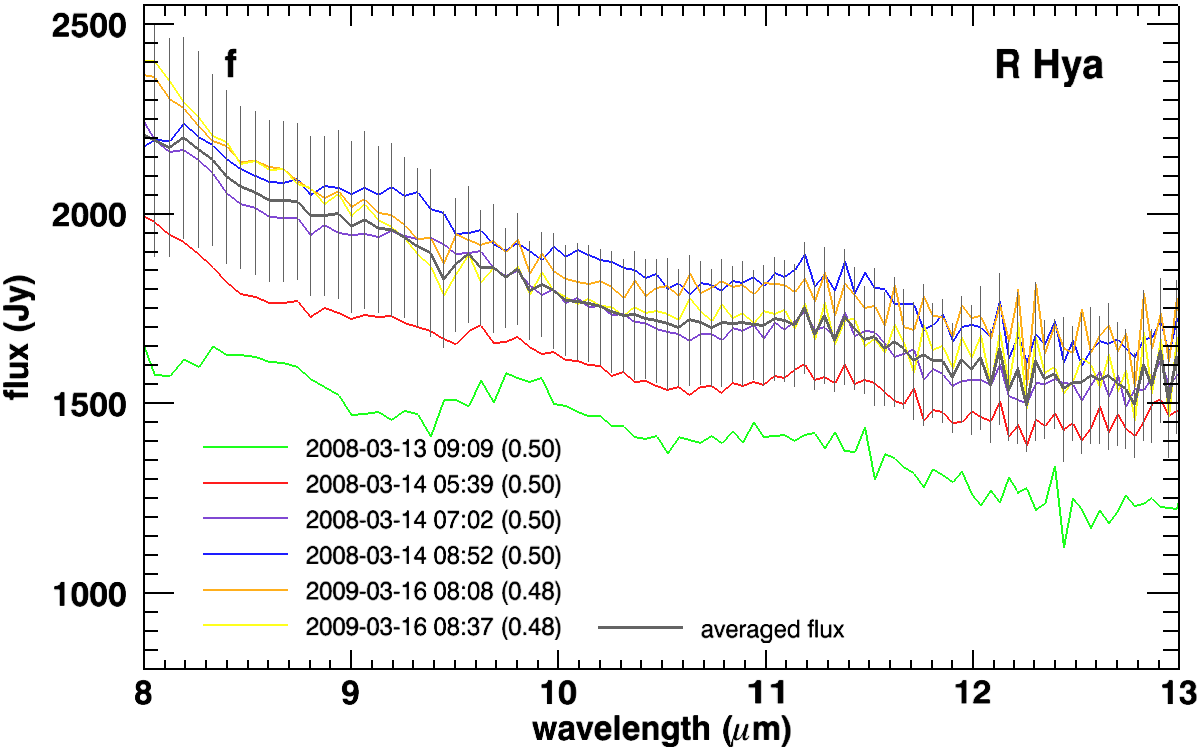}
    \includegraphics[width=0.44\linewidth]{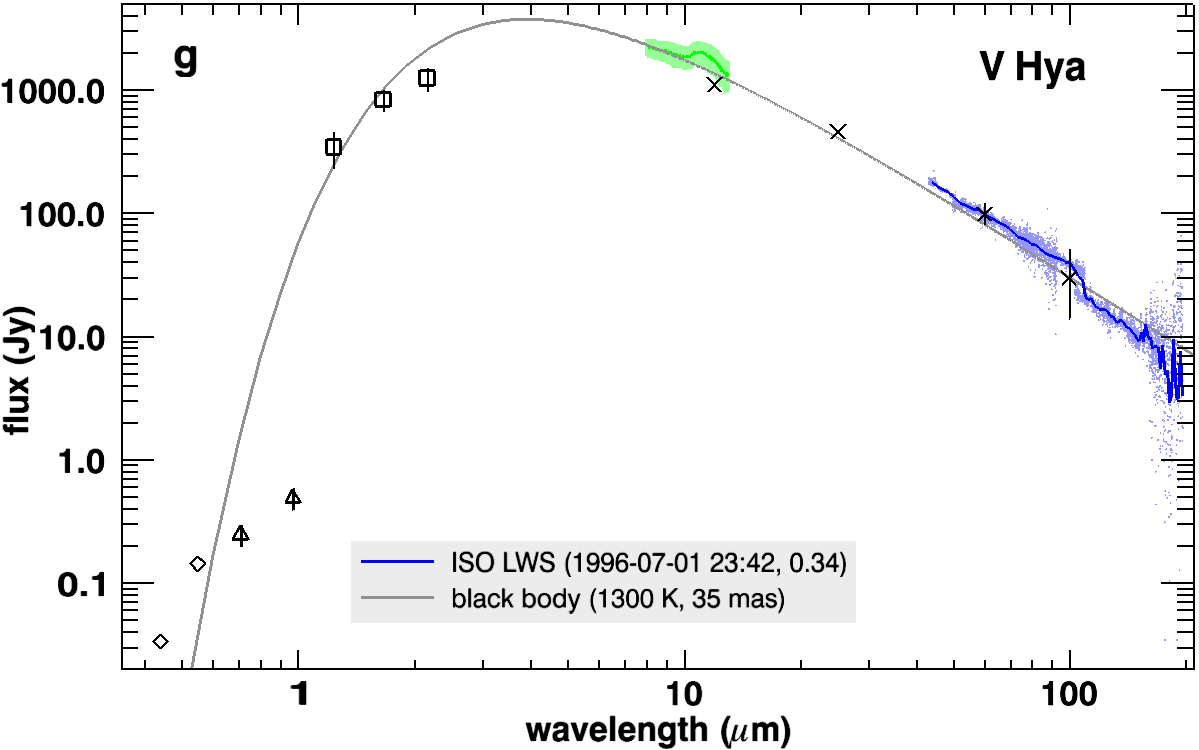}
    \hspace{0.2cm}
    \includegraphics[width=0.44\linewidth]{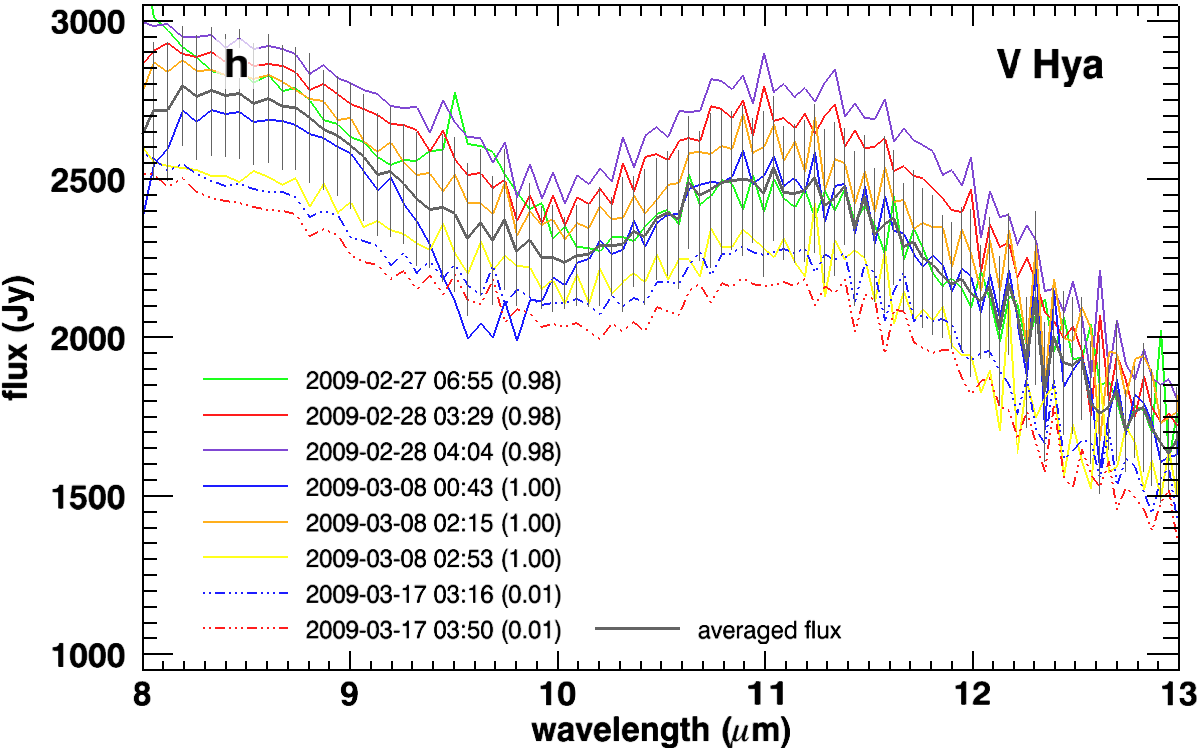}
    \caption{\textit{Left:} Spectral energy distribution for all stars (see text). The black body curve is only included for guidance. \textit{Right:} All available individual MIDI spectra around the visual maxima for all stars (including an average of these) except for R~Hya were the spectra around the visual minima are given. The visual light phases are given in brackets. Note that due to the telluric ozone feature the uncertainties at around 9.6~$\mu$m are increased. The corresponding plots for W~Hya can be found in paper~I (Fig.~5).}
    \label{FigSpectra}
   \end{figure*}
%---------------------------------------------------------------

%---------------------------------------------------------------
   \begin{figure*}
    \centering 
    \includegraphics[width=0.47\linewidth]{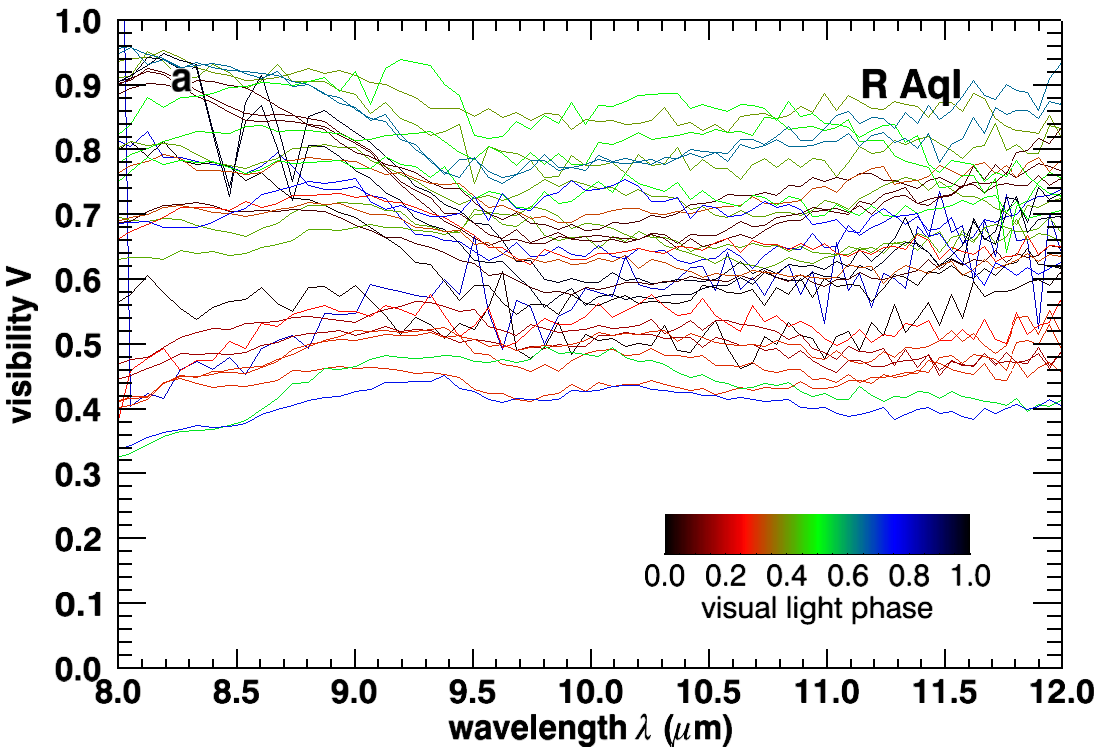}
    \hspace{0.2cm}
    \includegraphics[width=0.47\linewidth]{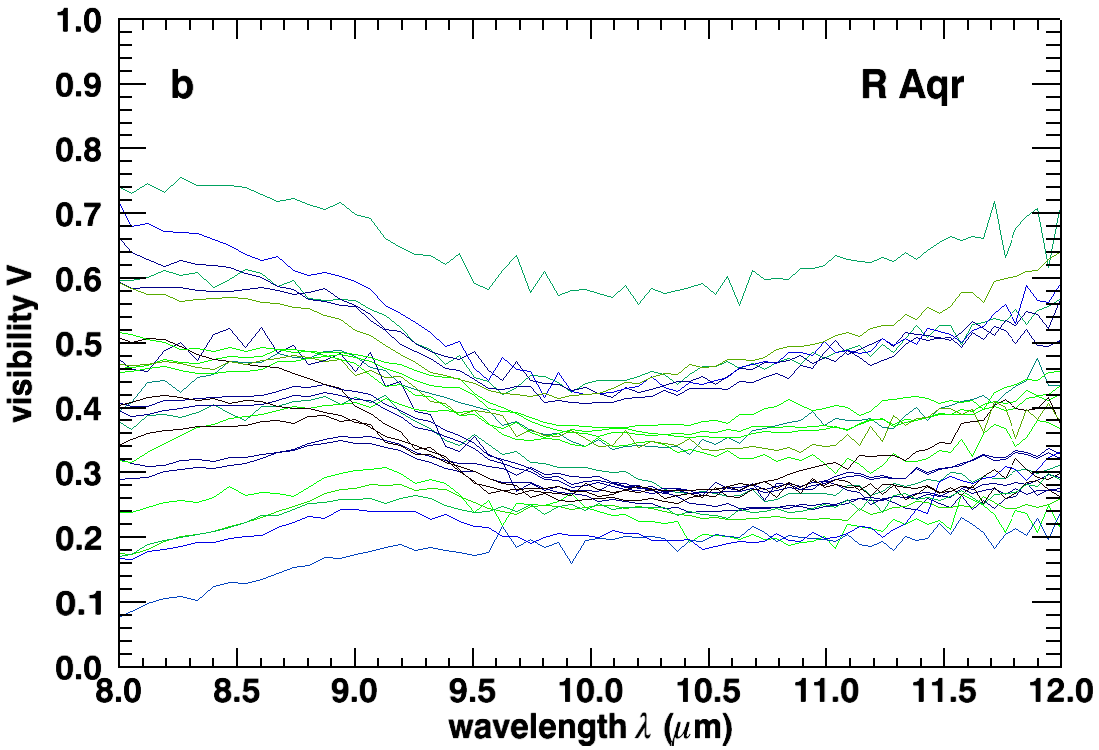}
    \includegraphics[width=0.47\linewidth]{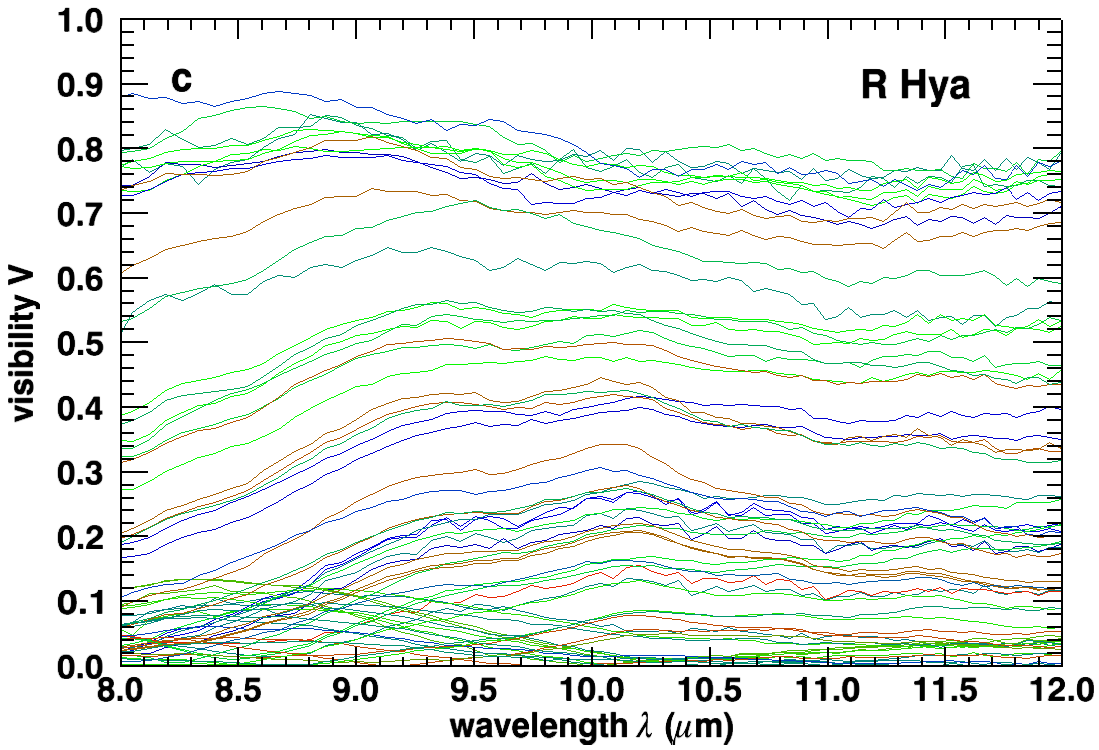}
    \hspace{0.2cm}
    \includegraphics[width=0.47\linewidth]{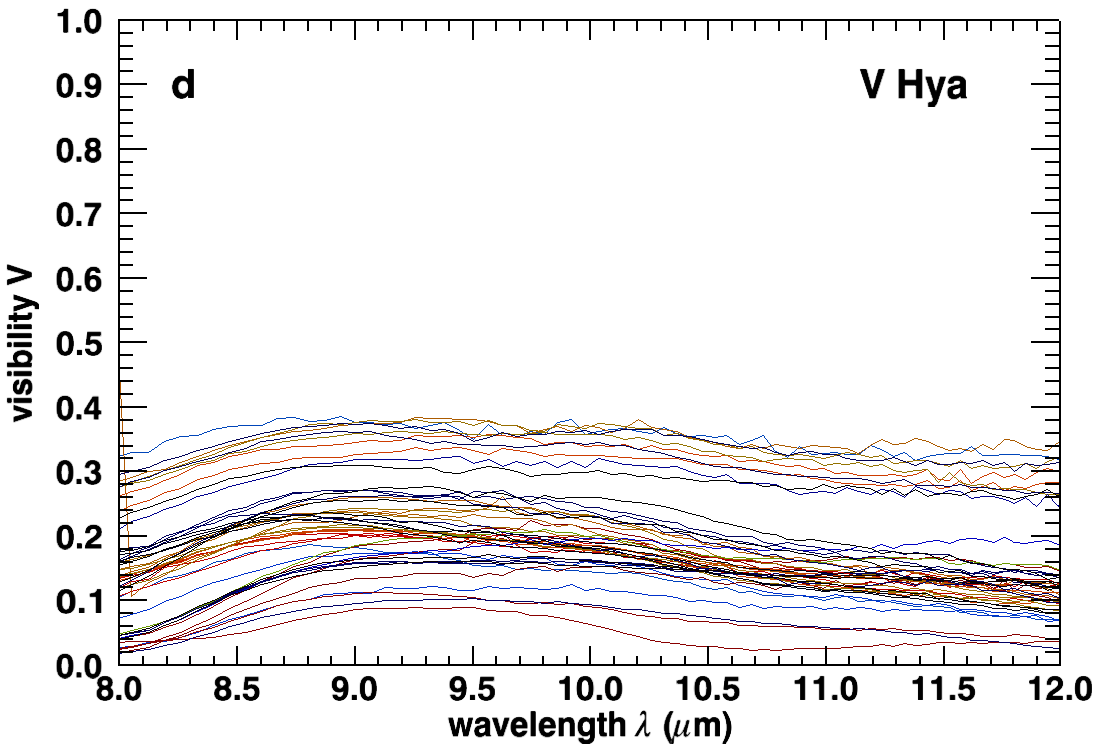}
    \caption{Calibrated visibilities as function of wavelength, color-coded by visual light phase. Errors are omitted for clarity but are given in Tables~\ref{TableResultsRAql} to~\ref{TableResultsVHya}. The corresponding plot for W~Hya can be found in paper~I (Fig.~1).}
    \label{FigVisibility}
   \end{figure*}
%---------------------------------------------------------------

   All calibrated MIDI spectra for each star are median averaged over all phases and cycles and are shown in the spectral energy distribution (SED) plots in the left hand panels of Fig.~\ref{FigSpectra}. The uncertainties are given by the standard deviation. The flux levels of ISO\footnote{\citet{Sloan2003} and http://iso.esac.esa.int/ida/} and IRAS\footnote{http://irsa.ipac.caltech.edu/Missions/iras.html} have been adjusted to coincide with the 12~$\mu$m flux. Photometric data from 2MASS and IRAS are plotted as well. A blackbody curve is over-plotted as guidance, assuming a diameter and temperature which best represents the data points by eye. However, due to the infrared excess, strong metallic oxide lines, molecule absorption and dust extinction, it is not expected that a blackbody curve fits the spectral data in an appropriate way. Individual spectra of the MIDI observations are given in the right hand panels of Fig.~\ref{FigSpectra}, showing the same features as the averaged MIDI spectra in the left hand panels.
    
   The ISO spectra of the O-rich stars are mainly dominated by absorption bands of \ce{H2O} between $2.5-3.0$~$\mu$m (stretching mode) and $5.0-8.0$~$\mu$m (bending mode), and an SiO absorption band between 8 and~9~$\mu$m ($\nu$~=~$1-0$). Distinct absorption lines of CO at around 2.4~$\mu$m, OH at $2.9-4.0$~$\mu$m, \ce{CO2} at 4.25~$\mu$m and \ce{SO2} at 7.4~$\mu$m can be seen in the spectra of some of the stars. From temperature investigations by e.g.~\citet{Justtanont2004} for W~Hya it could be derived that these absorptions originate from different molecular layers located in the close atmospheric environment of the star. OH and CO absorption bands arise mainly from a hot (about 3000~K), dense region very close to the stellar photosphere, where \ce{H2O} is still photodissociated by shocks. \ce{H2O} and \ce{CO2} absorption bands originate from a layer with a temperature of 1000~K, i.e.~a molecular layer (MOLsphere) farther out. The SiO molecule absorption arises in the same region where the \ce{H2O} molecular shell exists and where SiO is still not bound in dust grains.

   All O-rich stars in the sample show dust emission in their ISO spectrum with different peculiarities. The features between 10 and 20~$\mu$m are a combination of emission from amorphous silicates at 9.7~$\mu$m, amorphous \ce{Al2O3} at around 11.5~$\mu$m, and MgFeO at around 19~$\mu$m. Regarding the 13~$\mu$m emission feature, seen in R~Aql, R~Hya and W~Hya, it is still under debate whether it comes from spinel (\ce{MgAl2O4}) \citep[e.g.][]{Posch1999,Fabian2001,Heras2005} or corundum (crystalline \ce{Al2O3}, $\alpha$-\ce{Al2O3}) \citep[e.g.][]{Onaka1989,DePew2006}. W~Hya is the only star which clearly shows the molecular \ce{SO2} emission feature in the spectrum \citep[][]{Justtanont2004,ZhaoGeisler2011}, while R~Aqr is the only star which is completely dominated by amorphous silicate dust emission. In addition, R~Aqr has an overabundance of SiO molecules as discussed in \citet{Angeloni2007}. All stars also exhibit a remarkable infrared excess. In the dust emission scheme of \citet{Sloan1998Oxygen} R~Hya is classified as SE2t (broad oxygen-rich dust emission), R~Aql as SE5 (structured silicate emission), and R~Aqr and W~Hya as SE7 and SE8, respectively (classic narrow silicate emissions).

   If the individual and averaged MIDI spectra are compared with the ISO spectra (cf.~Fig.~\ref{FigSpectra}), it becomes obvious that the silicate emission at around 9.7~$\mu$m is not detected except for R~Aqr. This behavior can be attributed to instrumental characteristics. ISO has a much larger field of view (FoV) compared to MIDI. With a small FoV of about 1 to 2~arcsec\footnote{The exact value depends on the baseline used, the slit/mask position and other instrumental specifications.}, the emission of the extended silicate dust shell could therefore not be observed with MIDI.

   This applies to the nearby stars R~Hya and W~Hya. R~Aql and R~Aqr are located farther away and have nearly the same distance. Therefore, both stars should both either show or not show the silicate dust emission feature, but this apparent contradiction can be solved. In Sect.~\ref{secIntDis_OShell} it will be seen that the dust shell around R~Aqr is much closer due to the fact that R~Aqr is a symbiotic system probably containing a large amount of dust.

   However, this non-detection allows to derive a lower limit for the inner boundary of the silicate dust shell. Assuming a conservative value of the FoV of 1~arcsec, most of the silicate dust emission originates from a shell with an inner diameter larger than about 100~$\theta_{\mathrm{phot}}$\footnote{Photospheric diameter of the star, cf.~Sect.~\ref{secResSWaveDep}.} ($>$220~AU), 49~$\theta_{\mathrm{phot}}$ ($>$130~AU), and 28~$\theta_{\mathrm{phot}}$ ($>$90~AU) for R~Aql, R~Hya, and W~Hya, respectively. However, this does not mean that silicate dust does not exist closer to the star. Its abundance is just below the detection limit of MIDI. For R~Aqr it can be speculated that silicate dust must exist farther out, at a region with a diameter larger than 72~$\theta_{\mathrm{phot}}$ ($>$250~AU), since with the larger FoV of ISO more silicate emission could be revealed than with the smaller FoV of MIDI (Fig.~\ref{FigSpectra}c).

	The absolute flux levels for the MIDI spectra are lower than those of ISO/SWS and IRAS as it should be expected since the FoV of MIDI is smaller than those of ISO/SWS and IRAS. Also the 3.5~$\mu$m region of the ISO/SWS spectra, considered as a pseudo continuum, does nicely overlap with the blackbody continuum. However, this is not the case for R~Hya. This might be due to a calibration problem at around 9~$\mu$m since the IRAS spectrum is flat around that point while it is not for the ISO/SWS one. At the same time, the flux level of the MIDI spectrum is slightly higher, but this is probably due to the large MIDI flux calibration uncertainties. A flux variation due to the pulsation can be ruled out since the ISO spectrum is taken at around visual maximum while the averaged MIDI spectrum of R~Hya is mainly an averages of spectra taken around visual minimum.

   V~Hya is the only carbon star in the sample. The MIDI spectrum shows very clearly the dust emission of SiC at 11.3~$\mu$m. Amorphous carbon, which makes up most of the dust, is featureless. Both dust species typically condensate at around 2~photospheric radii at a temperature of around 1500~K \citep[cf.~e.g.][]{Ohnaka07}. Molecular shells of \ce{C2H2} (bands at $8-9$~$\mu$m and $11-14$~$\mu$m) and HCN (band at $11-13$~$\mu$m) at different distances and temperatures from the star might be present as well, but are not resolved in the low resolution MIDI spectra. The absolute flux level of the MIDI spectrum is also reasonable and joins with the blackbody curve. In the scheme of \citet{Sloan1998Carbon}, V~Hya is categorized as Red, meaning that a strong continuum increase in intensity toward longer wavelengths is superposed on the dust features.

%###########################################################################################
%###########################################################################################
\subsection{Visibility modeling results}\label{secObsSubVis}

   The most straight-forward way of interpreting sparsely sampled interferometric data (visibilities) is by fitting the Fourier transform of an assumed brightness distribution of the object. Simple size estimations can be obtained from elementary geometrical models with only a few free parameters. As mentioned in the introduction, the definition of a diameter is difficult because of its strong wavelength dependence as well as of phase-to-phase and long-term cycle-to-cycle variations. On the other hand, the size and its dependence on wavelength and pulsation phase can give some constraints on the chemical and physical mechanisms which are responsible for this appearance, but also which layer of the atmosphere is actually observed. This work focuses on interpreting the data using simple geometric models. In this section, all measurements over all pulsation phases, cycles and position angles are taken into account, knowing that this potentially biases the interpretation, but will give a first description of the size of the structures. In Sect.~\ref{secIntDis_Intra}, measurements are interpreted through each available pulsation phase allowing to put constraints on the dynamical mechanisms involved. Self-consistent hydrodynamic models will be required for a more realistic analysis.

   The low surface gravity results in an extended atmosphere and temperature structure and the formation of molecular layers around late-type stars. Therefore, no sharp transition between the star and the circumstellar environment exists. These stars can usually be modeled by using a uniform disk (UD) to account for the star and a uniform ring to represent a molecular layer (or simpler with two uniform disks). However, this would increase the numbers of parameters to be fitted, and the later discussed pulsation dependence studies (Sect.~\ref{secIntDis_Intra}) would not be possible. Therefore, a simple fully limb-darkened disk (FDD) will be used to represent the stellar photosphere and the close molecular environment. One can assume that the fitted FDD diameter will consequently be located in between the stellar photosphere and the outer boundary of the close molecular environment detectable with MIDI.
   
   Uniform disk (UD) model fits are given for comparison where appropriate. For some of the stars, a Gaussian function is added to constrain the size of the extended dust shell. For the stars where this was not possible, i.e.~where the dust shell was overresolved, only the flux contribution of a surrounding dust shell were determined by not forcing the visibility function to be~1 at zero spatial frequency (referred to as relative FDD/UD). The total visibility function, $V$, is written as
   \begin{eqnarray}
      V &=& |\,\epsilon_{\mathrm{FDD}} \; \mathcal{V}_{\mathrm{FDD}} + (1-\epsilon_{\mathrm{FDD}}) \; \mathcal{V}_{\mathrm{Gaussian}}\,|\;\mathrm{,}
   \end{eqnarray}
   with $\epsilon_{\mathrm{FDD}}$ the flux contribution of the FDD and $\mathcal{V}$ the complex visibilities \citep{ZhaoGeisler2010}. For an overresolved dust shell, $\mathcal{V}_{\mathrm{Gaussian}}$ is equal to zero.

%---------------------------------------------------------------
   \begin{figure*} %5 stars: 0.36 and 0.36 and \hspace{1cm}
    \centering
    \includegraphics[width=0.45\linewidth]{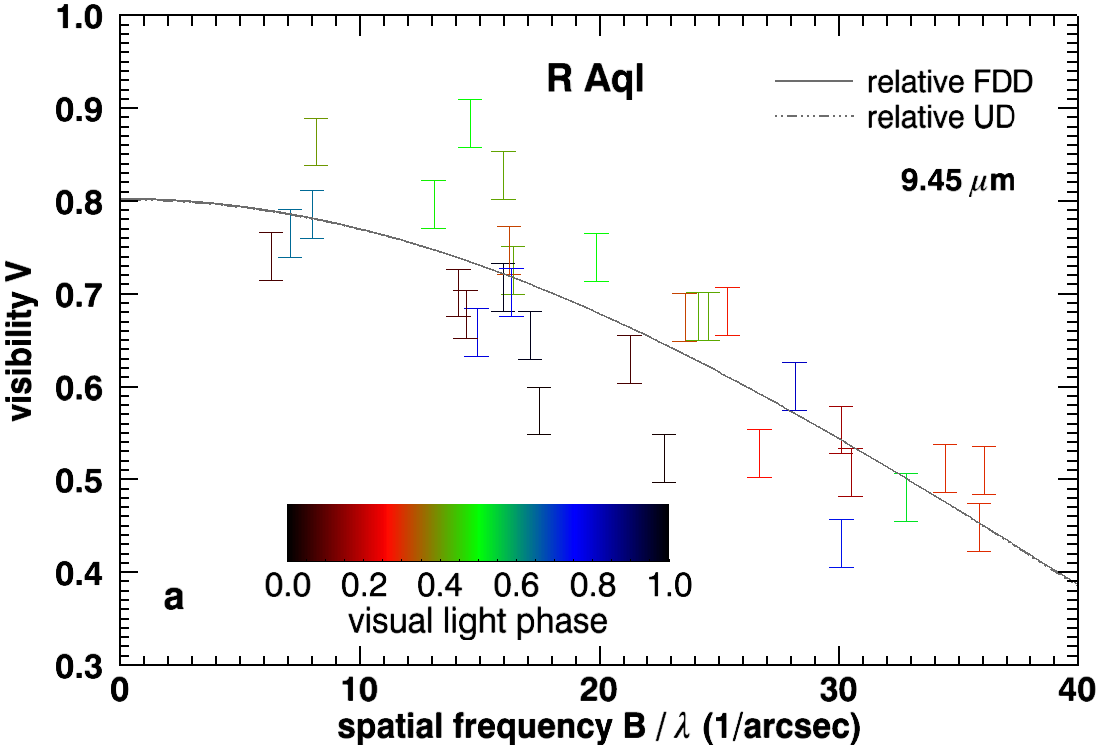}
    \hspace{0.2cm}
    \includegraphics[width=0.45\linewidth]{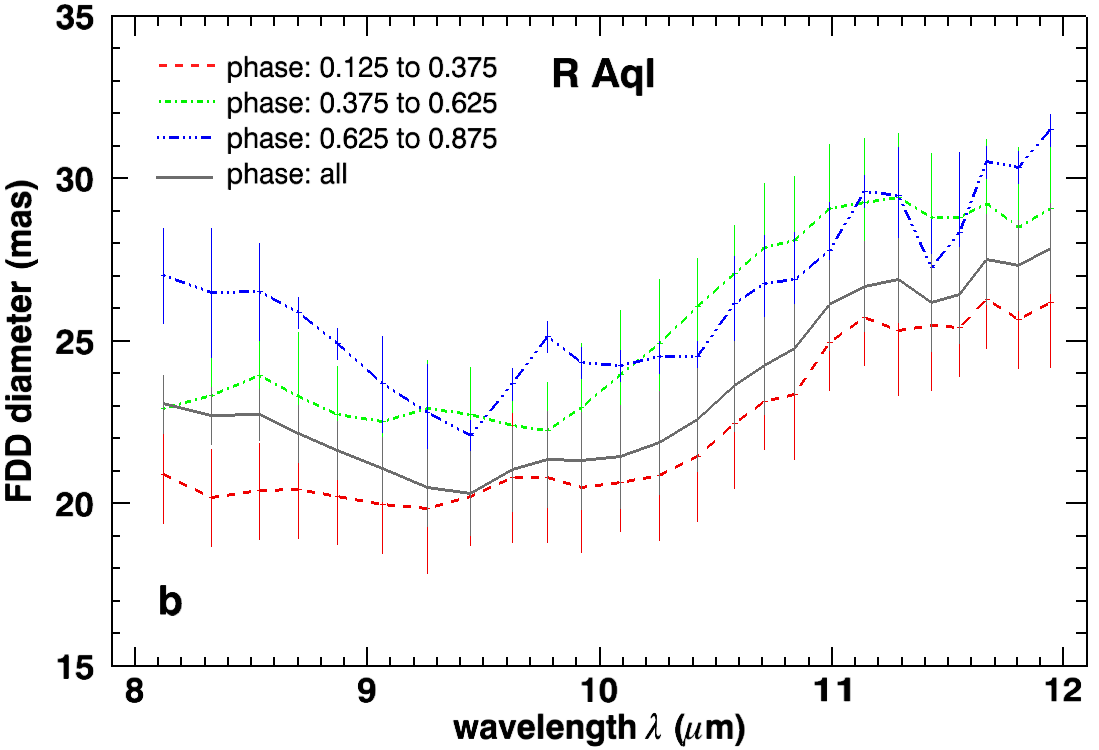}
    \includegraphics[width=0.45\linewidth]{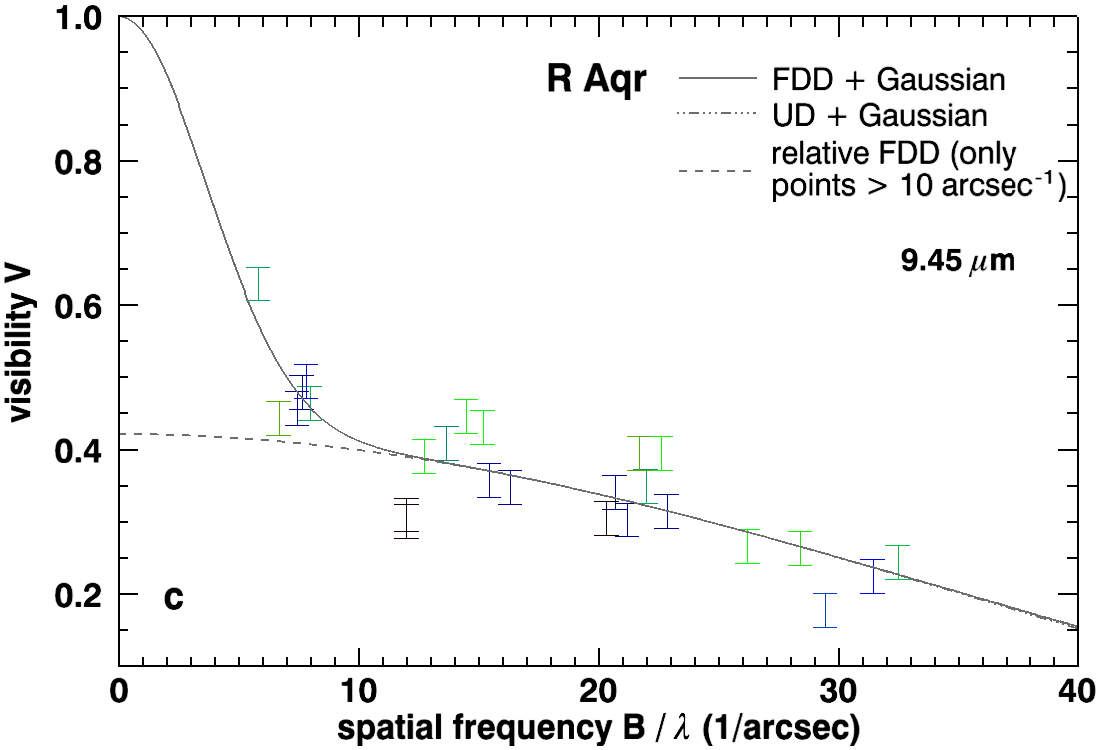}
    \hspace{0.2cm}
    \includegraphics[width=0.45\linewidth]{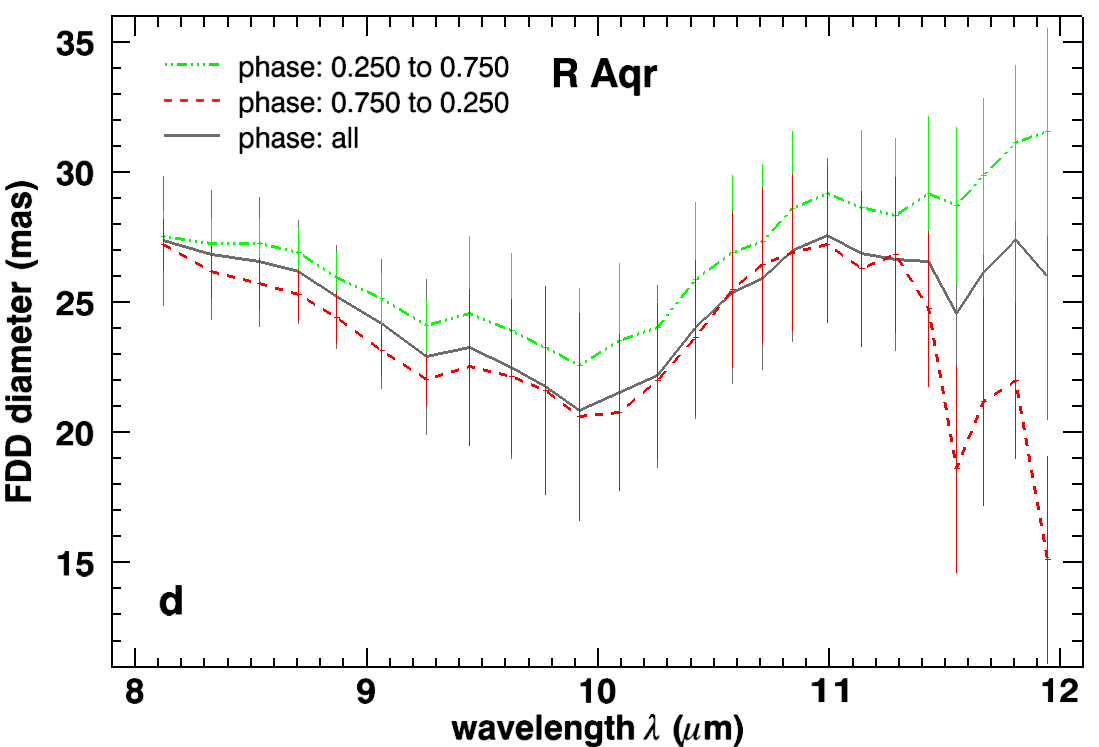}
    \includegraphics[width=0.45\linewidth]{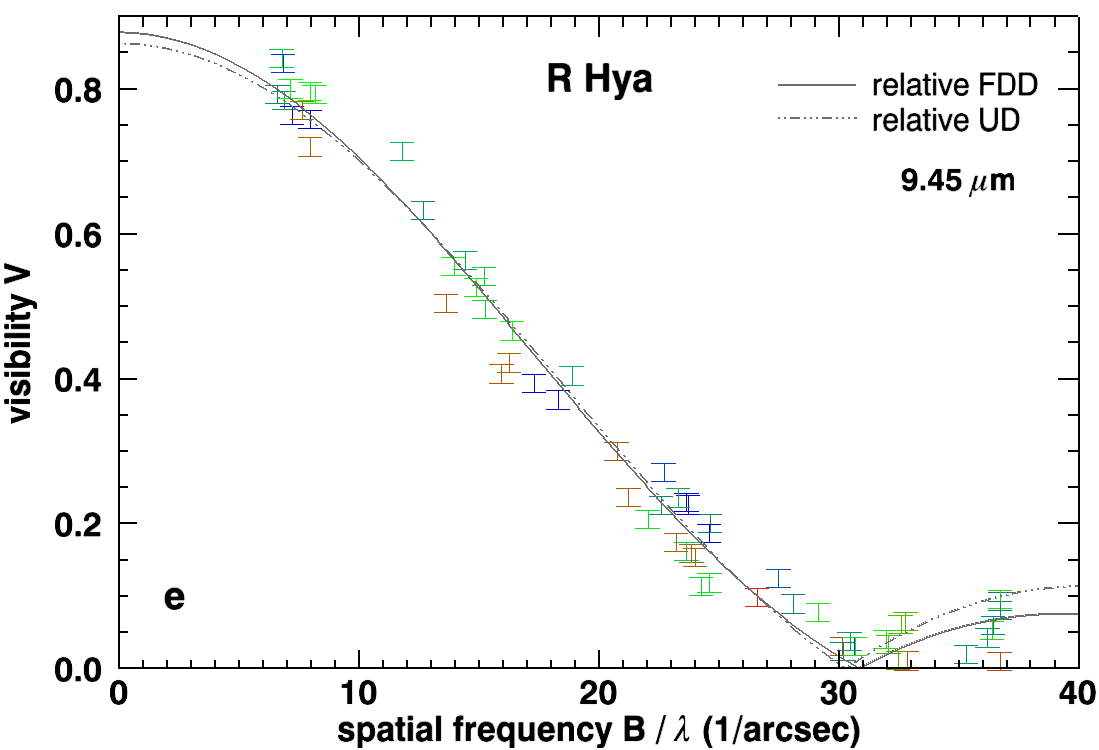}
    \hspace{0.2cm}
    \includegraphics[width=0.45\linewidth]{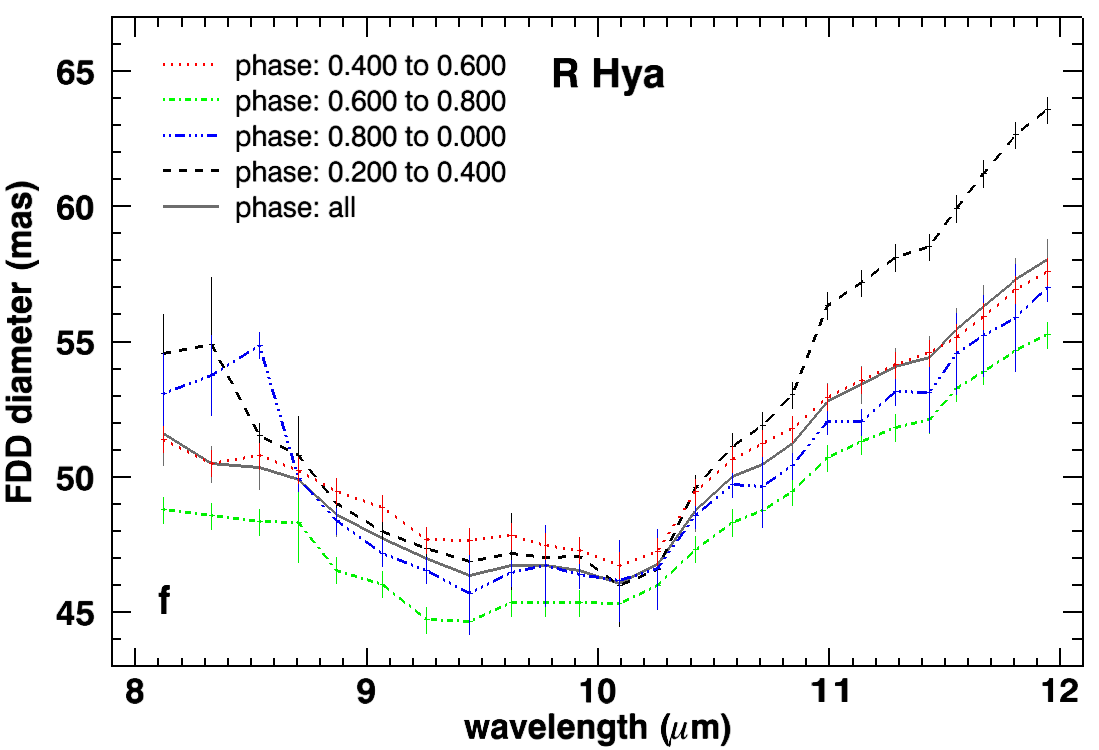}
    \includegraphics[width=0.45\linewidth]{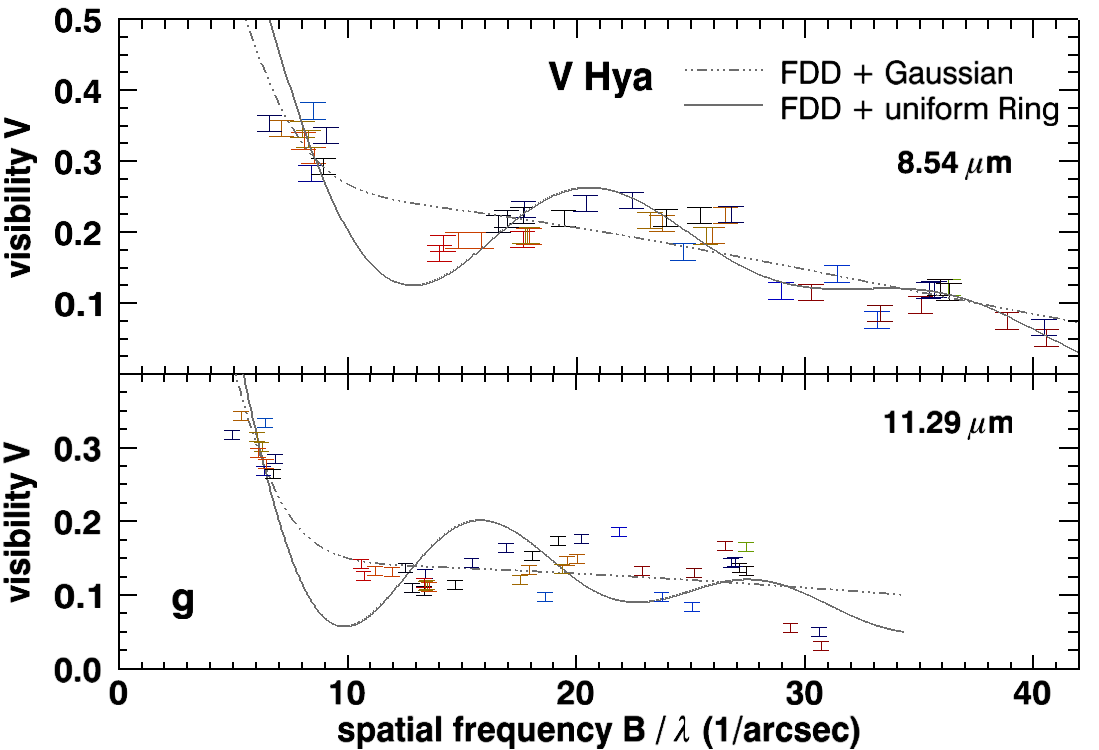}
    \hspace{0.2cm}
    \includegraphics[width=0.45\linewidth]{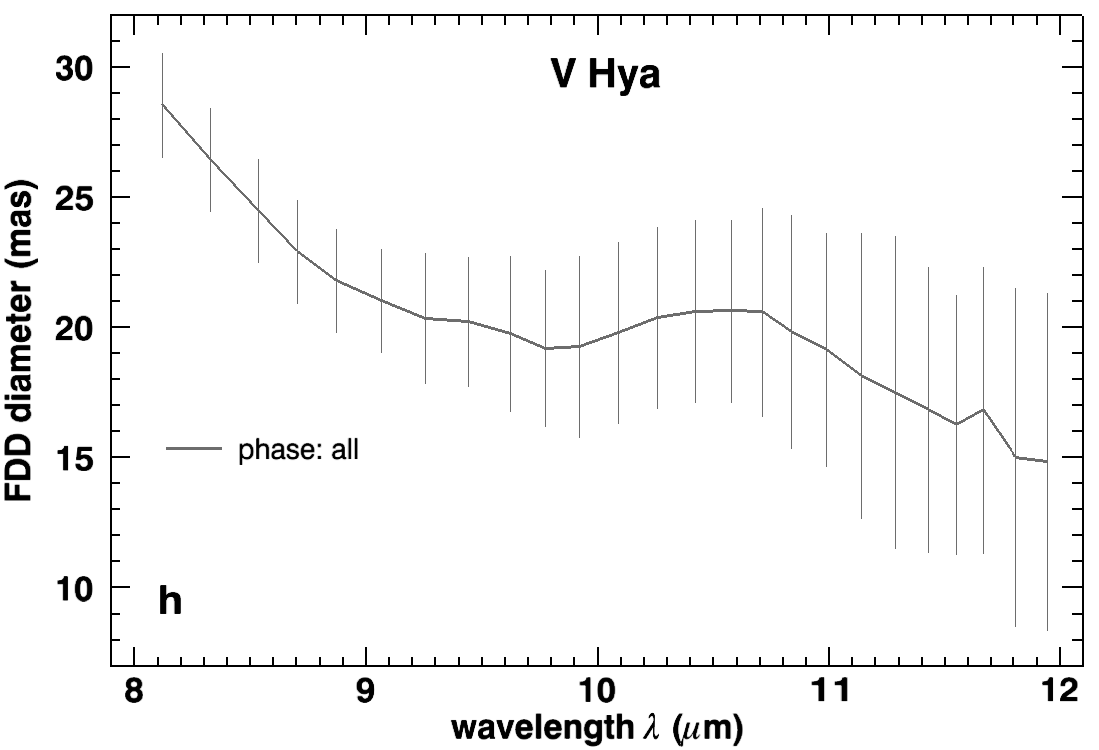}
    \caption{\textit{Left:} Fits of different models to the visibility measurements at a representative wavelength. The visibility data are plotted versus spatial frequency and are color-coded by visual light phase. \textit{Right:} The fitted circular fully limb-darkened disk diameter, $\theta_{\mathrm{FDD}}$, as function of wavelength for the full data set and selected pulsation phases. The corresponding plots for W~Hya can be found in paper~I (Fig.~6 and 8, respectively).}
    \label{FigVisFit}
   \end{figure*}
%---------------------------------------------------------------

   The final calibrated visibilities are shown in Fig.~\ref{FigVisibility} as function of wavelength, whereas the assigned visibility errors are given in Tables~\ref{TableResultsRAql} to~\ref{TableResultsVHya}. As described in paper~I, the visibility uncertainties are assumed to be the same within each wavelength bin to compensate for difficulties in determining them and to ensure a better fitting. The best model fits are displayed in the left hand panels of Fig.~\ref{FigVisFit} for each star. The fit and the data, color-coded by visual light phase, are plotted as function of spatial frequency (projected baseline divided by wavelength) for a representative wavelength. It is important to note that different spatial frequencies probe different regions, i.e.~high spatial frequencies are sensitive to small regions and low spatial frequencies to extended structures.

%---------------------------------------------------------------
   \begin{figure*}
    \centering
    \includegraphics[width=0.48\linewidth]{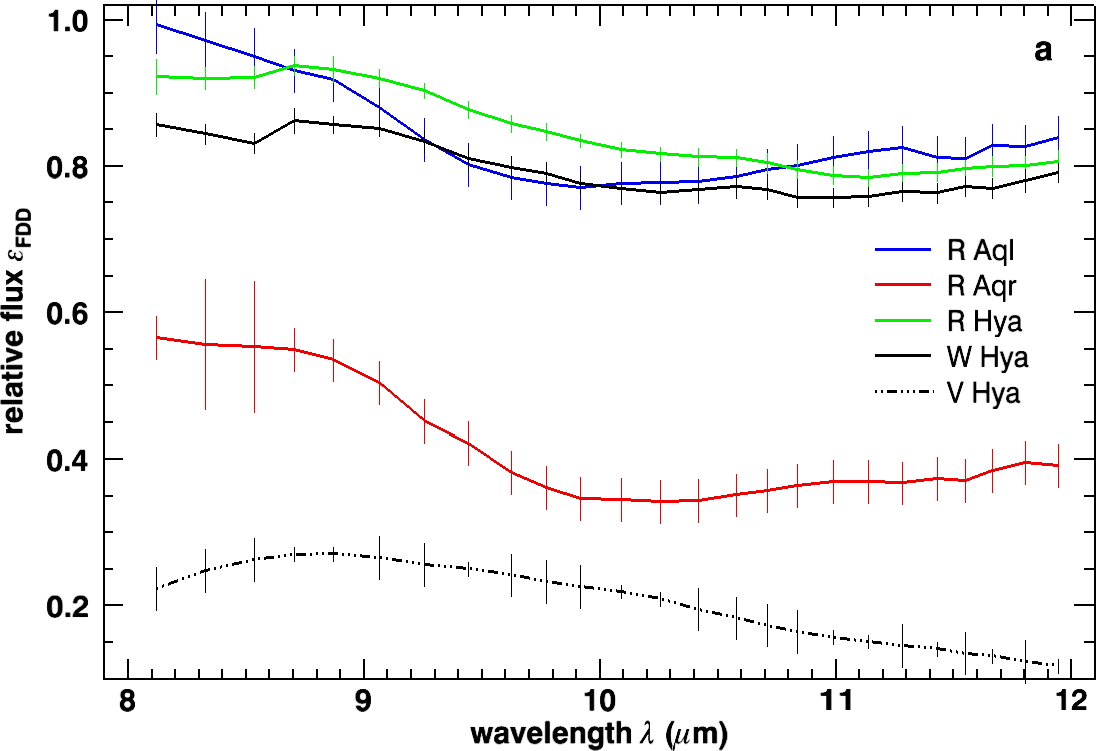}
    \hspace{0.1cm}
    \includegraphics[width=0.48\linewidth]{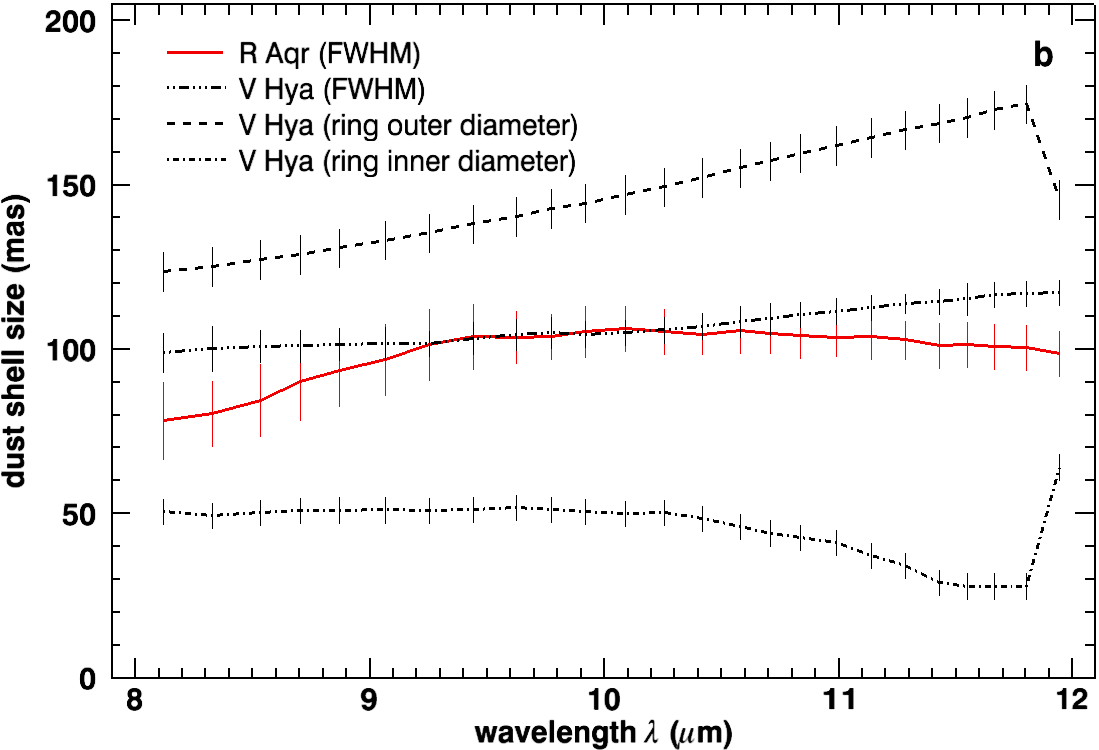}
    \caption{\textit{Left:} The relative flux contributions, $\epsilon_{\mathrm{FDD}}$, of the circular FDD obtained for the fits to the full data sets. \textit{Right:} Gaussian FWHM, $\theta_{\mathrm{G}}$, and ring, $\theta_{\mathrm{R,in}}$ and $\theta_{\mathrm{R,out}}$, dust shell sizes obtained from the fit to the full data set for two of the five stars as function of wavelength.}
    \label{FigFluxDust}
   \end{figure*}
%---------------------------------------------------------------

   The best model parameters with errors are derived by performing the Levenberg-Marquardt least-squares minimization procedure programmed for the interactive data language IDL as \texttt{MPFIT} by C.~B.~Markwardt\footnote{http://cow.physics.wisc.edu/$\sim$craigm/idl/idl.html} as described in paper~I. The FDD diameter as function of wavelength is shown in the right hand panels of Fig.~\ref{FigVisFit} for the full data set and selected pulsation phase bins (details on the dynamic behavior are discussed in Sect.~\ref{secIntDis_Intra}). The according numerical values are listed in Tables~\ref{TableResultsRAql} to~\ref{TableResultsVHya}. The relative flux contribution of the FDDs and the dust shell sizes as function of wavelength are plotted in Fig.~\ref{FigFluxDust}. The results for each star are described in the following in more detail and are summarized in Table~\ref{Table_Summary}.

%---------------------------------------------------------------
% bei 5 Sternen: width=0.260 and width=0.450 and \hspace{1.2cm}
% bei 4 Sternen: width=0.335 and width=0.590
   \begin{figure*}
    \centering
    \includegraphics[width=0.29\linewidth]{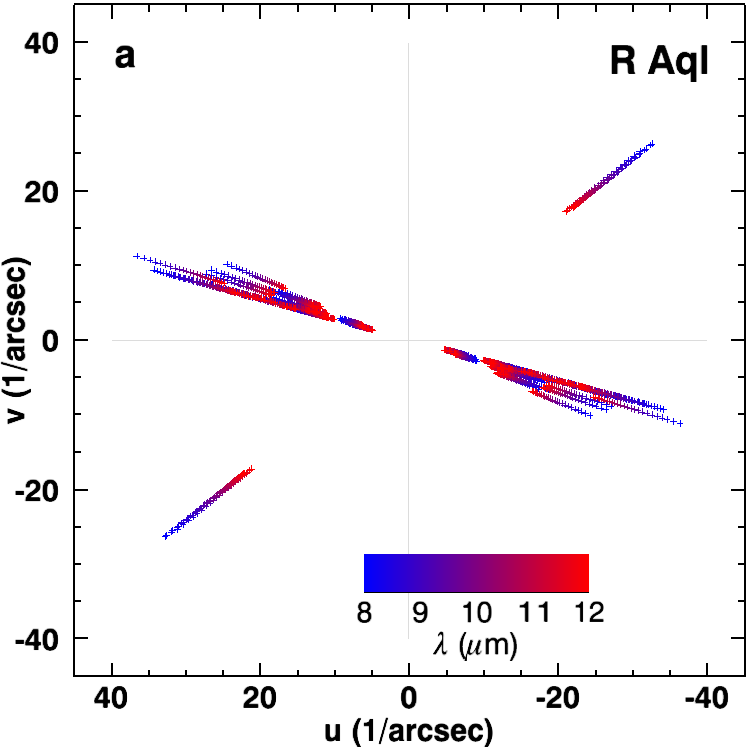}
    \hspace{1.2cm}
    \includegraphics[width=0.52\linewidth]{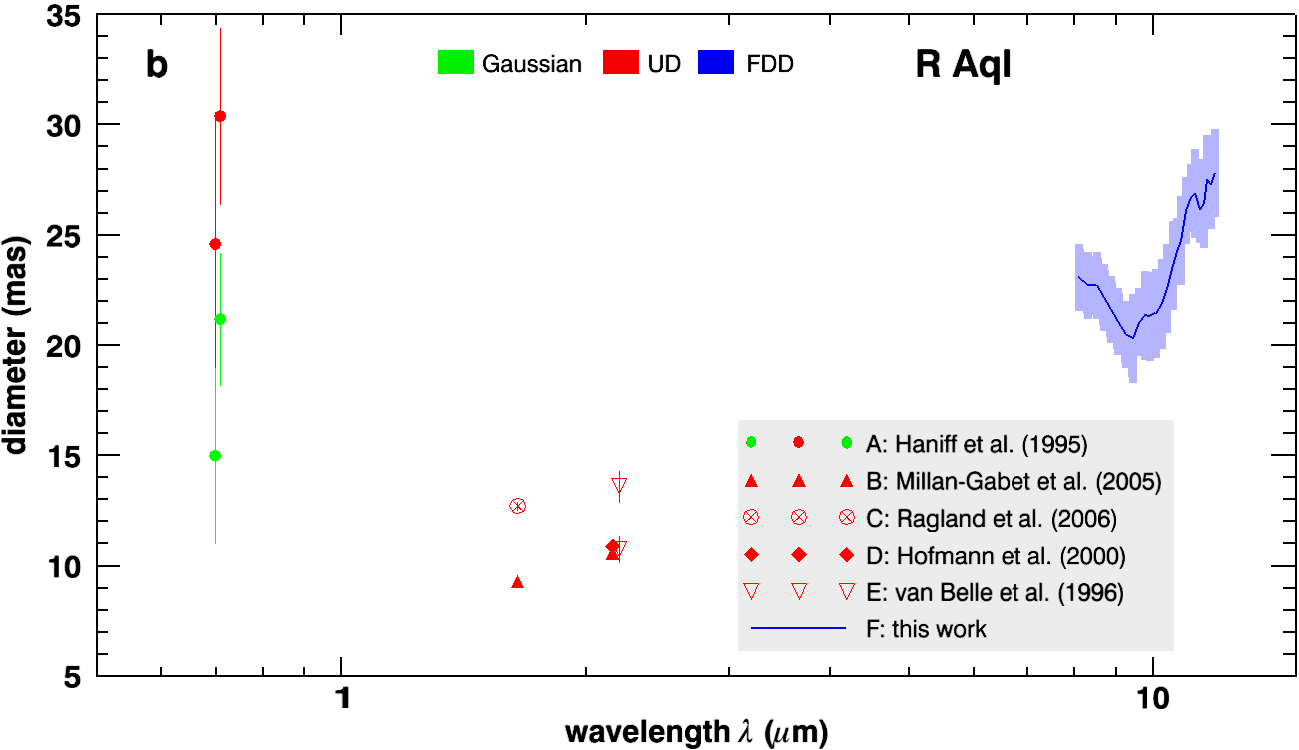}
    \hspace{3.2cm}
    \includegraphics[width=0.29\linewidth]{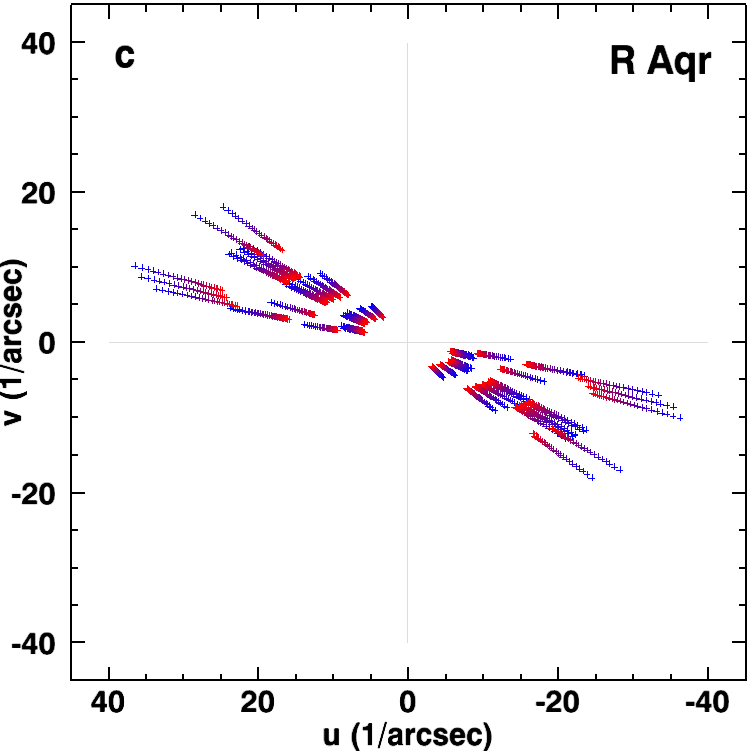}
    \hspace{1.2cm}
    \includegraphics[width=0.52\linewidth]{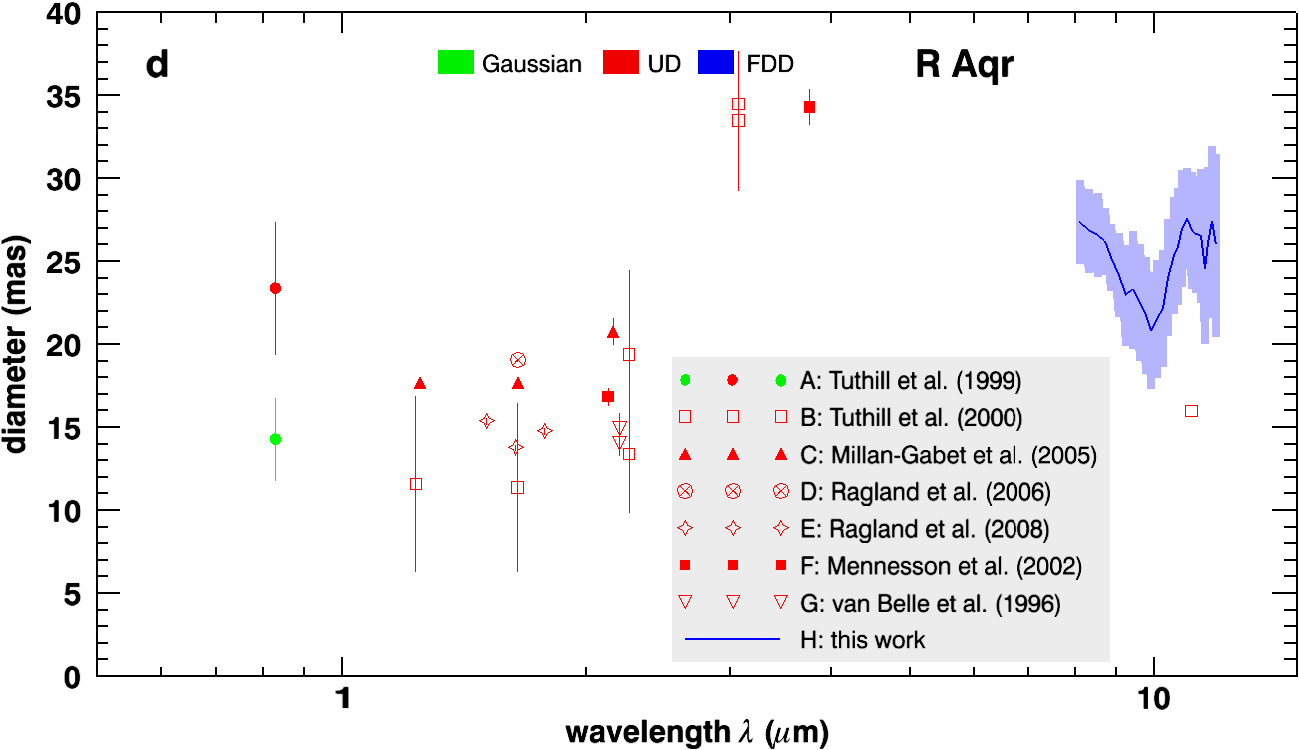}
    \hspace{3.2cm}
    \includegraphics[width=0.29\linewidth]{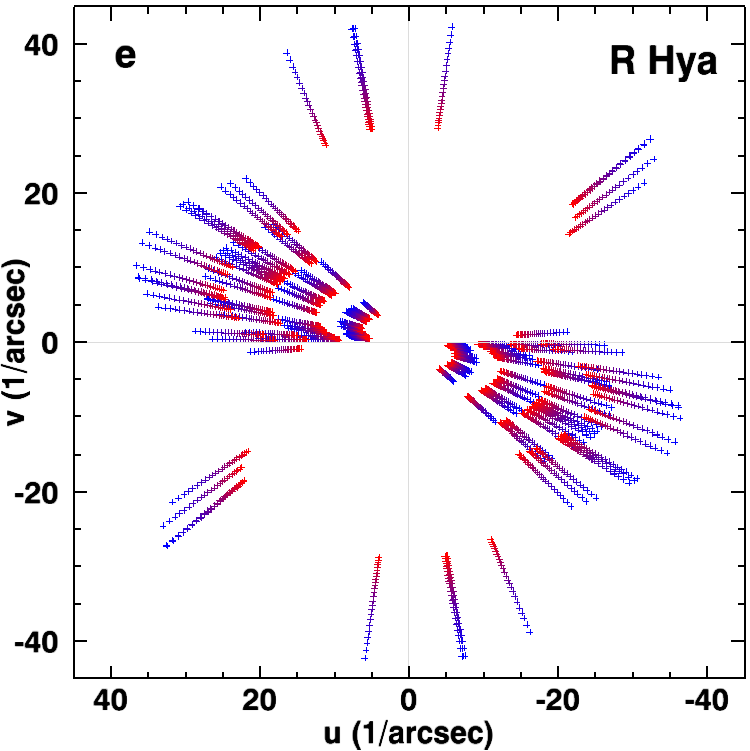}
    \hspace{1.2cm}
    \includegraphics[width=0.52\linewidth]{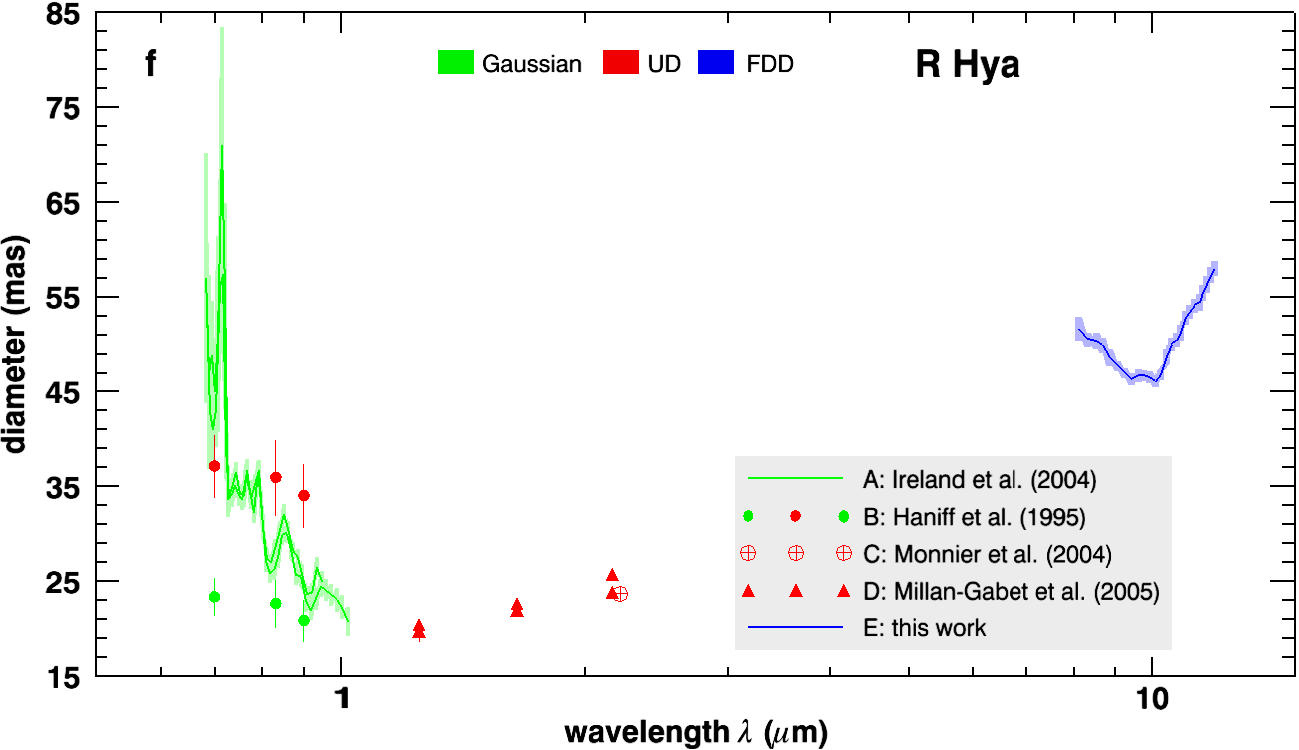}
    \hspace{3.2cm}
    \includegraphics[width=0.29\linewidth]{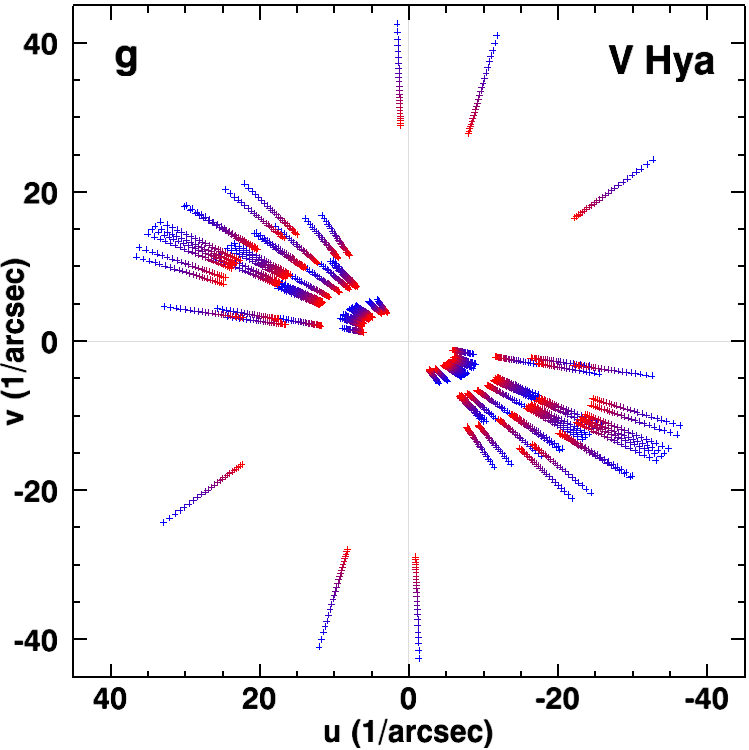}
    \hspace{1.2cm}
    \includegraphics[width=0.52\linewidth]{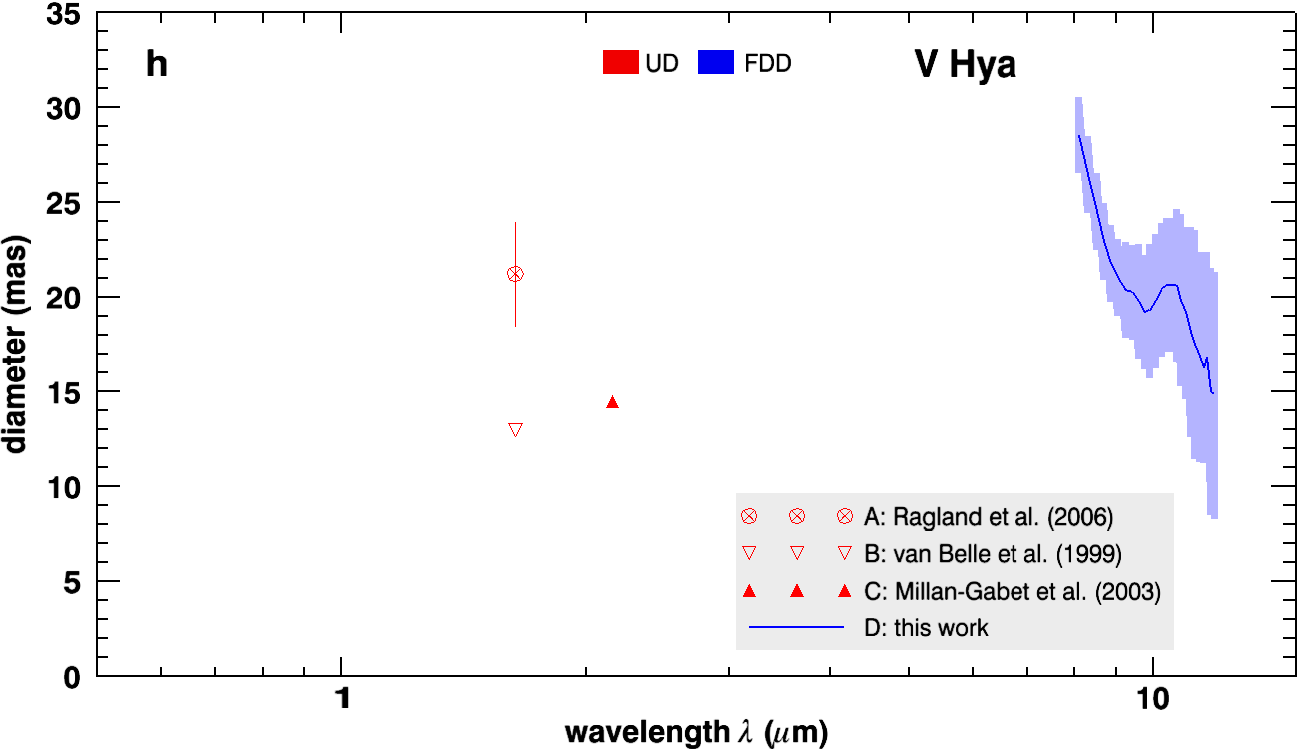}
    \caption{\textit{Left:} UV-coverage of all used interferometric observations. The visibility spectra are binned into 25 wavelength bins. \textit{Right:} Diameter measurements over a wide wavelength range with different models reported by various authors (FDD~$\approx$ 1.15~UD~$\approx$ 1.68~Gaussian). See Sect.~\ref{secResSWaveDep} and Table~\ref{TableDia_Targets} (Appendix~\ref{App_TableDia}) for more details. The corresponding UV-plot for W~Hya can be found in paper~I (Fig.~1) and an updated diameter plot for W~Hya is shown in Fig.~\ref{FigDiaWHya}.}
    \label{FigUVdia}
   \end{figure*}
%---------------------------------------------------------------

\subsubsection{R~Aql and R~Aqr}\label{secResSSubRAqlAqr}

   The individual visibility measurements of R~Aql and R~Aqr (Figs.~\ref{FigVisFit}a+c) have a relatively high scatter. This is caused mainly by the low N-band flux of the target and the calibrator and therefore a low signal-to-noise ratio of the fringe signal, but also due to the fact that the plot contains observations made at different pulsation phases and pulsation cycles. Intrinsic asymmetries of the objects might have an influence on the scatter as well.

   The distribution of the visibilities suggests that two components contribute to the total visibility, since for both stars the values are significantly below one at low spatial frequencies. This can be attributed to the presence of a dust shell, which was already inferred from the MIDI spectra and its comparison with the ISO spectra. However, from Figs.~\ref{FigVisFit}a+c, it is clear that the size of the dust shell can only be estimated for R~Aqr. Due to the lack of measurements at very low spatial frequencies this is not possible for R~Aql (a lower limit is obtained from the limited FoV, cf.~Sect.~\ref{secObsSubSpec}).
   
   Even though the first zero is not present, the clear decline with increasing spatial frequency makes it possible to fit a circular FDD or circular UD as second component to the data. A UD is in general best suited for a first diameter estimation of a star and is therefore often used in the literature. However, the investigation of the O-rich stars R~Hya and W~Hya, presented below, suggests that an equally simple geometrical FDD describes the observed visibilities in O-rich stars much better in the second lobe. However, both models are not distinguishable in the case of R~Aql and R~Aqr (Figs.~\ref{FigVisFit}a+c).

   The behavior of the FDD diameter, $\theta_{\mathrm{FDD}}$, as function of wavelength in the N-band is different for both stars (Figs.~\ref{FigVisFit}b+d and Tables~\ref{TableResultsRAql}+\ref{TableResultsRAqr}). While for R~Aql, $\theta_{\mathrm{FDD}}$ only slightly decreases from (23.1~$\pm$~1.5)~mas to (21.3~$\pm$~2.0)~mas, going from 8~$\mu$m to 10~$\mu$m, the relative decrease for R~Aqr, from (27.4~$\pm$~2.5)~mas to (20.9~$\pm$~3.5)~mas, is much higher. The following relative increase, $\theta_{12 \mu\mathrm{m}}$/$\theta_{10 \mu\mathrm{m}}$, beyond 10~$\mu$m is however similar, being (31~$\pm$~15)\% for R~Aql and (24~$\pm$~34)\% for R~Aqr. This apparent diameter increase from about 21~mas to (27.8~$\pm$~2.0)~mas and about 21~mas to (26.0~$\pm$~5.5)~mas is equivalent to an increase from 4.6~AU to 6.2~AU and 5.3~AU to 6.5~AU at the assumed distance of R~Aql and R~Aqr, respectively.

   In comparison, the relative flux contribution of the FDD component, $\epsilon_{\mathrm{FDD}}$, decreases from (0.99~$\pm$~0.02) to (0.80~$\pm$~0.02) and from (0.57~$\pm$~0.03) to (0.39~$\pm$~0.03) for R~Aql and R~Aqr, respectively, going to longer wavelengths (Fig.~\ref{FigFluxDust}a). This decrease reflects the increased flux contribution from the colder surrounding dust shell. In contrast to R~Aql, the relative mid-IR flux contribution of the star/molecule layer of R~Aqr is considerably lower because of the larger amount of circumstellar dust bound in the symbiotic system.

   For R~Aqr, measurements at spatial frequencies lower than 10~arcsec$^{-1}$ have larger visibilities than expected for a pure FDD. This is illustrated in Fig~\ref{FigVisFit}c by fitting a FDD to the measurements excluding the ones shortward of 10~arcsec$^{-1}$. This gives the possibility to constrain the inner extension of the silicate dust shell interferometrically. However, the Gaussian fit is dominated by one point at around 6~arcsec$^{-1}$, and therefore the result should be taken with caution. On the other hand, this single measurement was taken close in time to the points at around 7.5~arcsec$^{-1}$ and has a position angle not too far off.

   The fitted Gaussian FWHM diameter for R~Aqr steadily increases from (78~$\pm$~12)~mas to (104~$\pm$~7)~mas between 8 and 10~$\mu$m, and stays around this value between 10 and 12~$\mu$m (Fig.~\ref{FigFluxDust}b). This is equivalent to an increase from 20~AU to 26~AU and sets the characteristic silicate dust shell radius at 10~$\mu$m at a distance of about 7.6~$\pm$~1.5~times the photospheric radius. The constant or even slightly declining FWHM at wavelengths longward of 10~$\mu$m could be a hint that the dust shell is truncated due to the gravitational conditions set by the orbiting WD with a semi-major axis of around 15~AU (60~mas). (cf.~Sect.~\ref{secPropSSubRAqr}). With the sudden decline of the dust density, cold dust in outer regions, probed by wavelengths longer than 10~$\mu$m, is less abundant and can therefore not be detected. This would be also consistent with the low mass-loss rate reported for R~Aqr.

\subsubsection{R~Hya and W~Hya}\label{secResSSubRHyaWHya}

   The relatively high flux of the target and the calibrator entails a comparatively small scatter in the visibility measurements of R~Hya and W~Hya (Fig.~\ref{FigVisFit}e, and Fig.~6 in paper~1, respectively). However, there is still a considerable spread due to including observations obtained at different pulsations phases, pulsation cycles and position angles. From the investigation of W~Hya in paper~I (Fig.~6) and the plot for R~Hya in Fig.~\ref{FigVisFit}e it can be seen that a circular FDD model can well reproduce the visibility amplitude in the second lobe. % A UD would give a higher second lobe in Fourier space.

   The FDD diameter as function of wavelength in the N-band is similar for both stars (Fig.~\ref{FigVisFit}f, and Fig.~8 in paper~1). They are qualitatively also comparable to R~Aql. While the FDD diameter of R~Hya decreases only marginally from (51.6~$\pm$~1.2)~mas to (46.6~$\pm$~0.6)~mas between 8 and 10~$\mu$m, $\theta_{\mathrm{FDD}}$ stays almost constant at a value of about (80~$\pm$~1.2)~mas for W~Hya. At wavelengths longer than 10~$\mu$m the apparent diameters gradually increase again and reach (58.0~$\pm$~0.8)~mas and (105~$\pm$~1.2)~mas at 12~$\mu$m, corresponding to a relative increase, $\theta_{12 \mu\mathrm{m}}$/$\theta_{10 \mu\mathrm{m}}$, of (24~$\pm$~2)\% and (31~$\pm$~3)\% for R~Hya and W~Hya, respectively. The apparent diameter increase from 47~mas to 58~mas and 80~mas to 105~mas is equivalent to an increase from 6.1~AU to 7.5~AU and 7.1~AU to 9.5~AU at the distance of R~Hya and W~Hya, respectively. 

   In order to account for the flux contribution of an extended dust shell, the visibility function is not forced to be~1 at zero spatial frequency. Figure~\ref{FigFluxDust}a shows that with longer wavelengths the relative flux contribution of the FDD, $\epsilon_{\mathrm{FDD}}$, decreases from (0.92~$\pm$~0.03) to about (0.80~$\pm$~0.02) and from (0.85~$\pm$~0.02) to about (0.77~$\pm$~0.02) for R~Hya and W~Hya, respectively, reflecting the increased flux contribution from the colder surrounding dust shell. This is very similar to R~Aql.

\subsubsection{V~Hya}\label{secResSSubVHya}

   The scatter in the visibility measurements in the only C-rich AGB star in the sample is acceptable, since the target and the calibrator have a comparatively high flux (Fig.~\ref{FigVisFit}g). However, there is a considerable spread notably at higher spatial frequencies. This can probably be attributed to the presence of the temporally and spatially changing high velocity outflows (cf.~Sect.~\ref{secPropSSubVHya}).

   A model consisting of two components is necessary to describe the visibility measurements. One component is needed to explain the moderate decrease of the visibilities from 10 to 40~arcsec$^{-1}$. As for the other stars, this could be either a UD or an FDD. Both geometrical models cannot be distinguished by these observations. As for the O-rich stars, the FDD will be preferred in the following.

   The second component is essential to account for the points with high visibilities at spatial frequencies around 7~arcsec$^{-1}$. Similar to O-rich stars, a Gaussian, representing a carbon-rich (AMC + SiC) dust shell, could be an adequate function. However, the sinusoidal visibility variation between 10 and 20~arcsec$^{-1}$ suggests a brightness distribution with steep edges. A uniform circular ring with a sharp inner and outer edge could be a proper function describing a dust shell and causing a sinusoidal modulation. Since the sinusoidal effect is only weak at longer wavelengths the dust shell is better represented by a Gaussian in that wavelength regime.

   This is shown in Fig.~\ref{FigVisFit}g in more detail. At shorter wavelengths the FDD~+~ring model fits apparently better (upper panel, 8.5~$\mu$m), while at longer wavelengths the FDD~+~Gaussian model is better (lower panel, 11.3~$\mu$m). However, a comparison of the reduced chi square values,~$\chi^2_\mathrm{r}$, in Table~\ref{TableResultsVHya} reveals that the FDD~+~Gaussian model is actually the best representation across the whole N-band. Nevertheless, the $\chi^2_\mathrm{r}$'s are very similar between 8 and 9~$\mu$m for both models.
   
   Due to the lack of measurements at very low spatial frequencies the boundaries of the ring and the FWHM of the Gaussian are less well constrained, respectively. Since the fit included all measurements, morphological and variability effects are expected to have a non-negligible influence on the visibility measurements for this star. In particular, the ring dimensions should be taken with caution and the Gaussian FWHM should be used as a rough estimation of the inner dust shell boundary.
   
   The trend of the FDD diameter as function of wavelength in the N-band is shown in Fig.~\ref{FigVisFit}h. It can be inferred from this plot (and Table~\ref{TableResultsVHya}) that $\theta_{\mathrm{FDD}}$ decreases from (28.5~$\pm$~2.0)~mas to (14.9~$\pm$~6.5) mas between 8 and 12~$\mu$m with a local maximum at around 10.6~$\mu$m. This corresponds to a relative decrease $\theta_{12 \mu\mathrm{m}}$/$\theta_{8 \mu\mathrm{m}}$ of (48~$\pm$~24)\%. V~Hya is the only star in the sample exhibiting a strong diameter decrease from short to long wavelengths. The diameter decrease from 28.5~mas to 14.9~mas is equivalent to a decrease from 10.3~AU to 5.4~AU at the distance of V~Hya.

   The relative flux contribution of the FDD, $\epsilon_{\mathrm{FDD}}$, decreases from a maximum of (0.27~$\pm$~0.03) at 8.8~$\mu$m to (0.12~$\pm$~0.02) at 12~$\mu$m, reflecting the increased flux contribution from the colder surrounding AMC and SiC dust shell detected at longer wavelengths (Fig.~\ref{FigFluxDust}a and Sect.~\ref{secIntDis_CShell}). This decrease is similar to the O-rich stars in the sample. As for R~Aqr, the flux contribution of the star/molecule layer is considerably lower. This is probably again related to the large amount of dust bound in this system. In addition, the SiC dust shell with a spectral feature in the MIDI spectrum at around 11.3~$\mu$m could also explain part of the decrease of the relative flux contribution at longer wavelengths.

   The Gaussian FWHM diameter,~$\theta_{\mathrm{G}}$, on the other hand steadily increases from (99~$\pm$~6)~mas to (117~$\pm$~4) mas between 8 and 12~$\mu$m (Fig.~\ref{FigFluxDust}b). This is equivalent to an increase from 36~AU to 42~AU and sets the characteristic AMC dust shell radius at 10~$\mu$m at a distance of 8.7~$\pm$~0.4~times the photospheric radius. This dust condensation radius is at a similar distance as for the symbiotic O-rich star R~Aqr.

   The increase of~$\theta_{\mathrm{G}}$ with longer wavelengths is consistent with the results for the FDD~+~ring model. Figure~\ref{FigFluxDust}b shows that the outer ring diameter, $\theta_{\mathrm{r,out}}$, increases with increasing wavelength as well. Notable is also that the inner ring diameter, $\theta_{\mathrm{r,in}}$, approaches the FDD diameter at longer wavelengths.

%---------------------------------------------------------------
   \begin{figure}
     \centering
     \includegraphics[width=0.99\linewidth]{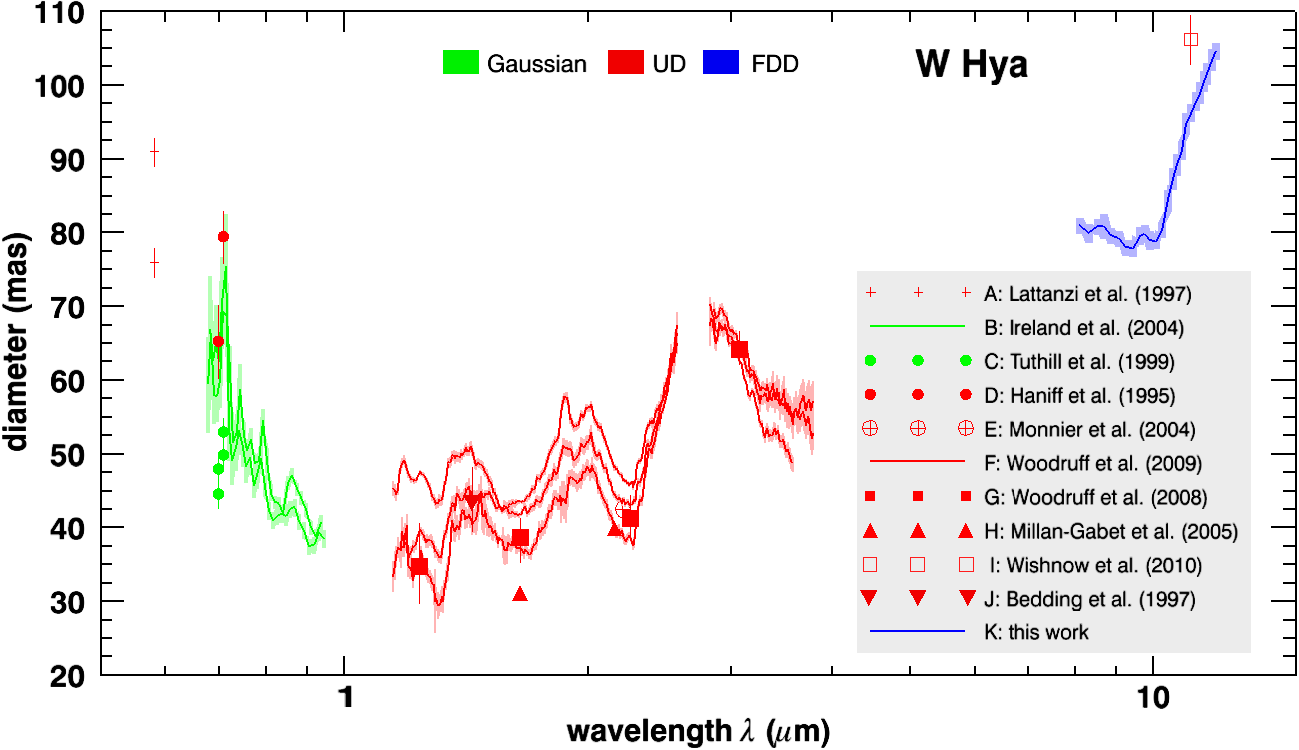}
     \caption{Same as right hand panels of Fig.~\ref{FigUVdia} but for W~Hya (updated version from paper~I, Fig.~9).}
     \label{FigDiaWHya}
   \end{figure}
%---------------------------------------------------------------

\subsection{Wavelength dependence of the diameter}\label{secResSWaveDep}

   In the following discussion, one has to keep in mind that the compared diameters and diameter ratios (summarized in Table~\ref{Table_Summary}) are averaged over the position angle, and are averages over the pulsation phase or may represent only a certain pulsation phase. It is also important to recognize that the FDD diameters obtained with MIDI describe a region whose exact location depends on the flux contribution of all constituents (continuum photosphere, atmospheric molecular layers and nearby dust shells; see next section) as function of wavelength and pulsation phase, and not only the photosphere of the star.

   The right hand panels of Fig.~\ref{FigUVdia} and Fig.~\ref{FigDiaWHya} show the obtained FDD diameters in relation to interferometric angular diameter determinations reported by various authors from the visual to the mid-IR ($0.6-12$~$\mu$m). They are obtained by fitting a Gaussian, a uniform disk or a fully limb-darkened disk to the data. A conversion between the models is not performed, since the various diameter determinations depend on the number of visibility measurements and their spatial frequency distribution the authors used to fit their model\footnote{Empirical conversion factors are approximately: FDD $\approx$ 1.15~UD $\approx$ 1.68~FWHM.}. Information on visual phases and position angles (if applicable) of these observations can be found in Table~\ref{TableDia_Targets} (Appendix~\ref{App_TableDia}).

   The observed apparent diameter changes dramatically within the given wavelength range due to the strong wavelength dependent opacity of the atmospheric constituents \citep{Baschek1991,Scholz2001}. For the O-rich stars, the measured diameters in the optical are sensitive to TiO bands. The largest variations are around the strongest bands at 712~nm and 670~nm with apparent diameter enlargements of up to a factor of two. Additionally, light scattered by dust \citep{Norris2012} might also cause a large increase in apparent diameter towards the blue \citep{Ireland2004a}.

   In the near-IR, predominantly \ce{H2O} and CO in different layers are responsible for the wavelength dependence of the diameter \citep{Hofmann1998,Jacob2000}. By comparing the observations conducted at J, H, K and L~band (1.25, 1.65, 2.16 and 3.8~$\mu$m, respectively), one can again infer from the right hand panels of Fig.~\ref{FigUVdia} and Fig.~\ref{FigDiaWHya} that also in the near-IR diameters vary by up to a factor of two, and in particular for W~Hya that there is a complex diameter dependence on pulsation cycle and pulsation phase \citep[see][for a more detailed discussion on this]{Woodruff2009}. This has been reported for a large number of AGB stars \citep[e.g.][]{Weiner2003}. Unfortunately, for V~Hya, only H and K-band observations have been published. Diameter determinations for C-stars in these bands are mainly influenced by the presence of molecular shells consisting of \ce{C2H2}, HCN, CN and CO \citep{GautschyLoidl2004,Paladini2009}.

   Emission from multiple layers and contamination by nearby continuum emission have a strong influence on determining the true photospheric extension of an AGB star through interferometric measurements. However, a reasonable estimate for the photospheric diameter, $\theta_{\mathrm{phot}}$, can be obtained from line free measurements at K-Band. The UD diameter at K-band is approximately 1.2~times the true photospheric diameter \citep[cf.~e.g.][]{MillanGabet2005}\footnote{However, for C-rich AGB stars, the K-band is probably not a good tracer for the true photosphere.}. The estimated $\theta_{\mathrm{phot}}$ used in this study are listed in Table~\ref{Table_Summary}. They were derived by averaging the K-band diameters shown in the right hand panels of Fig~\ref{FigUVdia} and Fig.~\ref{FigDiaWHya} (cf.~also Table~\ref{Table_Phenomenology} and Table~\ref{TableDia_Targets}) and dividing them by 1.2 ($\theta_{\mathrm{UD,K band}}$/1.2). The errors are obtained from the standard deviation.

   Maser observations of different molecules give additional diameter constraints. For O-rich stars, SiO and \ce{H2O} masers probe inner regions where the molecular layers are present and the first dust formation takes place, while OH masers trace wind regions farther out. Ring diameters for SiO masers, $\theta_{\mathrm{SiO}}$, were measured at 43.1 and 42.8~GHz for R~Aql, R~Aqr and W~Hya by the authors given in Sect.~\ref{secObsSubTar}. The averages of these diameters are listed in Table~\ref{Table_Summary}. A comparison of the location of the SiO masers with the photospheric diameters shows that for all three stars the SiO masers occur at approximately the same distance from the star. The ratios, $\theta_{\mathrm{SiO}}\,$/$\,\theta_{\mathrm{phot}}$, are between 2.2 and 2.5 (Table~\ref{Table_Summary}).

   Except for R~Aqr and V~Hya, a similar ratio of 2.2 can be found if the FDD diameters at 10~$\mu$m are compared with the photospheric diameters ($\theta_{\mathrm{FDD},10\mu\mathrm{m}}\,$/$\,\theta_{\mathrm{phot}}$). The SiO masers are therefore almost co-located with the region characterized by the FDD diameter (cf.~next section). 

   For R~Aqr and V~Hya, the determined FDD diameters lie closer to the true photometric diameter. This might be again related to the fact that both stars are close binary systems with a large dust content and apparently a closer formation of constituents traced in the N-band.

   In addition, it is notable from Fig.~\ref{FigUVdia}d that the apparent diameters of R~Aqr at 3.1~$\mu$m and within the L-band are higher compared to the K and N-band. This might be caused by an extended halo of OH as a result of the ionizing radiation of the compact close companion or the existence of hot circumstellar dust \citep{Tuthill2000}.
   
   \citet{MillanGabet2003} obtained for the AMC dust shell in V~Hya in the K-band a FWHM diameter, $\theta_{\mathrm{G},2.2\mu\mathrm{m}}$, of (35~$\pm$~3)~mas (with a flux contribution of 0.63~$\pm$~0.02). This is 3.0~$\pm$~0.3 times smaller compared to the FWHM diameter, $\theta_{\mathrm{G},10\mu\mathrm{m}}$, determined at 10~$\mu$m, and is therefore consistent with the fact that at the N-band colder circumstellar material farther out is probed.
   
   In addition, the AMC dust shell with a FWHM of (113~$\pm$~4)~mas (and a flux contribution of 0.85~$\pm$~0.01) determined with MIDI for V~Hya is in good agreement with the inner Gaussian dust shell diameter of (98~$\pm$~2)~mas (and a flux contribution of the inner and outer shell of 0.78~$\pm$~0.01) found by \citet{Townes2011} at 11.15~$\mu$m (narrow bandwidth) with the Infrared Spatial Interferometer (ISI).

   It should be also noted that the FDD diameter measured for W~Hya of (96.3~$\pm$~1.2)~mas at around 11.15~$\mu$m, tracing the \ce{Al2O3} dust shell (see next section), is consistent with the average UD diameter of (106.2~$\pm$~3.4)~mas of the inner dust shell determined by \citet{Wishnow2010} with ISI. The inner shell flux contribution (including the stellar flux) is with 0.64~$\pm$~0.02 similar to the value of 0.85~$\pm$~0.01 (Table~\ref{TableResultsVHya}) measured with MIDI as well.

%###########################################################################################
%###########################################################################################
\section{Interpretation and discussion}\label{secIntDis}

%---------------------------------------------------------------
    \begin{table*}
      \caption{Summary of the results.}           
      \label{Table_Summary}           
      \centering           
      \begin{tabular}{lC{22mm}C{22mm}C{22mm}C{22mm}c}           
%          \noalign{\smallskip}           
          \hline           
%          \noalign{\smallskip}           
          \noalign{\smallskip}           
        Target:                                            & R Aql           & R Aqr           & R Hya           & W Hya          & V Hya          \\
          \noalign{\smallskip}           
          \hline           
%          \noalign{\smallskip}           
          \noalign{\smallskip}           
        Best model:                                        & relative        & FDD +           & relative        & relative       & FDD +          \\
                                                           & FDD             & Gaussian        & FDD             & FDD            & Gaussian       \\
          \noalign{\smallskip}           
        $\epsilon_{\mathrm{FDD}}$ layer$^{\mathrm{a}}$     & 0.77~$\pm$~0.02 & 0.35~$\pm$~0.03 & 0.84~$\pm$~0.01 & 0.78~$\pm$~0.01& 0.23~$\pm$~0.03\\
          \noalign{\smallskip}           
        $\theta_{\mathrm{FDD}}$ layer$^{\mathrm{a}}$ (mas):& 21.3~$\pm$~2.0  & 20.9~$\pm$~3.5  & 46.6~$\pm$~0.6  & 79.0~$\pm$~1.2 & 19.3~$\pm$~3.5 \\
                                                           & (4.7 AU)        & (5.2 AU)        & (6.1 AU)        & (7.1 AU)       & (6.9 AU)       \\
          \noalign{\smallskip}           
        $\theta_{\mathrm{G}}$ dust$^{\mathrm{a,b}}$ (mas): & $>$~1000        & 105~$\pm$~8     & $>$~1000        & $>$~1000       & 105~$\pm$~4    \\
                                                           & ($>$~220~AU)    & (26.3 AU)       & ($>$~130~AU)    & ($>$~90~AU)    & (37.8 AU)      \\
          \noalign{\smallskip}           
          \hline           
%          \noalign{\smallskip}           
          \noalign{\smallskip}
        assumed $\theta_{\mathrm{phot}}^{\mathrm{c}}$ (mas):& 9.6~$\pm$~1.2  & 13.8~$\pm$~2.5  & 20.4~$\pm$~1.0  & 35.7~$\pm$~2.7 & 12.1~$\pm$~0.3 \\
        assumed $\theta_{\mathrm{SiO}}^{\mathrm{d}}$  (mas):& 24.2~$\pm$~2.7  & 31.7~$\pm$~1.2  & $-$  & 77.3~$\pm$~10.5 & $-$ \\
          \noalign{\smallskip}           
          \hline           
%          \noalign{\smallskip}           
          \noalign{\smallskip}           
        $\theta_{\mathrm{FDD},12\mu\mathrm{m}}\,$/$\,\theta_{\mathrm{FDD},10\mu\mathrm{m}}$: & (31~$\pm$~15)\%   & (24~$\pm$~34)\%   & (24~$\pm$~2)\%   & (31~$\pm$~3)\% &  (18~$\pm$~5)\%$^{\mathrm{e}}$   \\
%          \noalign{\smallskip}           
          \noalign{\smallskip}           
        $\theta_{\mathrm{SiO}}\,$/$\,\theta_{\mathrm{phot}}$:    & 2.5~$\pm$~0.4 & 2.3~$\pm$~0.4 & $-$ & 2.2~$\pm$~0.3 & $-$ \\
%          \noalign{\smallskip}           
          \noalign{\smallskip}           
$\theta_{\mathrm{FDD},10\mu\mathrm{m}}\,$/$\,\theta_{\mathrm{phot}}$:& 2.2~$\pm$~0.3 & 1.5~$\pm$~0.4 & 2.3~$\pm$~0.1 & 2.2~$\pm$~0.2 & 1.6~$\pm$~0.3\\
%          \noalign{\smallskip}           
          \noalign{\smallskip}           
$\theta_{\mathrm{G},10\mu\mathrm{m}}\,$/$\,\theta_{\mathrm{phot}}^{\mathrm{b}}$: & $>$~100 & 7.6~$\pm$~1.5 & $>$~49 & $>$~28 &8.7~$\pm$~0.4 \\
          \noalign{\smallskip}           
          \hline           
          \noalign{\smallskip}           
         Phase-to-phase variations: &  &  &  &  & \\
$\overline{\theta}_{\mathrm{FDD,max}}\,$/$\,\overline{\theta}_{\mathrm{FDD,min}}$: & (18~$\pm$~12)\% & (14~$\pm$~19)\% & (8~$\pm$~3)\% & (13~$\pm$~4)\% & $-$ \\
$\overline{\epsilon}_{\mathrm{FDD,max}}\,$/$\,\overline{\epsilon}_{\mathrm{FDD,min}}$: & (13~$\pm$~10)\% & (18~$\pm$~23)\% & (10~$\pm$~10)\% & (7~$\pm$~10)\% & $-$ \\
          \noalign{\smallskip}           
          \hline           
%          \noalign{\smallskip}           
          \noalign{\smallskip}           
        Cycle-to-cycle variations:  & $-$ & $-$ & (5~$\pm$~3)\% & (5~$\pm$~4)\%  &  $-$ \\
          \noalign{\smallskip}           
          \hline           
%          \noalign{\smallskip}           
          \noalign{\smallskip}           
        Asymmetry:                 & no & yes & no            & yes            &  yes   \\
          \noalign{\smallskip}           
          \hline           
%          \noalign{\smallskip}           
      \end{tabular}           
      \newline 
      \begin{flushleft}
        \textbf{Notes. }
        $^{\mathrm{a}}$~At 10~$\mu$m. 
        $^{\mathrm{b}}$~Lower limits of the dust shell diameter (not Gaussian) for R~Aql, R~Hya and W~Hya are obtained from FoV restrictions (cf.~Sect.~\ref{secObsSubSpec}). 
        $^{\mathrm{c}}$~Assumed photospheric diameter derived from the K-band diameter (Table~\ref{Table_Phenomenology}) divided by 1.2 \citep[cf.~e.g.][]{MillanGabet2005}. 
        $^{\mathrm{d}}$~Assumed SiO ring diameter (average of the values given in Sect.~\ref{secObsSubTar}, but see also Sect.~\ref{secResSWaveDep}). 
        $^{\mathrm{e}}$~For V~Hya, $\theta_{\mathrm{G},12\mu\mathrm{m}}\,$/$\,\theta_{\mathrm{G},8\mu\mathrm{m}}$ is given.
      \end{flushleft}
    \end{table*}
%---------------------------------------------------------------

   Some of the important results from the last sections are summarized in Table~\ref{Table_Summary} and show that the MIDI observations sample the region above the extended pulsating atmosphere where the molecular layers are present and probably first seed particles for dust formation originate. This study can therefore also help to understand the wind acceleration processes in AGB stars.

   The obtained angular diameters as function of wavelength in the N-band are significantly different between the four O-rich stars and the one C-rich star in the sample. R~Aql, R~Aqr, R~Hya and W~Hya show a moderate to low diameter decrease from 8 to 10~$\mu$m and a strong diameter increase from 10 to 12~$\mu$m, while the FDD diameter of V~Hya essentially decreases from 8 to 12~$\mu$m. This is again shown in more detail in Fig.~\ref{FigDiaTemp}a. Since different constituents cause these differences in the shapes, both underlying chemistries are discussed separately in the following. General conclusions will be derived for oxygen and carbon-rich stars concerning their molecular shells and dust production. In addition, phase-to-phase variations and asymmetries are investigated.

   Possible physical mechanism and chemical processes in the probed region are constrained by the temperature among other things. In particular, the temperature indicates if dust can already condensate at the determined FDD distance. A rough estimate of the equivalent blackbody temperature can be derived from the total flux~$S_{\nu}$ (in Jy), the flux ratio~$\epsilon_{\mathrm{FDD}}$ and the diameter~$\theta_{\mathrm{FDD}}$ (in rad) measured with MIDI via
   \begin{eqnarray}
       T_b &=& \frac{h\nu}{k_{\mathrm{B}}} \, \frac{1}{\mathrm{ln}\left(\frac{2h\nu^3}{\epsilon_{\mathrm{FDD}} I_\nu c^2} + 1\right)} \; \mathrm{,} \label{Eq_brightT}
   \end{eqnarray}
   with $I_\nu\;=\;4 S_\nu / (10^{26}\pi\theta_{\mathrm{FDD}}^2)$ the spectral brightness, $\nu$ the frequency of the observation, $h$ the Planck constant, $k_{\mathrm{B}}$ the Boltzmann constant and $c$ the speed of light. The obtained temperatures are plotted in Fig.~\ref{FigDiaTemp}b for all stars. Due to the use of a simple FDD, giving a diameter which lies between the stellar photosphere and the outer boundary of the close molecular environment, the real extension of the molecular structure is under-evaluated. Therefore, this approximation may overestimate the temperature of the close molecular environment.

   R~Aql, R~Hya and W~Hya show very similar trends with temperatures between 1500 and 2000~K and an error on the order of 200~K. R~Aqr and V~Hya exhibit a temperature increase longward of 11~$\mu$m, but have considerably higher errors on the order of 500~K. The largest error comes from the uncertainty of the total flux calibration. However, the temperatures between 10 and 12~$\mu$m, i.e.~where a larger FDD diameter were measured, are sufficiently low for the O-rich stars so that dust particles are able to form (see the following sections).

%%%%%%%%%%%%%%%%%%%%%%%%%%%%%%%%%%%%%%%%%%%%%%%%%%%%%%%%%%%%%%%%%%%%%%%%%%%%%%%%%%%%%%%%%%%%
\subsection{Dust and molecular shells in O-rich AGB stars}\label{secIntDis_OShell}

\subsubsection{Spectral features and detectability}

   The discussion on the spectra in Sect.~\ref{secObsSubSpec} has shown that several molecules are present in the upper atmosphere. In the N-band, between 8 and 13~$\mu$m, strong pure-rotation lines of \ce{H2O} are expected. In addition, SiO exhibits fundamental bands between 8 and 10~$\mu$m \citep[e.g.][]{Decin2000}. Such quasi-static, warm and dense molecular layers close to the star, at typically $2-3$~photospheric radii ($R_{\mathrm{phot}}$), have been detected in O-rich AGB stars \citep[e.g.][]{Mennesson2002,Perrin2004,Ireland2004d,Woodruff2004,Fedele2005} and red supergiant stars \citep[e.g.][]{Perrin07}. These layers were introduced earlier to explain spectroscopic observations \citep[e.g.][]{Hinkle1979,Tsuji1997,Yamamura1999}.

   Besides the classical amorphous silicate feature at 9.7~$\mu$m, spectral dust features in the N-band probably originate from amorphous aluminum oxide (\ce{Al2O3}) at around 11.5~$\mu$m \citep{Begemann97} and spinel (\ce{MgAl2O4}) at around 13~$\mu$m \citep[e.g.][]{Posch1999,Fabian2001}. In particular, amorphous \ce{Al2O3} provides significant opacity for wavelengths longwards of 10~$\mu$m \citep{Koike1995,Begemann97,Posch1999,Woitke2006a,Ireland2006,Robinson_Maldoni2010}. In addition, features of crystalline aluminum oxide (corundum, $\alpha$-\ce{Al2O3}) at 12.7~$\mu$m and several modifications of titanium oxide (e.g.~rutile, \ce{TiO2}) might be present as well \citep[e.g.][]{Posch1999,Posch2002}.
   
   It is expected that in O-rich atmospheres oxides condensate first since silicon in silicates have a comparable low electron affinity for oxygen \citep[cf.][]{Stencel1990} and have therefore in general a lower condensation temperature due to the lower binding energy. The condensation sequence probably starts with \ce{TiO2} before \ce{Al2O3}, \ce{MgAl2O4} and \ce{Mg_{0.1}Fe_{0.9}O} forms \citep{Tielens1990,GailSedlmayr1998,Jeong1999,Blommaert2006,Lebzelter2006,Verhoelst09}. If most of the available high electron affinity metals (primarily Al, Mg, Fe, Ca) have been oxidized, silicates, like forsterite (\ce{Mg2SiO4}) or olivine (\ce{Mg_{0.8}Fe_{1.2}SiO4}), may condense and will be absorbed onto the seed particles. However, iron-free silicates, nearly transparent at the stellar flux maximum at around 1~$\mu$m, can survive at higher temperatures, thus also condensing at small radii \citep{Norris2012}, and are eventually able to trigger the mass-loss through photon scattering \citep{Hoefner08}. Iron-rich silicates, which are not transparent to the stellar radiation, thus absorbing the radiation and heat up, can only exist at larger radii from the star.

% For the wind formation process more important are the condensation radii of each dust species. While transparent iron-free silicates can survive at higher temperatures, thus smaller radii and eventually trigger the mass-loss through scattering, iron-rich silicates, not transparent to the stellar radiation, thus absorbing the radiation and heat up, can only exists at larger radii from the star.

   Concerning the detectability of the spectral features, two characteristics have to be considered. First, if the mass loss is relatively low the abundance of silicates will remain low (since most of the oxygen is bound in oxides) and the N-band will not be dominated by the classical silicate feature \citep{Sogawa_Kozasa1997}. This is also related to the geometrical thickness of the dust shell as described in \citet{Egan_Sloan2001} (referred to as silicate dust sequence): spectra showing the classical narrow 10~$\mu$m feature arise from optically thin shells dominated by amorphous silicates while spectra with broad low-contrast emission peaking around 11$-$12~$\mu$m arise from optically and geometrically thin shells composed primarily of alumina dust. This might be also evidence for different evolutionary stages of O-rich AGB stars \citep{Stencel1990,Posch2002,Lebzelter2006}. In earlier evolutionary stages the dust mass-loss rates are low and aluminum oxide dust dominates while at later stages, when effective mass loss has set in, iron-rich silicates will form in large amounts farther away from the star and will dominate the dust spectra \citep[e.g.][]{Woitke2006a}.

   The second point concerns the fact that an interferometer always acts as a spatial filter. The measured visibility at a given baseline, and therefore the determined diameter, is a function of the flux contribution (or opacity) of all emitting components at that spatial frequency. In addition, only flux contribution of components within the field of view (FoV) are measurable.

%---------------------------------------------------------------
   \begin{figure*}
   \centering
   \includegraphics[width=0.48\linewidth]{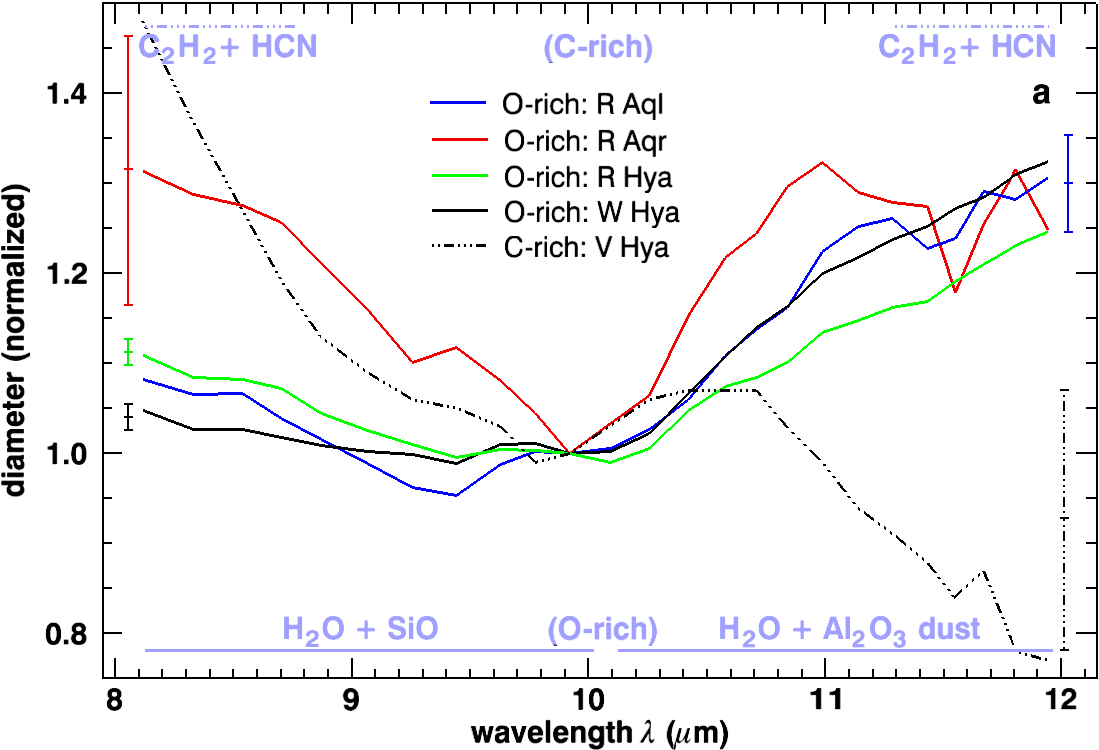}
   \hspace{0.1cm}
   \includegraphics[width=0.48\linewidth]{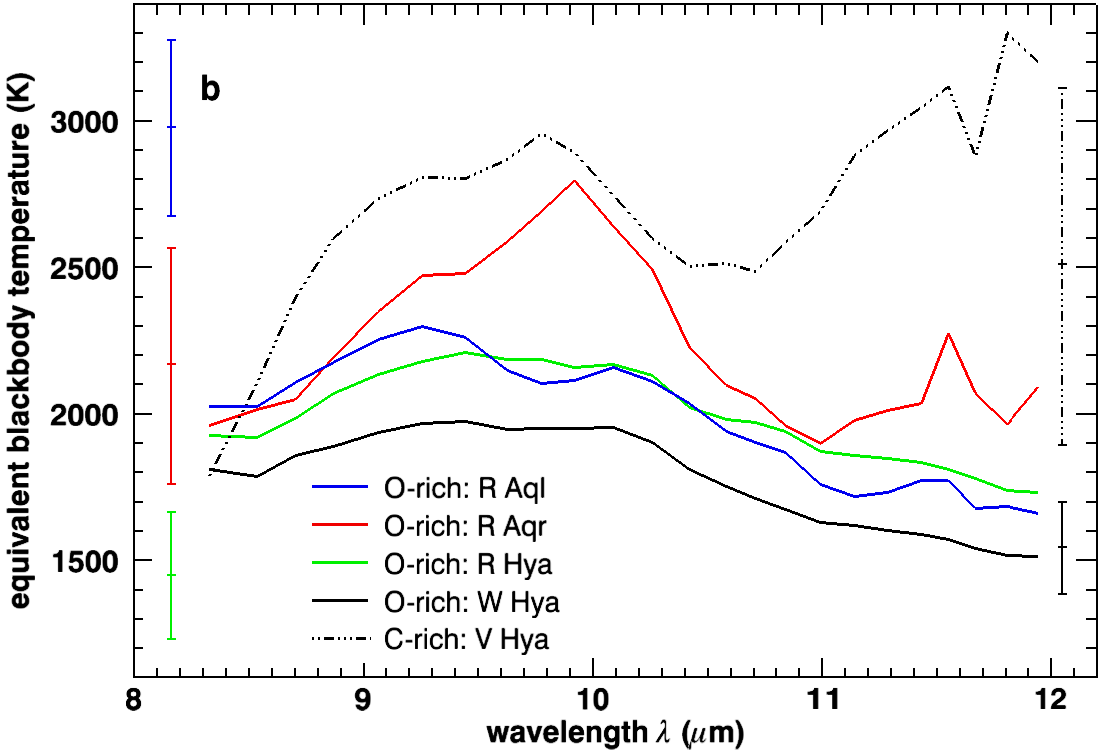}
   \caption{\textit{Left:} Normalized FDD diameters of the AGB stars studied in this work (from the right hand panels of Fig.~\ref{FigVisFit}). The diameters are normalized to one at 10~$\mu$m. Errors are approximately the same for all wavelengths and are given on the side. \textit{Right:} Equivalent blackbody temperature of all five AGB stars as function of wavelength (cf.~Eq.~\ref{Eq_brightT}).}
   \label{FigDiaTemp}
   \end{figure*}
%---------------------------------------------------------------

\subsubsection{The molecular water shell and close \ce{Al2O3} dust shell}

   The observations made in this work, in particular the trend of the apparent FDD diameter as function of wavelength, summarized in Fig.~\ref{FigDiaTemp}a, can be understood qualitatively. If they are additionally compared with the results of \citet{Ohnaka05} and \citet{Wittkowski07} for RR~Sco and S~Ori, respectively, as described in detail for W~Hya in paper~I, the following conclusions can be drawn: The overall larger diameter in the mid-IR originates from a warm molecular layer of \ce{H2O}, and the apparent gradual increase longward of 10~$\mu$m arises most likely from the presence of very close \ce{Al2O3} dust emitting in this wavelength range (labeled in Fig.~\ref{FigDiaTemp}a).
   
   However, also any other kind of dust compound thermally stable in the very close environment of the star and emitting in the mid-IR could be responsible for this. The iron-rich silicate dust emission does not have any influence since its emitting region is over-resolved at the relevant baselines. The comparison with the models for RR~Sco \citep{Ohnaka05} and S~Ori \citep{Wittkowski07} suggests that the partially resolved molecular layers are optically thick and that the nearby \ce{Al2O3} dust shell is optically and geometrically thin \citep[cf.~also][]{Egan_Sloan2001}.

   The formation of \ce{Al2O3} dust at these short distances from the stellar surface would be consistent with the empirical results by e.g.~\citet{LorenzM_Pompeia2000}. The interpretation of the apparent diameter increase beyond 10~$\mu$m as a result of the presence of an \ce{Al2O3} dust shell above the molecular water vapor shell is supported by the fact that in R~Aql and R~Hya the spectra are characterized by broad oxygen-rich dust emission, meaning that oxides are the dominant dust species. In addition, the determined equivalent blackbody temperature (Fig.~\ref{FigDiaTemp}b) indicates as well that the temperatures are low enough that dust composed of titanium and aluminum oxides is stable in the probed region but iron-rich silicate dust not.

   For R~Aqr, \ce{Al2O3} grains are probably present as well, but their emission is smeared out by the large amount of amorphous silicate dust in the MIDI FoV mainly emitting in the N-band. Extinction effects are probably the reason why the diameter increase longward of 10~$\mu$m is first strong but reaches a constant value beyond 11~$\mu$m. In addition, the ionizing radiation of the compact companion prevents the formation of large amounts of water vapor, meaning that~\ce{H2O} cannot sufficiently form from OH and H in the reaction equilibrium. Therefore, the FDD diameter at around 10~$\mu$m is only marginally larger than the photospheric diameter.
   
\subsubsection{The contribution of SiO}

   The only O-rich star in the sample which does not fit very well into the previous interpretation is R~Aqr. An overabundance of SiO \citep{Angeloni2007} and a strong silicate dust emission could already be inferred from the spectrum and can be attributed to the gravitational attraction of the WD in this symbiotic system. The high abundance of SiO molecules and the gravitational attraction leads to an effective formation of silicate dust close to the star and results in a detectable silicate emission feature in the MIDI spectrum and a high flux contribution of this dust shell. 
   
   At shorter wavelengths, the FDD diameter increases due to the presence and increased flux contribution of a SiO molecular shell emitting in the 8 to 9~$\mu$m range. This is similar to the apparent enlargement for the RSG star $\alpha$~Orionis in this wavelength regime \citep{Perrin07}. Even the other O-rich stars show a small diameter enlargement shortwards of 10~$\mu$m due to a close SiO molecular shell. A possible relevance of SiO at shorter wavelengths is supported by the occurrence of SiO masers in the region of the water vapor layer and inner boundary of the putative \ce{Al2O3} dust shell. Since specific physical conditions are necessary in order to exhibit maser emission and these conditions are solely present at a certain distance from the star, the SiO maser shell in R~Aqr is also located at a similar distance from the star as for the other O-rich stars in the sample.

\subsubsection{The outer silicate dust shell}

   Since, except for R~Aqr, the silicate emission feature is not visible in the MIDI spectra, but seen in the ISO spectra, it can be assumed that there is a surrounding dust shell consistent of mainly the more abundant classical silicates, and that even outside the FoV of MIDI dust formation is still ongoing. From visibility modeling and the FoV some limits on the size of the silicate dust shells were obtained. These lower limits are summarized in Table~\ref{Table_Summary} and show that the amorphous iron-rich silicate dust is located fairly far away from the star.

   However, this could be related for R~Aql and R~Hya to their recent thermal pulse with a short period of enhanced mass loss. Such an outer detached shell may contribute significantly to the total silicate dust emission. In addition, W~Hya is known to have a very extended dust shell \citep{Hawkins1990}. Since the mass-loss rates for these three stars are comparably low, the dust shells are not very prominent and contribute only weakly to the total flux in the mid-IR. In contrast, the emission of the silicate dust in the symbiotic system R~Aqr contributes substantially to the total flux and the characteristic dust condensation radius could be determined to approximately 7.6~$R_{\mathrm{phot}}$.

\subsubsection{On the wind formation and mass-loss mechanism}

   The details of the pulsation-enhanced dust-driven wind mass-loss mechanism in O-rich AGB stars are still under investigation \citep{Hoefner2003,Woitke2006a,Hoefner08,Norris2012}. It is believed that the wind is initialized by absorbing the momentum of the outward-directed stellar radiation by dust and re-emitting it in all directions. The required increase of the scale height of the atmosphere is naturally done by the pulsation and the formation of shock fronts, accelerating the gas to reach the location where it is able to condense into dust grains. Since amorphous \ce{Al2O3} has only a moderate abundance, it can probably not be responsible to initiate the mass-loss \citep[cf.~e.g.][]{Woitke2006a}. \ce{Al2O3} can exist close to the star without inducing mass loss. Unfortunately, the observations presented here cannot give a conclusive answer which dust species are responsible for triggering a high mass loss.

   As described in paper~I, the scattering off large iron-free silicate grains, as proposed by \citet{Hoefner08}, cannot be verified either, since they were not detected. However, this might be related to their low abundance making their emission smeared out by \ce{H2O}, SiO, and \ce{Al2O3}. The low abundance of iron-free silicates is not critical for driving the wind since a small fraction of Si condensed into \ce{Mg2SiO4} would be already sufficient \citep{Hoefner08}. Recent observations in scattered light in the near-IR by \citet{Norris2012} support the existence of large iron-free silicates at around 2~stellar radii. Since for this first study no radiative transfer modeling associated to self-consistent dynamic modeling was done, no strong conclusions about the grain mixture and wind formation in the close environment can be drawn.
   
   Not many other wind acceleration mechanisms come in mind. There might be the possibility that small amounts of carbon grains play a role as speculated by \citet{HoefnerAndersen2007}, but with the side effect that this would lead to a too high infrared excess. Despite its very low abundance, TiO$_2$ might have some relevance as well \citep{Posch2002}, but is not detectable with these observations. In this context, it might be interesting to know if scattering on \ce{Al2O3} grains is important. Also the role of large amounts of water vapor in molecular shells and the radiation pressure on water molecules may need more detailed calculations. In addition, metallic Fe might have some effects as well \citep{McDonald2010}.

   It is clear that quantitative modeling is necessary to support the above findings, in particular, if the derived constituents of the close molecular and dust layers (\ce{H2O}, SiO, \ce{Al2O3}) can really provide sufficient opacities to explain the observed diameter dependence on wavelength, in particular if \ce{Al2O3} could cause the apparent diameter increase beyond 10~$\mu$m. In order to quantify the results, it will be necessary to apply dynamic atmospheric models of e.g.~Ireland \& Scholz \citep{Ireland2006,Ireland2011} or H{\"o}fner et al.~\citep[see][]{Lebzelter2010,Sacuto2011}. Even if \ce{Al2O3} dust is not consistently included in these models so far, and has often to be added ad hoc, this will give more detailed insight into the physical processes at work there.

%%%%%%%%%%%%%%%%%%%%%%%%%%%%%%%%%%%%%%%%%%%%%%%%%%%%%%%%%%%%%%%%%%%%%%%%%%%%%%%%%%%%%%%%%%%%
\subsection{Dust and molecular shells in C-rich AGB stars}\label{secIntDis_CShell}
  
\subsubsection{Spectral features}
  
   The study of the oxygen-rich stars in the sample has shown that dense molecular layers, extending to up to 2~photospheric radii, are present in these stars in agreement with findings in other O-rich AGB stars. In contrast, the physical properties of the outer atmosphere of carbon stars and their temporal variations have not yet been probed well. Hydrostatic and dynamic model atmospheres of carbon-rich Miras as well as non-Mira carbon stars fail to explain observed spectra longward of about 5~$\mu$m \citep{Jorgensen2000,GautschyLoidl2004}. These models predict very strong absorption due to \ce{C2H2} and HCN at 7 and 14~$\mu$m, but the observed ISO spectra of carbon stars show only weak absorption due to these molecular species. However, \citet{Aoki1998} and \citet{Aoki1999} proposed that emission from extended warm molecular layers, containing \ce{C2H2} and HCN, may be responsible for the discrepancy between the ISO spectra and the models. Both species are among the most abundant in C-rich atmospheres.

   While in the K and H band the spectrum is contaminated with CO bands, absorption features of \ce{C2H2} and HCN, as well as emission from SiC at 11.3~$\mu$m \citep{Kozasa1996} and featureless amorphous carbon (AMC) dust, play an important role in the N-band. In addition, molecules such as silane (11~$\mu$m) and ammonia (10.7~$\mu$m) may also provide significant opacities in C-rich stars \citep[][and references therein]{Monnier2000}.

   The only observations with MIDI of classical C-rich stars were conducted by \citet{Ohnaka07} for V~Oph and \citet{Sacuto2011} for R~Scl. \citet{Ohnaka07} modeled V~Oph with a dust shell, consisting of AMC (85\%) and SiC (15\%), with a multi-dimensional Monte Carlo code \citep{Ohnaka2006}, and added a polyatomic molecular layer, consisting of \ce{C2H2} and HCN, where the opacities of these molecules were calculated with an appropriate band model assuming LTE. \citet{Sacuto2011} applied a self-consistent dynamic model atmosphere to their data \citep{Hoefner2003,Mattsson2008,Mattsson2010}. With this model, molecular shells of \ce{C2H2} and HCN above the stellar photosphere could be probed. However, the dusty environment could not be adequately sampled and a complementary model was applied (a hydrostatic stellar model plus a dust radiative transfer code).

\subsubsection{The dust and molecular shells}

%Since the geometrical model for V~Hya presented here separates the stellar part (the FDD fit) from the extended circumstellar environment (the Gaussian fit to the dust shell), only the Gaussian FWHM (Fig.~\ref{FigFluxDust}b) can be directly compared with the equivalent UD diameters for V~Oph and R~Scl (Fig.~2d in \citet{Ohnaka07} and Fig.~3 \citet{Sacuto2011}, respectively), tracing the molecular and dust shells. The chromatic shape of the FWHM for V~Hya and the equivalent UD for V~Oph and R~Scl are very similar. The apparent diameters increase longwards of 10~$\mu$m due to the presence of SiC dust surrounding the star and emitting in the N-band. The increase is about 20\% in V~Hya.
%In addition, a simultaneous diameter enlargement shortward of 8.5~$\mu$m and longwards of 11.5~$\mu$m due to \ce{C2H2} and HCN can be seen in V~Oph and R~Scl. The contribution of HCN in these wavelength regions is weaker and \citet{Ohnaka07} gave only upper limits of its column density. This increase could not been seen in the chromatic shape of the Gaussian FWHM for V~Hya.

	The FWHM of the Gaussian distribution in V~Hya (Fig.~\ref{FigFluxDust}b) slightly increases by about 20\% longwards of 10~$\mu$m probably due to the presence of SiC dust surrounding the star and emitting in the N-band. The FDD diameter as function of wavelength (Fig.~\ref{FigVisFit}h) shows a strong increase of 30\% shortward of 9~$\mu$m which could be related to the presence of molecular layers of \ce{C2H2} and \ce{HCN} \citep[see][]{Ohnaka07,Sacuto2011}. These molecular layers would then lie at about 2~photospheric radii. Noteworthy, the equivalent blackbody temperature shortward of 9~$\mu$m gets low enough ($\approx$~2000~K, Fig.~\ref{FigDiaTemp}b), supporting the fact that molecules are able to form.

	There is, however, a discrepancy of the FDD diameter at wavelengths longward of 11~$\mu$m. With the presence of the mentioned molecular shells one would also expect some increase of the FDD diameter in this wavelength part. There are possibly several reasons for this discrepancy. First of all, the uncertainties in fitting a FDD (determined by the visibilities with spatial frequencies larger than 10~arcsec$^{-1}$) are much higher in this wavelength regime due to a much higher scatter of the visibilities (cf.~Fig.~\ref{FigVisFit}h and Table~\ref{TableResultsVHya}).
	
	This scatter comes from combining observations done at different epochs and position angles. For example, asymmetries are very pronounced in V~Hya (cf.~Sect.~\ref{secIntDis_Asy}). The position angles where the observations were made are similar to the position angles of the high velocity wind \citep[cf.~Fig.~\ref{FigUVdia}g and e.g.][]{Hirano2004} and temporal fluctuations of the density and emissivity of the massive outflows can be expected.

   Therefore, it can only be concluded that SiC and certainly AMC dust is present at the very close circumstellar environment. The dust-driven wind formation scenario in C-rich stars is to a certain amount satisfactorily solved \citep[e.g.][]{GailSedlmayr1987,Hoefner2003,GautschyLoidl2004,Nowotny2005,Woitke2006b,Wachter2008}. Radiation pressure on carbon grains is efficient enough to initiate the wind.

   The presence of \ce{C2H2} and HCN in V~Hya cannot be confirmed with certainty due to measurement uncertainties and applying only a simple geometric model. However, also the visibility as function of wavelength (Fig~\ref{FigVisibility}d) is very similar to the ones of V~Oph and R~Scl (Fig.~1 in \citet{Ohnaka07} and Fig.~3 \citet{Sacuto2011}, respectively), suggesting the same atmospheric conditions and the presence of those layers. %Since the equivalent blackbody temperature is low enough in the extended atmosphere (Fig.~\ref{FigDiaTemp}b), it is plausible that polyatomic molecules may form behind shock fronts and that molecular layers will be established. %Since no reliable photospheric angular diameter was found in the literature, it cannot be verified if the emission from a potential \ce{C2H2} layer, together with the dust emission, is responsible for a N-band angular diameter significantly larger than the photospheric diameter, and that \ce{C2H2} plays a role similar to \ce{H2O} in oxygen-rich environments as proposed for V~Oph \citep{Ohnaka07} and consistently modeled for R~Scl \citep{Sacuto2011}.

   V~Hya shows some similarities to R~Aqr except for the underlying chemistry. As for R~Aqr, the average N-band FDD diameter is only about 1.6~times larger than the photospheric diameter, and the dust condensation radius is at about 8~times the photospheric radius. However, it is not expected that the K-band diameter is a good tracer for the true photospheric radius in C-rich stars, and the condensation of AMC dust could occur at smaller radii, i.e.~at higher temperatures.
   
   As for the oxygen-rich stars, it is necessary to use dynamic model atmospheres \citep[e.g.][]{Sacuto2011}, computed with stellar parameters corresponding to V~Hya, for a better understanding of the physical processes responsible for the molecule and dust formation close to the star.

%%%%%%%%%%%%%%%%%%%%%%%%%%%%%%%%%%%%%%%%%%%%%%%%%%%%%%%%%%%%%%%%%%%%%%%%%%%%%%%%%%%%%%%%%%%%
\subsection{Dynamic behavior of the close environment}\label{secIntDis_Intra}

   As a result of the intrinsic pulsation, mid-IR flux variations on the order of 20\%, with a phase offset of about 0.15, were detected in this study except for R~Aql (Sect.~\ref{secObsSubLC}). This section investigates how this relates to changes of the apparent angular sizes and flux contributions of the close stellar atmosphere, and how this influences or triggers the dust formation and mass loss. 
  
%---------------------------------------------------------------
   \begin{figure*}
   \centering
   \includegraphics[width=0.48\linewidth]{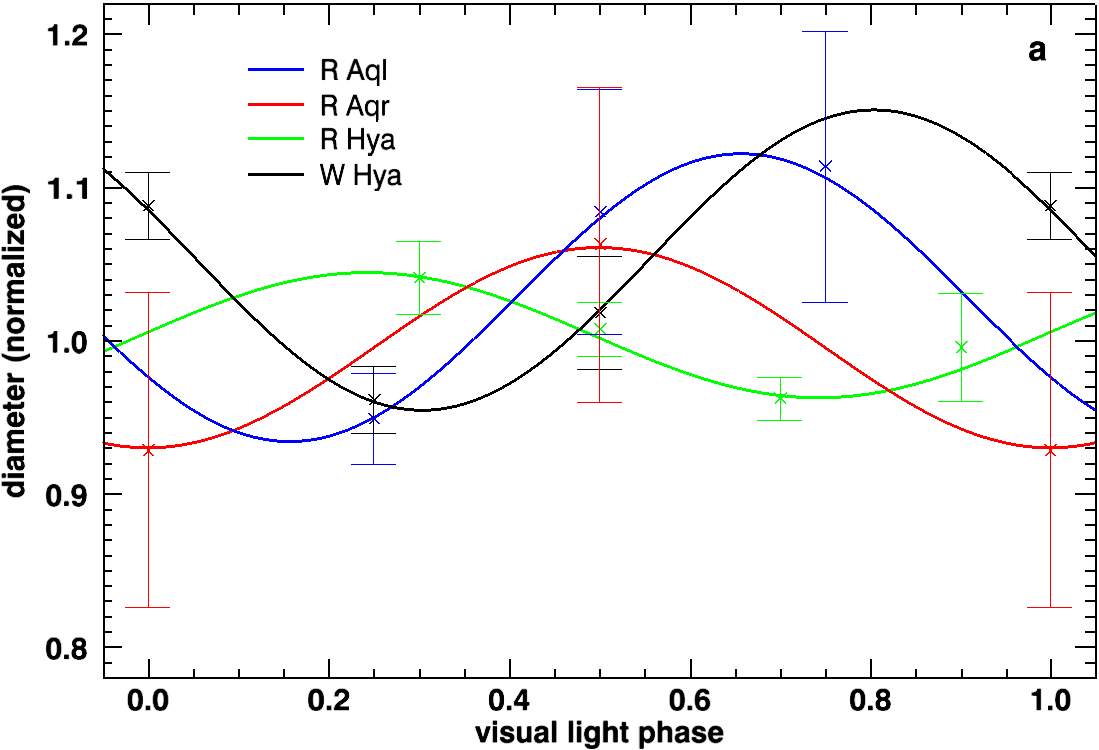}
   \hspace{0.1cm}
   \includegraphics[width=0.48\linewidth]{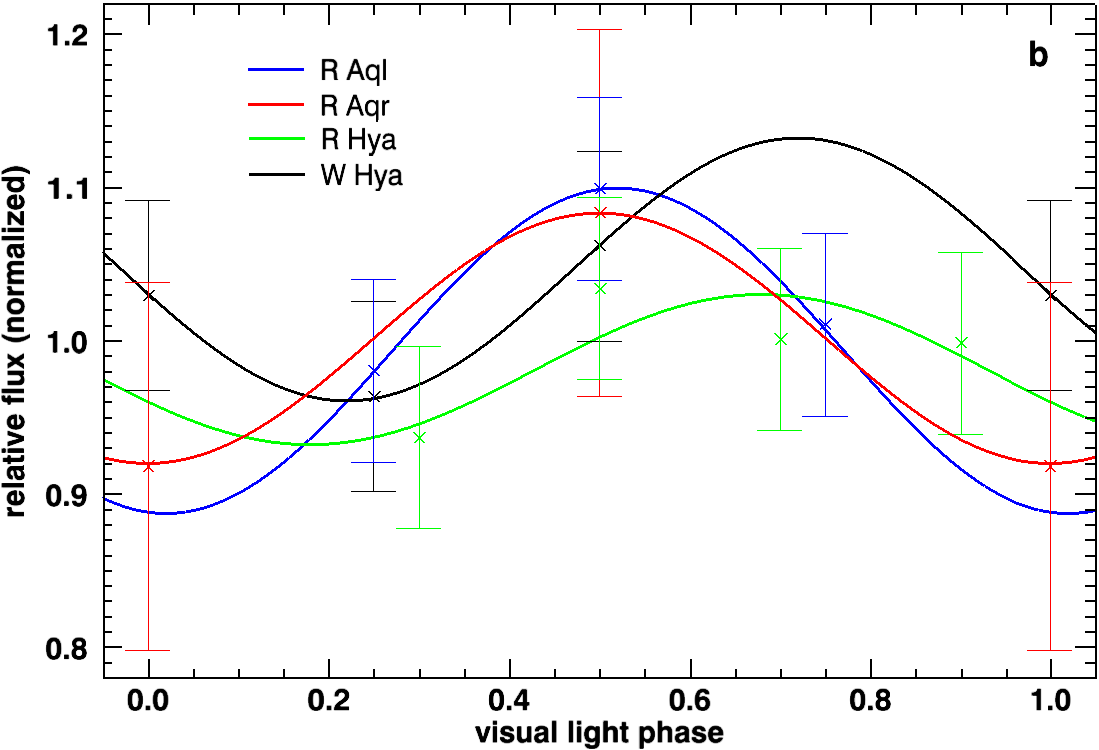}
   \caption{Phase-to-phase variations of all oxygen-rich stars in the sample. The left penal shows the normalized FDD diameter as function of visual light phase and the right panel the normalized relative flux contribution of the FDD as function of visual light phase (normalized in respect to the full data set). It was not possible to obtain reliable results for the C-rich star V~Hya.}
   \label{FigCycleAll}
   \end{figure*}
%---------------------------------------------------------------

%---------------------------------------------------------------
   \begin{table}[ht]
     \caption{Phase-to-Phase Variations.}
     \label{TableInCycleVar}
     \centering
     \begin{tabular}{ccccc}
       \hline
       \noalign{\smallskip}
       Phase                & Phase & No of & Norm.$^{\mathrm{b}}$ FDD & Norm.$^{\mathrm{b}}$ Flux \\
       ~Bin$^{\mathrm{a}}$  & Range &  Obs. &        Diameter          &   Contribution               \\
       \noalign{\smallskip}
       \hline
       \noalign{\smallskip}
       \multicolumn{5}{l}{\textbf{~~R~Aql:}}                           \\
         1   &  0.875$-$0.125 &  8 & --$^{\mathrm{c}}$ & --$^{\mathrm{c}}$ \\
         2   &  0.125$-$0.375 &  9 & 0.95~$\pm$~0.03 & 0.98~$\pm$~0.06 \\
         3   &  0.375$-$0.625 &  9 & 1.08~$\pm$~0.08 & 1.10~$\pm$~0.06 \\
         4   &  0.625$-$0.875 &  6 & 1.11~$\pm$~0.09 & 1.01~$\pm$~0.06 \\
       \noalign{\smallskip}
       \hline
       \noalign{\smallskip}
       \multicolumn{5}{l}{\textbf{~~R~Aqr:}}                           \\
         1   &  0.750$-$0.250 & 12 & 0.93~$\pm$~0.10 & 0.92~$\pm$~0.12 \\
         2   &  0.250$-$0.750 & 14 & 1.06~$\pm$~0.10 & 1.08~$\pm$~0.12 \\
       \noalign{\smallskip}
       \hline
       \noalign{\smallskip}
       \multicolumn{5}{l}{\textbf{~~R~Hya:}}                           \\
         1   &  0.000$-$0.200 &  0 &      --            &     --       \\
         2   &  0.200$-$0.400 & 14 & 1.04~$\pm$~0.02 & 0.94~$\pm$~0.06 \\
         3   &  0.400$-$0.600 & 27 & 1.01~$\pm$~0.02 & 1.04~$\pm$~0.06 \\
         4   &  0.600$-$0.800 & 20 & 0.96~$\pm$~0.01 & 1.00~$\pm$~0.06 \\
         5   &  0.800$-$0.000 &  3 & 1.00~$\pm$~0.04 & 1.00~$\pm$~0.06 \\
       \noalign{\smallskip}
       \hline
       \noalign{\smallskip}
       \multicolumn{5}{l}{\textbf{~~W~Hya:}}                           \\
         1   &  0.875$-$0.125 & 23 & 1.09~$\pm$~0.02 & 1.03~$\pm$~0.06 \\
         2   &  0.125$-$0.375 & 42 & 0.96~$\pm$~0.02 & 0.96~$\pm$~0.06 \\
         3   &  0.375$-$0.625 & 10 & 1.02~$\pm$~0.04 & 1.06~$\pm$~0.06 \\
         4   &  0.625$-$0.875 &  0 &      --            &     --       \\
       \noalign{\smallskip}
       \hline
%       \noalign{\smallskip}
%       \hline
     \end{tabular}
     \begin{flushleft}
       \textbf{Notes. }
       $^{\mathrm{a}}$~Cf.~right hand panels of Fig.~\ref{FigLightPhase}. $^{\mathrm{b}}$~Normalized in respect to the full data set. $^{\mathrm{c}}$~No reliable fit could be obtained.
     \end{flushleft}
   \end{table}
%---------------------------------------------------------------

\subsubsection{Size and flux variations}\label{secIntDis_Intra_Intra}

   Cycle-to-cycle variations were studied for all oxygen-rich stars for the phase ranges given in the left hand panels of Fig.~\ref{FigLightPhase}. For this, the best geometrical model was fitted to the visibilities with the relative fluxes fixed to the value obtained for the full data set (cf.~Fig.~\ref{FigFluxDust}a). No cycle-to-cycle investigation could be done for V~Hya due to the fact that observations are not repeated at the same phase in consecutive cycles.

   However, reliable cycle-to-cycle variations could be obtained for R~Hya and W~Hya (for W~Hya see paper~I). Both stars show a maximum variation of the N-band averaged and normalized FDD diameter, $\overline{\theta}_{\mathrm{FDD}}$ (normalized in respect to the full data set), on the order of (5~$\pm$~4)\%, thus much lower than phase-to-phase variations as shown below. This verifies that the folding of consecutive pulsation cycles into one cycle is an acceptable assumption for the following phase-to-phase analysis. For R~Aql and R~Aqr, the uncertainties are higher and no reliable trend could be found.

	For studying the phase-to-phase behavior of the diameter and the relative flux contribution of the FDD, the pulsation phase of each star is divided into several phase bins as shown in the right hand panels of Fig.~\ref{FigLightPhase} and listed in Table~\ref{TableInCycleVar}. As for the cycle-to-cycle investigation, the best geometrical model is then fitted to the data within each bin. For the fit to R~Aqr, the Gaussian FWHM is fixed to the value obtained for the full data set (cf.~Fig.~\ref{FigFluxDust}b). No reliable fits could be obtained for V~Hya because of a very unfavorable spatial frequency coverage. In several phase bins the FDD diameter could not been fitted or the uncertainties were too high due to potential asymmetries and unknown cycle-to-cycle variations (cf.~Sect.~\ref{secIntDis_Asy}) even after reducing the number of phase bins. Therefore, V~Hya is not considered in the following discussion.
   
   For a simple comparison between phase bins, the obtained FDD diameters, $\theta_{\mathrm{FDD}}$ (cf.~right hand penals of Fig.~\ref{FigVisFit}), and relative flux contributions, $\epsilon_{\mathrm{FDD}}$, are averaged over the N-band wavelength regime (8$-$12~$\mu$m). Since the chromatic trend of $\theta_{\mathrm{FDD}}$ and $\epsilon_{\mathrm{FDD}}$ does not change considerably between phases, and in general other error sources are more prominent (see below), this approach can be appropriate for a first investigation. The similar chromatic trend found for each subsequent phase bin indicates also that the probed different layers behave similarly as function of pulsation phase.

   The N-band averaged diameters, $\overline{\theta}$, and relative flux contributions, $\overline{\epsilon}$, are plotted as function of visual phase in Fig.~\ref{FigCycleAll}, and are listed in Table~\ref{TableInCycleVar} in comparison to the full data set values. The errors are in general relatively high because of the low number of visibility points used for these fits. In addition, the large phase binning reduces the true amplitude of the variation somewhat. It should be also noted that for R~Aqr only two phase bins are used and that there are no fit results and observations available for R~Aql and R~Hya at visual maximum, respectively.

   Except for R~Hya, the FDD diameter is largest at or after the visual minimum (Fig.~\ref{FigCycleAll}a and Table~\ref{Table_Summary}). The increase, $\overline{\theta}_{\mathrm{FDD,max}}$/$\overline{\theta}_{\mathrm{FDD,min}}$, between minimum and maximum is on the order of 15\%. While the layer representing the photosphere in a Mira variable is in the mid-infrared largest near the stellar luminosity maximum \citep{Weiner2003}, the more outer pulsating layers reach their maximum extension with an increasing phase delay. This is in agreement with the movement of a mass-shell at around 2~stellar radii as modeled by \citet{Nowotny2010} (their Fig.~1 and 2) for a carbon star. However, for R~Hya a different trend was found making this interpretation uncertain. The large error bars also reflecting this uncertainty.
   
   In contrast, the variation of the relative flux contribution as function of visual light phase shows a more consistent behavior. The flux contribution is lower at around visual maximum than at around visual minimum (Fig.~\ref{FigCycleAll}b and Table~\ref{Table_Summary}). The increase, $\overline{\epsilon}_{\mathrm{FDD,max}}$/$\overline{\epsilon}_{\mathrm{FDD,min}}$, between minimum and maximum is on the order of 10 to 15\%, but also with large uncertainties and therefore not strongly significant.
   
   The location of the outer silicate dust shell as function of visual phase was studied for the symbiotic system R~Aqr. For this the Gaussian FWHM was not fixed to the full data set value. At visual minimum, the averaged Gaussian FWHM, $\overline{\theta}_{\mathrm{G}}$, was larger than at visual maximum by about (24~$\pm$~19)\%.

\subsubsection{Dust and wind formation}

   Taking the above findings at face value, one infers that new dust forms around or after the visual minimum when the pulsating layer at around 2~photospheric radii reaches its maximum extension. Since a higher relative flux contribution of the FDD around the visual minimum is approximately canceled out by the lower absolute flux, the temperature in this region will be lower too.
   
   Calculating the equivalent blackbody temperature according to Eq.~\ref{Eq_brightT}, the temperatures around the visual minimum are on average (150~$\pm$~100)~K lower and at around the visual maximum on average (150$\pm$~100)~K higher than the temperatures averaged over the cycles shown in the right panel of Fig.~\ref{FigDiaTemp}. This means that in particular in the 11--12~$\mu$m range, where the putative \ce{Al2O3} dust shell is traced, the temperatures around the visual minimum are getting as low as the condensation temperature of \ce{Al2O3} of approximately 1400~K, showing that new \ce{Al2O3} dust could form.
   
   The phase behavior would also be consistent with the theoretical predictions of \citet{Ireland2006}. In their model, the largest rate of grain growth occurs between phases 0.3 and 0.7, similar to what is predicted for carbon-rich stars \citep[cf.~Fig.~2 in][]{Nowotny2010}.

   While the observations made in this work show that \ce{Al2O3} dust might newly form around the visual minimum, it cannot be determined conclusively how this relates to the acceleration of the wind and the mass-loss rate. As mentioned before and in paper~I, \ce{Al2O3} can exist in the upper atmosphere without being a wind-driver. No conclusion can be given concerning the relevance of micrometer-sized iron-free silicates for the wind formation since this needs the use of radiative transfer modeling associated to self-consistent dynamic modeling taking opacities of different realistic chemical compounds into account. However, it can be speculated that the relevant constituents that are responsible for the wind acceleration are formed along with \ce{Al2O3}.

%%%%%%%%%%%%%%%%%%%%%%%%%%%%%%%%%%%%%%%%%%%%%%%%%%%%%%%%%%%%%%%%%%%%%%%%%%%%%%%%%%%%%%%%%%%%
\subsection{Asymmetries}\label{secIntDis_Asy}
   
%---------------------------------------------------------------
   \begin{figure*}
   \centering
   \includegraphics[width=0.48\linewidth]{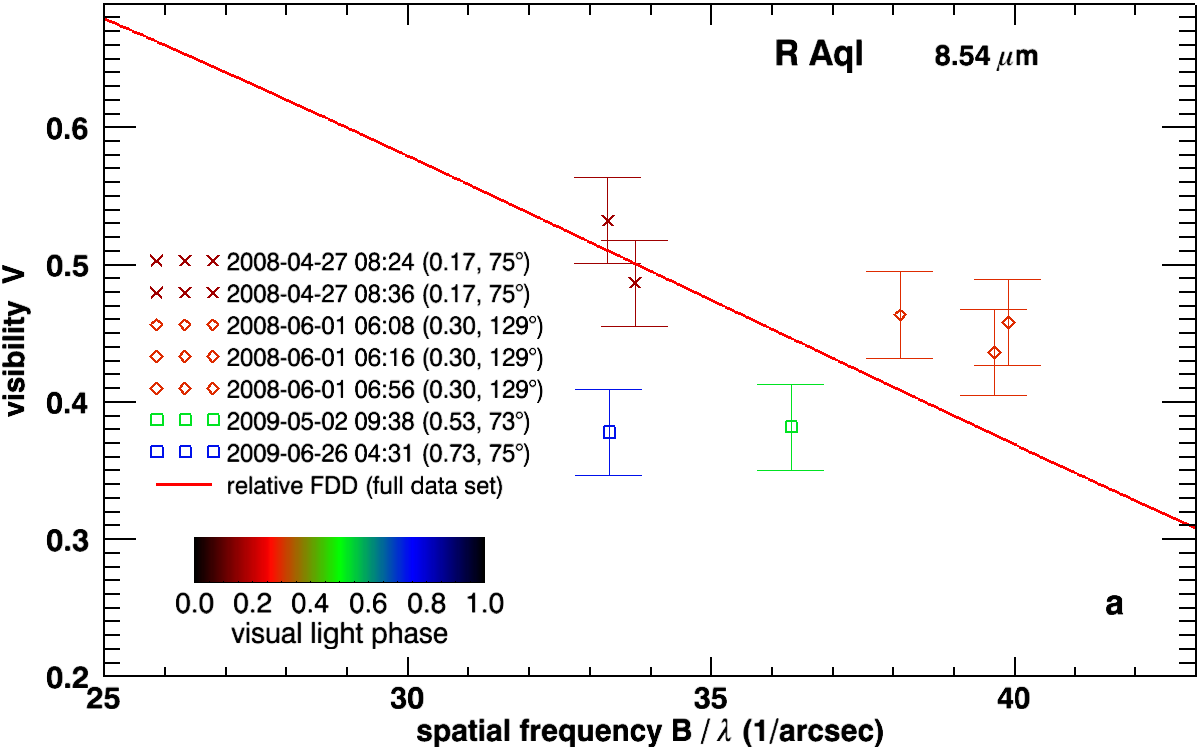}
   \hspace{0.1cm}
   \includegraphics[width=0.48\linewidth]{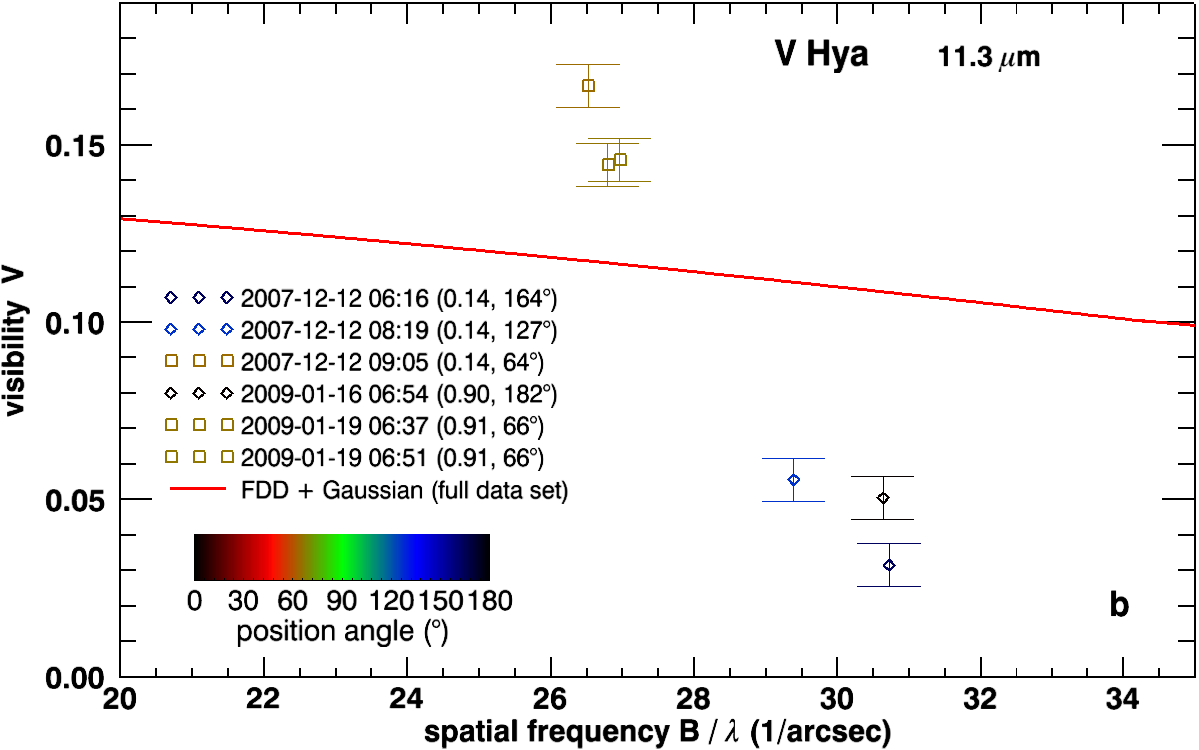}
   \caption{Selected visibility measurements to investigate the asymmetries in R~Aql \textit{(left)} and V~Hya \textit{(right)}. The visual light phase and position angle are given in parenthesis for each data point. See text for more details.}
   \label{FigAsymmetry}
   \end{figure*}
%---------------------------------------------------------------

   The good uv-coverage for R~Hya and W~Hya, shown in the left hand panels of Fig.~\ref{FigUVdia} and Fig.~1 in paper~I, makes it possible to directly fit a elliptical FDD to the data. For R~Aql, R~Aqr and V~Hya, the uv-planes are not filled well. Departure from spherical symmetry was here studied by simply comparing visibility measurements at different position angles taken closely in time and with similar projected baselines. Differential phases were not used but will be investigated in more detail in future studies. % Only for R~Aqr it was not possible to find such pairs for a reliable investigation.

   As described in detail in paper~I, W~Hya shows a small elongation in the mid-IR. The position angle (PA) and axis ratio of an elliptical FDD were determined to (11~$\pm$~20)$^\circ$ and 0.87~$\pm$~0.07, respectively. The asymmetry was explained by a possible enhanced dust concentration along an N-S axis. As for W~Hya, an elliptical FDD was fitted to the full data set of R~Hya, but also to subsets of the full data set to test for time dependencies. However, no departure from spherical symmetry could be revealed within the measurement uncertainties. Any intrinsic deviation from sphericity is less then 10\%.

   While for W~Hya contradictory position angles were found in the literature, the non-detection of an asymmetry in R~Hya is consistent with the results of \citet{Ireland2004a} and \citet{Monnier2004}. Both authors did not find any departures from symmetry within the measurement errors in the optical and near-IR (680$-$940~nm and 2.26~$\mu$m, respectively).

   Because both stars have similar intrinsic properties but a different asymmetric appearance, they are excellent targets for studying the origin and formation of asymmetries in more detail, e.g.~if they are related to different evolutionary phases. By comparing both stars, it could be investigated why W~Hya shows an asymmetry and R~Hya not, and in particular why in W~Hya asymmetries appear different at different scales and how they are connected.
   
   Figure~\ref{FigAsymmetry} shows selected visibility measurements for R~Aql and V~Hya to investigate their asymmetry. In order to do so, groups of points with similar spatial frequencies are compared with the best model fit to the full data set serving as reference\footnote{Note that a comparably large phase and spatial frequency range is used. This allows a more reliable analysis due to comparing groups of points instead of only single measurements with large errors. However, it can not be excluded that this is biased by the fact that different spatial frequencies probe different parts of the envelope. Therefore, only a qualitative analysis is given here.}. In R~Aql, it can be seen that the visibilities measured at visual light phase 0.17 with a PA of 75$^\circ$ (crosses) lie on the full data set fit as well as basically also the measurements taken at an almost perpendicular PA of 129$^\circ$ (diamonds). Therefore, it can be concluded that there is no departure from symmetry in the close vicinity of the star at around 2.2~photospheric radii within the measurement uncertainties. In contrast, it can be inferred that at a distinct different visual light phase the visibilities (squares) are smaller and therefore probing a region appearing larger (cf.~Sect.~\ref{secIntDis_Intra_Intra}). 

   In V~Hya an opposite effect can be seen. Points with a PA around 65$^\circ$ (squares) have a higher visibility than expected for a circular FDD and points with PAs from 130 to 180$^\circ$ a lower visibility. This seems not be related to the pulsation since at different light phases but similar PAs the visibilities are similar. This means that a strong asymmetry already sets in in the very close stellar environment at less then 2~photospheric radii with an elongation along a North-South direction, therefore, a similar orientation as found by e.g.~\citet{Tsuji1988,Kahane1988,Kahane1996,Sahai2003} and particular also by \citet{Lagadec2005} in the mid-infrared (cf.~Sec.~\ref{secPropSSubVHya}). In contrast to R~Aql, where the symmetry is seen over the whole N-band wavelength regime, the asymmetry in V~Hya appears stronger at longer wavelengths (10 to 12~$\mu$m). 
   
   For R~Aqr, only one set of visibilities could be compared. In the shorter wavelength regime (8 to 10~$\mu$m), the visibilities taken at 2007-10-07 05:54 at a PA of 79$^\circ$ (phase~0.87, B$_{\mathrm{proj}}$~=~40.3~m) are larger than the ones taken at 2007-10-07 01:16 (phase~0.87, B$_{\mathrm{proj}}$~=~41.3~m) at a PA of 59$^\circ$ (cf.~Table~\ref{TableApp_log}). This is a hint that the distribution of SiO strongly deviates from spherical symmetry. From this investigation it follows that three out of the five stars observed show asymmetries. %,and it can be concluded that asymmetries with different peculiarities are common.

%###########################################################################################
%###########################################################################################
\section{Summary}\label{secConc}

   Five AGB stars, namely R~Aql, R~Aqr, R~Hya, W~Hya, and V~Hya, were observed over a period of nearly three years covering several pulsation cycles. All stars were for the first time monitored in the mid-IR (8$-$13~$\mu$m) with the interferometric instrument MIDI. These high-resolution observations are sensitive to the structure of the stellar atmosphere, consisting of the continuum photosphere and overlying molecular layers, as well as to the properties of the dust shell.

   The results and conclusions can be summarized as follows:
   \begin{itemize}
      \item Except for R~Aql, a periodic flux variation is found in the N-band. The typical semi-amplitudes are on the order of 20\% to~30\% for the Miras and 10\% for the SR/Mira variable W~Hya. The amplitude is probably also related to the dust content in the system. The mid-IR maximum occurs always after the visual maximum at visual phase 0.15~$\pm$~0.05.

      \item In the MIDI spectrum of R~Aqr, the silicate dust emission feature could be clearly identified, while it was not present in the spectra of the other O-rich stars. This has been attributed to the field-of-view of MIDI smaller than the one of ISO. The expected SiC dust emission feature is detected in the MIDI spectrum of the carbon star V~Hya.

      \item The obtained visibility data were best fitted by a fully limb-darkened disk (FDD), representing the central star and the close atmospheric layers, and a spherical Gaussian distribution, representing the extended dusty environment.

      \item The trend of the FDD diameter as function of wavelength is similar for all O-rich stars in the sample. The apparent size is almost constant between 8 and 10~$\mu$m (21.3~$\pm$~2.0~mas, 20.9~$\pm$~3.5~mas, 46.6~$\pm$~0.6~mas, and 79.0~$\pm$~1.2~mas, for R~Aql, R~Aqr, R~Hya, and W~Hya, respectively) and gradually increases at wavelengths longer than 10~$\mu$m. The enlargement is for all stars between 25\% and 30\%. In contrast, the relative flux contribution of the FDD decreases, reflecting the increased flux contribution from the surrounding dust shell.

      \item The measured FDD diameters at 10~$\mu$m are about 2.2~times larger than the photospheric diameters (except for R~Aqr and V~Hya). This ratio is consistent with observations of other AGB stars. The smaller ratio of about 1.6 for R~Aqr and V~Hya can be understood from the gravitational interaction with a compact companion.

      \item The obtained diameter trends for the O-rich stars are compared with the findings of \citet{Ohnaka05} and \citet{Wittkowski07} for RR~Sco and S~Ori, respectively. It can be concluded that the overall larger diameter in the mid-IR originates from a warm molecular layer of \ce{H2O}, and the gradual increase longward of 10~$\mu$m can be most likely attributed to the contribution of a spatially resolved close \ce{Al2O3} dust shell. A contribution of SiO shortward of 10~$\mu$m can be seen as well, maybe except for W~Hya.
      
      \item These observations of a larger sample of stars than available before confirm previous results, and emphasize the need for dynamic models able to treat the formation of dust in a self-consistent way. In particular, the close dust shell composed of \ce{Al2O3} well below the distance at which the iron-rich silicate dust shell is traced, could now be revealed for a significant number of stars.

      \item A comparison with available SiO maser ring diameters shows that the SiO maser emission is co-located with the region probed with MIDI.

      \item For the carbon rich star V~Hya it can be concluded that SiC and AMC dust is present in the very close circumstellar environment. The chromatic shape of the Gaussian FWHM shows an increase around the SiC feature (11.3~$\mu$m). The apparent FDD diameter as function of wavelength shows a strong increase of more than 30\% shortward of 9~$\mu$m, which could be a hint that close molecular layers consisting of \ce{C2H2} and HCN at about 2~photospheric radii are probed. However, a simultaneous increase at wavelengths longward of 11~$\mu$m could not been detected, but this might be related to large fitting uncertainties. %\citep{Ohnaka07,Sacuto2011}

      \item It is found that the iron-rich silicate dust shells are located fairly far away from the star, at radii larger than 30~times the photospheric radius. Only the silicate dust shell of R~Aqr is located closer to the star at around 8~photospheric radii, which is most probably related to the interaction with a close companion.

      \item Phase-to-phase variations of the FDD diameter and the relative flux contribution of the O-rich stars are on the order of 15\% and 10--15\%, respectively. It was found that the relative flux contribution was highest at around visual minimum, while there was no clear behavior detectable for the FDD diameter as function of visual light phase. Due to the large uncertainties this is not strongly significant and no firm conclusions can be drawn concerning the mass-loss mechanism. Nevertheless, one can speculate that new dust forms around visual minimum.

      \item Maximal diameter changes due to cycle-to-cycle variations are on the order of 6\%, and are therefore lower than phase-to-phase variations. This could only be determined for R~Hya and W~Hya.

      \item The close environment geometry could be investigated for all stars. R~Aql and R~Hya do not show any deviation from sphericity, while R~Aqr, W~Hya and V~Hya do.
   \end{itemize}

   Most of the proposed goals could be accomplished. However, a few items remain open. First, no clear connection could be established between the dust condensation and wind formation. Second, the size of the iron-rich silicate dust shell could not be satisfactorily constrained for 3 of the 4 oxygen-rich stars due to the lack of observations at short baselines. Third, the pulsation phase coverage allowed only for some of the stars an analysis of the apparent diameter with time.

   The majority of the above conclusions rely on fits of a relatively simple FDD to the observational data. However, this can only be an approximation, even if this model represents the data quite well. In order to obtain a more meaningful diameter and the exact composition, it is therefore necessary to compare with hydrodynamic model atmospheres of e.g.~Ireland \& Scholz \citep{Ireland2006,Ireland2011} and H{\"o}fner et al.~\citep[see][]{Lebzelter2010,Sacuto2011}.

   Another big reward can be gained if these results can be combined with kinematic information. Complementary high-resolution radio and mm observations, tracing the relevant constituents and sampling the dust formation zone probed with MIDI, are therefore necessary.

%###########################################################################################
%###########################################################################################

\begin{acknowledgements}
The first author would like to thank the \emph{International Max-Planck Research School} (IMPRS-HD) for its financial support with a fellowship. We acknowledge with thanks the variable star observations from the AAVSO International Database contributed by observers worldwide and used in this research. This research has made use of the SIMBAD database, operated at the CDS, France, the ISO/IRAS database and NASA's Astrophysical Data System. This work has been supported in part by a grant from the National Science Council (NSC~99-2112-M-003-003-MY3). We would also like to thank the referee for the very valuable comments.
\end{acknowledgements}

%###########################################################################################
%###########################################################################################

% references (literature):

%###########################################################################################
%###########################################################################################

% appendix

% \Online

\appendix

%----------------
\begin{table*}

\section{Visibility model fit results for R~Aql, R~Aqr, R~Hya and V~Hya}\label{App_VisFit}

\caption[Model fit results for R~Aql]{\textbf{R~Aql}: circular fully limb-darkened disk~(FDD) and circular uniform disk~(UD).}
\label{TableResultsRAql}                               
\centering                               
\begin{tabular}{cc|ccc|ccc}                               
%\noalign{\smallskip}                               
\hline                               
\noalign{\smallskip}                       
 &  & \multicolumn{3}{c}{circular FDD}         & \multicolumn{3}{c}{circular UD}         \\
$\lambda$ & $\Delta V^{\mathrm{a}}$ & $\theta_{\mathrm{FDD}}$   & Flux $\epsilon_{\mathrm{FDD}}$   & $\chi^\mathrm{2,b}_\mathrm{r}$ & $\theta_{\mathrm{UD}}$   & Flux $\epsilon_{\mathrm{UD}}$   & $\chi^\mathrm{2,b}_\mathrm{r}$ \\
($\mu\mathrm{m}$) &  & ($\mathrm{mas}$)   &    &  & ($\mathrm{mas}$)   &    &  \\
\noalign{\smallskip}                       
\hline                               
\noalign{\smallskip}                               
 8.12 & 0.034 & 23.1 $\pm$ 0.9 & 0.99 $\pm$ 0.04 & 4.7 & 20.4 $\pm$ 0.8 & 0.99 $\pm$ 0.04 & 4.7 \\
 8.33 & 0.033 & 22.7 $\pm$ 0.9 & 0.97 $\pm$ 0.04 & 5.0 & 20.1 $\pm$ 0.9 & 0.97 $\pm$ 0.03 & 5.1 \\
 8.54 & 0.031 & 22.7 $\pm$ 0.8 & 0.95 $\pm$ 0.04 & 3.8 & 20.1 $\pm$ 0.7 & 0.95 $\pm$ 0.03 & 3.8 \\
 8.71 & 0.031 & 22.2 $\pm$ 0.9 & 0.93 $\pm$ 0.03 & 3.5 & 19.6 $\pm$ 0.7 & 0.93 $\pm$ 0.03 & 3.5 \\
 8.87 & 0.031 & 21.6 $\pm$ 0.9 & 0.92 $\pm$ 0.03 & 3.5 & 19.2 $\pm$ 0.8 & 0.92 $\pm$ 0.03 & 3.5 \\
 9.07 & 0.029 & 21.1 $\pm$ 1.0 & 0.88 $\pm$ 0.03 & 4.2 & 18.7 $\pm$ 0.9 & 0.88 $\pm$ 0.03 & 4.2 \\
 9.26 & 0.029 & 20.5 $\pm$ 1.2 & 0.84 $\pm$ 0.03 & 5.0 & 18.2 $\pm$ 1.1 & 0.84 $\pm$ 0.03 & 5.0 \\
 9.45 & 0.026 & 20.3 $\pm$ 1.3 & 0.80 $\pm$ 0.03 & 6.7 & 18.1 $\pm$ 1.2 & 0.80 $\pm$ 0.03 & 6.7 \\
 9.63 & 0.028 & 21.1 $\pm$ 1.3 & 0.78 $\pm$ 0.03 & 5.3 & 18.7 $\pm$ 1.1 & 0.78 $\pm$ 0.03 & 5.3 \\
 9.78 & 0.027 & 21.4 $\pm$ 1.5 & 0.78 $\pm$ 0.03 & 6.9 & 19.0 $\pm$ 1.4 & 0.78 $\pm$ 0.03 & 6.9 \\
 9.92 & 0.027 & 21.3 $\pm$ 1.5 & 0.77 $\pm$ 0.03 & 6.5 & 18.9 $\pm$ 1.3 & 0.77 $\pm$ 0.03 & 6.5 \\
10.09 & 0.027 & 21.4 $\pm$ 1.6 & 0.78 $\pm$ 0.03 & 7.5 & 19.0 $\pm$ 1.4 & 0.78 $\pm$ 0.03 & 7.5 \\
10.26 & 0.026 & 21.9 $\pm$ 1.6 & 0.78 $\pm$ 0.03 & 7.7 & 19.4 $\pm$ 1.4 & 0.78 $\pm$ 0.03 & 7.7 \\
10.42 & 0.029 & 22.6 $\pm$ 1.6 & 0.78 $\pm$ 0.03 & 6.0 & 20.1 $\pm$ 1.5 & 0.78 $\pm$ 0.03 & 6.0 \\
10.58 & 0.031 & 23.6 $\pm$ 1.4 & 0.79 $\pm$ 0.03 & 4.8 & 21.0 $\pm$ 1.3 & 0.79 $\pm$ 0.03 & 5.8 \\
10.71 & 0.030 & 24.2 $\pm$ 1.5 & 0.80 $\pm$ 0.03 & 5.3 & 21.5 $\pm$ 1.3 & 0.79 $\pm$ 0.03 & 5.3 \\
10.84 & 0.030 & 24.8 $\pm$ 1.4 & 0.80 $\pm$ 0.03 & 4.6 & 22.0 $\pm$ 1.3 & 0.80 $\pm$ 0.03 & 4.6 \\
10.99 & 0.030 & 26.1 $\pm$ 1.4 & 0.81 $\pm$ 0.03 & 4.6 & 23.2 $\pm$ 1.3 & 0.81 $\pm$ 0.03 & 4.6 \\
11.14 & 0.027 & 26.7 $\pm$ 1.4 & 0.82 $\pm$ 0.03 & 5.2 & 23.7 $\pm$ 1.2 & 0.82 $\pm$ 0.03 & 5.2 \\
11.29 & 0.031 & 26.9 $\pm$ 1.3 & 0.83 $\pm$ 0.03 & 3.8 & 23.8 $\pm$ 1.1 & 0.82 $\pm$ 0.03 & 3.8 \\
11.43 & 0.030 & 26.2 $\pm$ 1.5 & 0.81 $\pm$ 0.03 & 4.3 & 23.2 $\pm$ 1.2 & 0.81 $\pm$ 0.03 & 4.4 \\
11.55 & 0.029 & 26.4 $\pm$ 1.5 & 0.81 $\pm$ 0.03 & 4.9 & 23.5 $\pm$ 1.3 & 0.81 $\pm$ 0.03 & 5.0 \\
11.67 & 0.026 & 27.5 $\pm$ 1.4 & 0.83 $\pm$ 0.03 & 5.6 & 24.4 $\pm$ 1.2 & 0.83 $\pm$ 0.03 & 5.6 \\
11.81 & 0.029 & 27.3 $\pm$ 1.4 & 0.83 $\pm$ 0.03 & 4.3 & 24.2 $\pm$ 1.2 & 0.83 $\pm$ 0.03 & 4.4 \\
11.95 & 0.029 & 27.9 $\pm$ 1.4 & 0.84 $\pm$ 0.03 & 4.2 & 24.7 $\pm$ 1.2 & 0.84 $\pm$ 0.03 & 4.2 \\
\noalign{\smallskip}                               
\hline                               
\end{tabular}   

\vspace{5mm}                         

\caption[Model fit results for R~Aqr]{\textbf{R~Aqr}: circular fully limb-darkened disk~(FDD) + circular Gaussian~(G) and circular uniform disk~(UD) + circular Gaussian~(G).}
\label{TableResultsRAqr}                               
\centering                               
\begin{tabular}{cc|cccc|cccc}                               
%\noalign{\smallskip}                               
\hline                               
\noalign{\smallskip}                               
 &  & \multicolumn{4}{c}{circular FDD + circular Gaussian}             & \multicolumn{4}{c}{circular UD + circular Gaussian}             \\
$\lambda$ & $\Delta V^{\mathrm{a}}$ & $\theta_{\mathrm{FDD}}$   & Flux $\epsilon_{\mathrm{FDD}}$   & $\theta_{\mathrm{G}}$   & $\chi^\mathrm{2,b}_\mathrm{r}$ & $\theta_{\mathrm{UD}}$   & Flux $\epsilon_{\mathrm{UD}}$   & $\theta_{\mathrm{G}}$   & $\chi^\mathrm{2,b}_\mathrm{r}$ \\
($\mu\mathrm{m}$) &  & ($\mathrm{mas}$)   &    & ($\mathrm{mas}$)   &  & ($\mathrm{mas}$)   &    & ($\mathrm{mas}$)   &  \\
\noalign{\smallskip}                               
\hline                               
\noalign{\smallskip}                               
8.12 & 0.025 & 27.4 $\pm$ 2.5 & 0.57 $\pm$ 0.03 & 78 $\pm$ 12 & 4.9 & 24.2 $\pm$ 2.0 & 0.56 $\pm$ 0.03 & 78 $\pm$ 11 & 4.9 \\
8.33 & 0.026 & 26.8 $\pm$ 2.5 & 0.56 $\pm$ 0.09 & 81 $\pm$ 10 & 4.9 & 23.7 $\pm$ 2.0 & 0.55 $\pm$ 0.09 & 80 $\pm$ 11 & 4.8 \\
8.54 & 0.025 & 26.6 $\pm$ 2.5 & 0.55 $\pm$ 0.09 & 85 $\pm$ 11 & 5.4 & 23.5 $\pm$ 2.0 & 0.55 $\pm$ 0.03 & 84 $\pm$ 12 & 5.4 \\
8.71 & 0.024 & 26.2 $\pm$ 2.0 & 0.55 $\pm$ 0.03 & 90 $\pm$ 12 & 4.7 & 23.2 $\pm$ 2.0 & 0.55 $\pm$ 0.03 & 90 $\pm$ 13 & 4.7 \\
8.87 & 0.022 & 25.2 $\pm$ 2.0 & 0.54 $\pm$ 0.03 & 94 $\pm$ 11 & 5.1 & 22.4 $\pm$ 2.0 & 0.53 $\pm$ 0.03 & 94 $\pm$ 12 & 5.1 \\
9.07 & 0.023 & 24.2 $\pm$ 2.5 & 0.50 $\pm$ 0.03 & 97 $\pm$ 11 & 4.4 & 21.5 $\pm$ 2.0 & 0.50 $\pm$ 0.03 & 97 $\pm$ 11 & 4.4 \\
9.26 & 0.021 & 22.9 $\pm$ 3.0 & 0.45 $\pm$ 0.03 & 102 $\pm$ 11 & 5.4 & 20.4 $\pm$ 2.5 & 0.45 $\pm$ 0.03 & 101 $\pm$ 10 & 5.3 \\
9.45 & 0.023 & 23.3 $\pm$ 3.5 & 0.42 $\pm$ 0.03 & 104 $\pm$ 10 & 4.4 & 20.7 $\pm$ 2.5 & 0.42 $\pm$ 0.03 & 104 $\pm$ 10 & 4.4 \\
9.63 & 0.025 & 22.5 $\pm$ 3.5 & 0.38 $\pm$ 0.03 & 104 $\pm$ 8 & 3.6 & 20.0 $\pm$ 3.0 & 0.38 $\pm$ 0.03 & 104 $\pm$ 9 & 3.6 \\
9.78 & 0.020 & 21.8 $\pm$ 3.5 & 0.36 $\pm$ 0.03 & 104 $\pm$ 7 & 4.6 & 19.4 $\pm$ 3.5 & 0.36 $\pm$ 0.03 & 104 $\pm$ 7 & 4.6 \\
9.92 & 0.019 & 20.9 $\pm$ 3.5 & 0.35 $\pm$ 0.03 & 105 $\pm$ 8 & 4.8 & 18.6 $\pm$ 3.0 & 0.35 $\pm$ 0.03 & 105 $\pm$ 8 & 4.8 \\
10.09 & 0.017 & 21.5 $\pm$ 3.5 & 0.35 $\pm$ 0.03 & 106 $\pm$ 7 & 5.3 & 19.2 $\pm$ 3.0 & 0.35 $\pm$ 0.03 & 106 $\pm$ 7 & 5.3 \\
10.26 & 0.017 & 22.2 $\pm$ 3.5 & 0.34 $\pm$ 0.03 & 105 $\pm$ 7 & 5.8 & 19.7 $\pm$ 3.0 & 0.34 $\pm$ 0.03 & 105 $\pm$ 7 & 5.8 \\
10.42 & 0.018 & 24.0 $\pm$ 3.5 & 0.34 $\pm$ 0.03 & 105 $\pm$ 6 & 4.9 & 21.4 $\pm$ 3.5 & 0.34 $\pm$ 0.03 & 105 $\pm$ 7 & 4.9 \\
10.58 & 0.021 & 25.4 $\pm$ 3.5 & 0.35 $\pm$ 0.03 & 106 $\pm$ 7 & 3.1 & 22.5 $\pm$ 3.0 & 0.35 $\pm$ 0.03 & 106 $\pm$ 6 & 3.1 \\
10.71 & 0.019 & 25.9 $\pm$ 3.5 & 0.36 $\pm$ 0.03 & 105 $\pm$ 6 & 4.3 & 23.0 $\pm$ 3.0 & 0.36 $\pm$ 0.03 & 105 $\pm$ 6 & 4.3 \\
10.84 & 0.020 & 27.0 $\pm$ 3.5 & 0.36 $\pm$ 0.03 & 104 $\pm$ 7 & 3.8 & 24.0 $\pm$ 3.0 & 0.36 $\pm$ 0.03 & 104 $\pm$ 6 & 3.9 \\
10.99 & 0.019 & 27.6 $\pm$ 3.0 & 0.37 $\pm$ 0.03 & 104 $\pm$ 6 & 4.1 & 24.5 $\pm$ 3.0 & 0.37 $\pm$ 0.03 & 104 $\pm$ 6 & 4.1 \\
11.14 & 0.022 & 26.9 $\pm$ 3.5 & 0.37 $\pm$ 0.03 & 104 $\pm$ 7 & 2.8 & 23.9 $\pm$ 3.0 & 0.37 $\pm$ 0.03 & 104 $\pm$ 6 & 2.8 \\
11.29 & 0.018 & 26.6 $\pm$ 3.5 & 0.37 $\pm$ 0.03 & 103 $\pm$ 6 & 5.0 & 23.6 $\pm$ 3.5 & 0.37 $\pm$ 0.03 & 103 $\pm$ 7 & 5.0 \\
11.43 & 0.022 & 26.6 $\pm$ 4.0 & 0.37 $\pm$ 0.03 & 101 $\pm$ 7 & 3.7 & 23.6 $\pm$ 3.5 & 0.37 $\pm$ 0.03 & 101 $\pm$ 6 & 3.7 \\
11.55 & 0.022 & 24.6 $\pm$ 4.5 & 0.37 $\pm$ 0.03 & 102 $\pm$ 7 & 4.0 & 21.8 $\pm$ 4.5 & 0.37 $\pm$ 0.03 & 101 $\pm$ 7 & 4.0 \\
11.67 & 0.022 & 26.1 $\pm$ 4.5 & 0.39 $\pm$ 0.03 & 101 $\pm$ 7 & 4.2 & 23.2 $\pm$ 4.0 & 0.38 $\pm$ 0.03 & 101 $\pm$ 7 & 4.2 \\
11.81 & 0.029 & 27.4 $\pm$ 4.5 & 0.40 $\pm$ 0.03 & 101 $\pm$ 7 & 2.7 & 24.3 $\pm$ 4.0 & 0.40 $\pm$ 0.03 & 100 $\pm$ 8 & 2.7 \\
11.95 & 0.032 & 26.0 $\pm$ 5.5 & 0.39 $\pm$ 0.03 & 99 $\pm$ 7 & 2.8 & 23.1 $\pm$ 5.0 & 0.39 $\pm$ 0.03 & 99 $\pm$ 7 & 2.8 \\
\noalign{\smallskip}                               
\hline                               
\end{tabular}                               
\begin{flushleft}
   \textbf{Notes. }
   $^{\mathrm{a}}$~mean visibility error used for the corresponding wavelength bin (cf.~Sect.~\ref{secObsSubVis}); $^{\mathrm{b}}$~reduced chi square
\end{flushleft}
\end{table*}        
%----------------

%----------------
\begin{table*}
\caption[Model fit results for R~Hya]{\textbf{R~Hya}: circular fully limb-darkened disk~(FDD) and circular uniform disk~(UD).}
\label{TableResultsRHya}                       
\centering                       
\begin{tabular}{cc|ccc|ccc}                       
%\noalign{\smallskip}                               
\hline                       
\noalign{\smallskip}                       
 &  & \multicolumn{3}{c}{circular FDD}         & \multicolumn{3}{c}{circular UD}         \\
$\lambda$ & $\Delta V^{\mathrm{a}}$ & $\theta_{\mathrm{FDD}}$   & Flux $\epsilon_{\mathrm{FDD}}$   & $\chi^\mathrm{2,b}_\mathrm{r}$ & $\theta_{\mathrm{UD}}$   & Flux $\epsilon_{\mathrm{UD}}$   & $\chi^\mathrm{2,b}_\mathrm{r}$ \\
($\mu\mathrm{m}$) &  & ($\mathrm{mas}$)   &    &  & ($\mathrm{mas}$)   &    &  \\
\noalign{\smallskip}                       
\hline                       
\noalign{\smallskip}                       
8.12 & 0.012 & 51.6 $\pm$ 1.2 & 0.92 $\pm$ 0.03 & 13 & 45.6 $\pm$ 1.2 & 0.91 $\pm$ 0.02 & 19 \\
8.33 & 0.012 & 50.5 $\pm$ 0.7 & 0.92 $\pm$ 0.02 & 13 & 43.4 $\pm$ 1.2 & 0.89 $\pm$ 0.03 & 18 \\
8.54 & 0.012 & 50.4 $\pm$ 0.8 & 0.92 $\pm$ 0.02 & 14 & 43.2 $\pm$ 0.8 & 0.89 $\pm$ 0.02 & 19 \\
8.71 & 0.011 & 49.9 $\pm$ 0.7 & 0.94 $\pm$ 0.01 & 13 & 42.9 $\pm$ 0.6 & 0.91 $\pm$ 0.02 & 18 \\
8.87 & 0.011 & 48.6 $\pm$ 0.8 & 0.93 $\pm$ 0.02 & 12 & 42.1 $\pm$ 0.6 & 0.91 $\pm$ 0.02 & 17 \\
9.07 & 0.012 & 47.7 $\pm$ 0.5 & 0.92 $\pm$ 0.01 & 10 & 41.1 $\pm$ 0.6 & 0.90 $\pm$ 0.01 & 13 \\
9.26 & 0.011 & 47.0 $\pm$ 0.5 & 0.90 $\pm$ 0.01 & 9 & 40.5 $\pm$ 0.6 & 0.88 $\pm$ 0.01 & 11 \\
9.45 & 0.013 & 46.4 $\pm$ 0.6 & 0.88 $\pm$ 0.01 & 7 & 40.3 $\pm$ 0.6 & 0.86 $\pm$ 0.02 & 9 \\
9.63 & 0.011 & 46.8 $\pm$ 0.6 & 0.86 $\pm$ 0.01 & 9 & 40.3 $\pm$ 0.6 & 0.84 $\pm$ 0.01 & 11 \\
9.78 & 0.011 & 46.7 $\pm$ 0.5 & 0.85 $\pm$ 0.01 & 9 & 40.3 $\pm$ 0.6 & 0.83 $\pm$ 0.01 & 11 \\
9.92 & 0.011 & 46.6 $\pm$ 0.6 & 0.84 $\pm$ 0.01 & 8 & 40.3 $\pm$ 0.6 & 0.82 $\pm$ 0.01 & 10 \\
10.09 & 0.011 & 46.1 $\pm$ 0.5 & 0.82 $\pm$ 0.01 & 8 & 40.3 $\pm$ 0.6 & 0.81 $\pm$ 0.01 & 10 \\
10.26 & 0.011 & 46.8 $\pm$ 0.5 & 0.82 $\pm$ 0.01 & 9 & 40.9 $\pm$ 0.8 & 0.81 $\pm$ 0.01 & 11 \\
10.42 & 0.009 & 48.8 $\pm$ 0.8 & 0.81 $\pm$ 0.01 & 14 & 42.2 $\pm$ 0.6 & 0.80 $\pm$ 0.01 & 18 \\
10.58 & 0.009 & 50.0 $\pm$ 0.6 & 0.81 $\pm$ 0.01 & 16 & 43.2 $\pm$ 0.6 & 0.79 $\pm$ 0.01 & 21 \\
10.71 & 0.010 & 50.5 $\pm$ 0.7 & 0.80 $\pm$ 0.01 & 16 & 43.7 $\pm$ 0.6 & 0.79 $\pm$ 0.01 & 20 \\
10.84 & 0.009 & 51.3 $\pm$ 0.8 & 0.80 $\pm$ 0.01 & 19 & 44.3 $\pm$ 0.6 & 0.78 $\pm$ 0.01 & 24 \\
10.99 & 0.008 & 52.8 $\pm$ 0.7 & 0.79 $\pm$ 0.01 & 21 & 45.5 $\pm$ 0.6 & 0.77 $\pm$ 0.01 & 28 \\
11.14 & 0.008 & 53.4 $\pm$ 0.7 & 0.78 $\pm$ 0.01 & 21 & 46.0 $\pm$ 0.8 & 0.77 $\pm$ 0.02 & 27 \\
11.29 & 0.010 & 54.1 $\pm$ 0.7 & 0.79 $\pm$ 0.01 & 16 & 46.6 $\pm$ 0.8 & 0.77 $\pm$ 0.01 & 21 \\
11.43 & 0.011 & 54.4 $\pm$ 0.8 & 0.79 $\pm$ 0.01 & 13 & 46.9 $\pm$ 0.8 & 0.77 $\pm$ 0.01 & 16 \\
11.55 & 0.011 & 55.5 $\pm$ 0.8 & 0.80 $\pm$ 0.01 & 14 & 47.8 $\pm$ 0.8 & 0.78 $\pm$ 0.01 & 17 \\
11.67 & 0.011 & 56.3 $\pm$ 0.8 & 0.80 $\pm$ 0.02 & 15 & 48.5 $\pm$ 1.0 & 0.78 $\pm$ 0.02 & 19 \\
11.81 & 0.008 & 57.3 $\pm$ 0.8 & 0.80 $\pm$ 0.01 & 29 & 49.3 $\pm$ 1.2 & 0.78 $\pm$ 0.02 & 37 \\
11.95 & 0.007 & 58.0 $\pm$ 0.8 & 0.81 $\pm$ 0.02 & 40 & 49.9 $\pm$ 1.0 & 0.79 $\pm$ 0.02 & 51 \\
\noalign{\smallskip}                       
\hline                       
\end{tabular}   

\vspace{5mm}                 

\caption[Model fit results for V~Hya]{\textbf{V~Hya}: circular fully limb-darkened disk~(FDD) + circular Gaussian~(G) and circular fully limb-darkened disk~(FDD) + circular uniform ring~(R).}                                   
\label{TableResultsVHya}                                   
\centering                                   
\begin{tabular}{cc|cccc|ccccc}                                   
%\noalign{\smallskip}                                   
\hline                                   
\noalign{\smallskip}                                   
 &  & \multicolumn{4}{c}{circular FDD + circular Gaussian}             & \multicolumn{5}{c}{circular FDD + circular ring}                 \\
$\lambda$ & $\Delta V^{\mathrm{a}}$ & $\theta_{\mathrm{FDD}}$   & Flux $\epsilon_{\mathrm{FDD}}$   & $\theta_{\mathrm{G}}$   & $\chi^\mathrm{2,b}_\mathrm{r}$ & $\theta_{\mathrm{FDD}}$   & Flux $\epsilon_{\mathrm{FDD}}$   & $\theta_{\mathrm{R,in}}$   & $\theta_{\mathrm{R,out}}$   & $\chi^\mathrm{2,b}_\mathrm{r}$ \\
($\mu\mathrm{m}$) &  & ($\mathrm{mas}$)   &    & ($\mathrm{mas}$)   &  & ($\mathrm{mas}$)   &    & ($\mathrm{mas}$)   & ($\mathrm{mas}$)   &  \\
\noalign{\smallskip}                                   
\hline                                   
\noalign{\smallskip}                                   
8.12 & 0.011 & 28.5 $\pm$ 2.0 & 0.22 $\pm$ 0.03 & 99 $\pm$ 6 & 8 & 35.3 $\pm$ 2.0 & 0.32 $\pm$ 0.04 & 51 $\pm$ 4 & 124 $\pm$ 6 & 14 \\
8.33 & 0.011 & 26.5 $\pm$ 2.0 & 0.25 $\pm$ 0.03 & 100 $\pm$ 7 & 8 & 31.9 $\pm$ 2.0 & 0.33 $\pm$ 0.04 & 49 $\pm$ 4 & 125 $\pm$ 6 & 12 \\
8.54 & 0.012 & 24.5 $\pm$ 2.0 & 0.26 $\pm$ 0.03 & 101 $\pm$ 5 & 8 & 30.0 $\pm$ 2.0 & 0.34 $\pm$ 0.04 & 50 $\pm$ 4 & 127 $\pm$ 6 & 12 \\
8.71 & 0.011 & 22.9 $\pm$ 2.0 & 0.27 $\pm$ 0.01 & 101 $\pm$ 5 & 9 & 28.7 $\pm$ 2.0 & 0.35 $\pm$ 0.04 & 51 $\pm$ 4 & 129 $\pm$ 6 & 12 \\
8.87 & 0.011 & 21.8 $\pm$ 2.0 & 0.27 $\pm$ 0.01 & 102 $\pm$ 5 & 10 & 27.9 $\pm$ 2.0 & 0.35 $\pm$ 0.04 & 51 $\pm$ 4 & 131 $\pm$ 6 & 12 \\
9.07 & 0.011 & 21.0 $\pm$ 2.0 & 0.27 $\pm$ 0.03 & 102 $\pm$ 5 & 11 & 27.7 $\pm$ 2.0 & 0.35 $\pm$ 0.04 & 51 $\pm$ 4 & 133 $\pm$ 6 & 15 \\
9.26 & 0.010 & 20.4 $\pm$ 2.5 & 0.26 $\pm$ 0.03 & 102 $\pm$ 4 & 13 & 27.7 $\pm$ 2.0 & 0.34 $\pm$ 0.04 & 51 $\pm$ 4 & 135 $\pm$ 6 & 17 \\
9.45 & 0.010 & 20.2 $\pm$ 2.5 & 0.25 $\pm$ 0.01 & 103 $\pm$ 5 & 14 & 27.9 $\pm$ 2.0 & 0.34 $\pm$ 0.04 & 51 $\pm$ 4 & 138 $\pm$ 6 & 20 \\
9.63 & 0.009 & 19.8 $\pm$ 3.0 & 0.24 $\pm$ 0.03 & 105 $\pm$ 5 & 18 & 28.1 $\pm$ 2.0 & 0.33 $\pm$ 0.04 & 52 $\pm$ 4 & 140 $\pm$ 6 & 26 \\
9.78 & 0.009 & 19.2 $\pm$ 3.0 & 0.23 $\pm$ 0.03 & 105 $\pm$ 5 & 19 & 28.1 $\pm$ 2.0 & 0.32 $\pm$ 0.04 & 51 $\pm$ 4 & 143 $\pm$ 6 & 29 \\
9.92 & 0.008 & 19.3 $\pm$ 3.5 & 0.23 $\pm$ 0.03 & 105 $\pm$ 4 & 20 & 28.6 $\pm$ 2.0 & 0.32 $\pm$ 0.04 & 51 $\pm$ 4 & 145 $\pm$ 6 & 30 \\
10.09 & 0.008 & 19.8 $\pm$ 3.5 & 0.22 $\pm$ 0.01 & 105 $\pm$ 4 & 25 & 29.3 $\pm$ 2.0 & 0.31 $\pm$ 0.04 & 50 $\pm$ 4 & 147 $\pm$ 6 & 39 \\
10.26 & 0.007 & 20.4 $\pm$ 3.5 & 0.21 $\pm$ 0.01 & 106 $\pm$ 4 & 26 & 30.4 $\pm$ 2.0 & 0.30 $\pm$ 0.04 & 50 $\pm$ 4 & 149 $\pm$ 6 & 41 \\
10.42 & 0.006 & 20.6 $\pm$ 3.5 & 0.20 $\pm$ 0.03 & 107 $\pm$ 4 & 33 & 31.0 $\pm$ 2.0 & 0.29 $\pm$ 0.04 & 49 $\pm$ 4 & 152 $\pm$ 6 & 56 \\
10.58 & 0.006 & 20.6 $\pm$ 3.5 & 0.18 $\pm$ 0.03 & 108 $\pm$ 5 & 35 & 31.0 $\pm$ 2.0 & 0.27 $\pm$ 0.04 & 46 $\pm$ 4 & 155 $\pm$ 6 & 65 \\
10.71 & 0.007 & 20.6 $\pm$ 4.0 & 0.17 $\pm$ 0.03 & 109 $\pm$ 5 & 29 & 31.1 $\pm$ 2.0 & 0.26 $\pm$ 0.04 & 44 $\pm$ 4 & 157 $\pm$ 6 & 57 \\
10.84 & 0.006 & 19.8 $\pm$ 4.5 & 0.17 $\pm$ 0.03 & 111 $\pm$ 4 & 36 & 30.7 $\pm$ 2.0 & 0.25 $\pm$ 0.04 & 43 $\pm$ 4 & 160 $\pm$ 6 & 73 \\
10.99 & 0.005 & 19.2 $\pm$ 4.5 & 0.16 $\pm$ 0.01 & 112 $\pm$ 4 & 41 & 30.4 $\pm$ 2.0 & 0.24 $\pm$ 0.04 & 41 $\pm$ 4 & 162 $\pm$ 6 & 85 \\
11.14 & 0.006 & 18.1 $\pm$ 5.5 & 0.15 $\pm$ 0.01 & 113 $\pm$ 4 & 37 & 29.2 $\pm$ 2.0 & 0.22 $\pm$ 0.04 & 37 $\pm$ 4 & 165 $\pm$ 6 & 77 \\
11.29 & 0.006 & 17.5 $\pm$ 6.0 & 0.15 $\pm$ 0.03 & 114 $\pm$ 3 & 33 & 28.6 $\pm$ 2.0 & 0.21 $\pm$ 0.04 & 34 $\pm$ 4 & 167 $\pm$ 6 & 70 \\
11.43 & 0.006 & 16.9 $\pm$ 5.5 & 0.14 $\pm$ 0.01 & 115 $\pm$ 4 & 33 & 27.7 $\pm$ 2.0 & 0.20 $\pm$ 0.04 & 29 $\pm$ 4 & 169 $\pm$ 6 & 71 \\
11.55 & 0.006 & 16.3 $\pm$ 5.0 & 0.14 $\pm$ 0.03 & 115 $\pm$ 5 & 33 & 27.8 $\pm$ 2.0 & 0.19 $\pm$ 0.04 & 28 $\pm$ 4 & 171 $\pm$ 6 & 73 \\
11.67 & 0.006 & 16.8 $\pm$ 5.5 & 0.13 $\pm$ 0.01 & 117 $\pm$ 4 & 30 & 28.8 $\pm$ 2.0 & 0.19 $\pm$ 0.04 & 28 $\pm$ 4 & 173 $\pm$ 6 & 69 \\
11.81 & 0.006 & 15.0 $\pm$ 6.5 & 0.12 $\pm$ 0.03 & 117 $\pm$ 4 & 34 & 28.9 $\pm$ 2.0 & 0.18 $\pm$ 0.04 & 28 $\pm$ 4 & 175 $\pm$ 6 & 79 \\
11.95 & 0.007 & 14.9 $\pm$ 6.5 & 0.12 $\pm$ 0.01 & 117 $\pm$ 4 & 30 & 25.5 $\pm$ 2.0 & 0.13 $\pm$ 0.04 & 64 $\pm$ 4 & 146 $\pm$ 6 & 70 \\
\noalign{\smallskip}                                   
\hline                                   
\end{tabular}                                   
\begin{flushleft}
   \textbf{Notes. }
   $^{\mathrm{a}}$~mean visibility error used for the corresponding wavelength bin (cf.~Sect.~\ref{secObsSubVis}); $^{\mathrm{b}}$~reduced chi square
\end{flushleft}
\end{table*}                                   
%----------------

%###########################################################################################
%###########################################################################################

%----------------   
    \begin{table*}

        \section{Diameter details}\label{App_TableDia}

        \caption{Details of the diameter measurements presented in the right hand panels of Fig.~\ref{FigUVdia} and Fig.~\ref{FigDiaWHya}.}        
        \label{TableDia_Targets}        
        \centering        
        \footnotesize         
        \begin{tabular}{L{40mm}C{10mm}C{23mm}p{85mm}}
%            \noalign{\medskip}        
            \hline        
            \noalign{\smallskip}        
            \textbf{R~Aql:} Author  &  Model  &  Visual phase  &  Comments \\
            \noalign{\smallskip}        
            \hline        
            \noalign{\smallskip}        
            A:  \citet{Haniff1995}      &  UD+G & 0.06         &  UD and Gaussian at 700 and 710~nm \\
            B:  \citet{MillanGabet2005} &  UD   & 0.90         &  UD diameter at H and K band \\
            C:  \citet{Ragland2006}     &  UD   & 0.70         &  UD diameter at H band \\
            D:  \citet{Hofmann2000}     &  UD   & 0.17         &  UD diameter at K band \\
            E:  \citet{vanBelle1996}    &  UD   & 0.90 \& 0.31 &  upper for phase 0.31 \& lower for phase 0.90 (K band) \\
            F:  this work               &  FDD  & $0.0-1.0$    &  full data set with error bars \\
            \noalign{\smallskip}        
            \hline        
            \noalign{\smallskip}        
              \textbf{R~Aqr:} Author  &  Model  &  Visual phase  &  Comments\\
            \noalign{\smallskip}        
            \hline        
            \noalign{\smallskip}        
            A:  \citet{Tuthill1999}     &  UD+G & $-$          &  UD and Gaussian at 830~nm \\
            B:  \citet{Tuthill2000}     &  UD   & 0.12 \& 0.68 &  phase 0.68 for upper/lower point at 2.2/3.1~$\mu$m \\
            C:  \citet{MillanGabet2005} &  UD   & 0.40         &  UD diameter at J, H and K band \\
            D:  \citet{Ragland2006}     &  UD   & 0.30         &  UD diameter at H band \\
            E:  \citet{Ragland2008}     &  UD   & $0.60-1.11$  &  UD diameters around the H band at different phases \\
            F:  \citet{Mennesson2002}   &  UD   & 0.41 \& 0.51 &  UD diameter at K and L band \\
            G:  \citet{vanBelle1996}    &  UD   & 0.34 \& 0.57 &  upper for phase 0.34 \& lower for phase 0.57 (K band) \\
            H:  this work               &  FDD  & $0.0-1.0$    &  full data set with error bars \\
            \noalign{\smallskip}        
            \hline        
            \noalign{\smallskip}        
              \textbf{R~Hya:} Author  &  Model  &  Visual phase  &  Comments\\
            \noalign{\smallskip}        
            \hline        
            \noalign{\smallskip}        
            A:  \citet{Ireland2004a}    &  G    & 0.62         &  upper for PA~$123^{\circ}$ \& lower for PA~$72^{\circ}$ \\
            B:  \citet{Haniff1995}      &  UD+G & 0.28         &  UD and Gaussian at 700, 833 and 902~nm \\
            C:  \citet{Monnier2004}     &  UD   & 0.50         &  UD diameter at K band \\
            D:  \citet{MillanGabet2005} &  UD   & 0.80         &  UD diameters at J, H and K band \\
            E:  this work               &  FDD  & $0.0-1.0$    &  full data set with error bars \\
            \noalign{\smallskip}        
            \hline        
            \noalign{\smallskip}        
              \textbf{W~Hya:} Author  &  Model  &  Visual phase  &  Comments\\
            \noalign{\smallskip}        
            \hline        
            \noalign{\smallskip}        
            A:  \citet{Lattanzi1997}    &  UD  & 0.64 & two perpendicular axes ($\eta$~$\approx$~0.86, PA~$\approx$~$143^{\circ}$, 583~nm) \\
            B:  \citet{Ireland2004a}    &  G   & 0.44 & upper curve for PA~$120^{\circ}$ and lower curve for PA~$252^{\circ}$ \\
            C:  \citet{Haniff1995}      &  UD  & 0.04 & UD diameter at 700 and 710~nm \\
            D:  \citet{Tuthill1999}     &  G   & 0.04 & elliptical Gaussian at 700 and 710~nm ($\eta$~$\approx$~0.94, PA~$\approx$~$93^{\circ}$)\\
            E:  \citet{Monnier2004}     &  UD  & 0.50 & UD reported as a bad fit to the data (K band) \\
            F:  \citet{Woodruff2009}    &  UD  & $0.58-1.53$  &  curves for phase 0.58~(middle), 0.79~(lower) and 1.53~(upper) \\
            G:  \citet{Woodruff2008}    &  UD  & $0.50-1.00$  &  mean of multiple measurements in given phase range (J, H, K, L)\\
            H:  \citet{MillanGabet2005} &  UD  & 0.60         &  UD diameter at H and K band \\
            I:  \citet{Wishnow2010}     &  UD  & $0.1-1.1$    &  inner UD dust shell diameter at 11.15~$\mu$m \\
            J:  \citet{Bedding1997}     &  UD  & 0.43         &  UD diameter at 1.45~$\mu$m \\
            K:  this work               &  FDD & $0.0-1.0$    &  full data set with error bars \\
            \noalign{\smallskip}        
            \hline        
            \noalign{\smallskip}        
              \textbf{V~Hya:} Author  &  Model  &  Visual phase  &  Comments\\
            \noalign{\smallskip}        
            \hline        
            \noalign{\smallskip}        
            A:  \citet{Ragland2006}     &  UD    & 0.10          &  UD diameter at H band \\
            B:  \citet{vanBelle1999}    &  UD    & 0.00          &  UD diameter at H band \\
            C:  \citet{MillanGabet2003} &  UD    & $-$           &  UD diameter at K band \\
            D:  this work               &  FDD   & $0.0-1.0$     &  full data set with error bars \\
            \noalign{\smallskip}        
            \hline        
        \end{tabular}

    \section{Observation log}\label{App_ObsLog}

    \caption[AT configurations]{AT stations used in this study (cf.~http://www.eso.org/sci/facilities/paranal/telescopes/vlti/configuration/index.html).}
    \label{TableSec3_usedbase}
    \centering
    \begin{tabular}{ccccc}
%            \noalign{\smallskip}
%            \noalign{\smallskip}
            \hline
%            \noalign{\smallskip}
            \noalign{\smallskip}
            %Name & AT stations & Ground length$^{\mathrm{a}}$ & Ground PA$^{\mathrm{a}}$ & Resolution$^{\mathrm{b}}$\\
            Name & AT stations & Ground length & Ground PA & Resolution (at 10~$\mu$m)\\
            \noalign{\smallskip}
            \hline
            \noalign{\smallskip}
                A & E0-G0 & 16.006 m &  71.020$^\circ$ & 128 mas\\
                B & G0-H0 & 31.998 m &  70.998$^\circ$ & 64 mas\\
         B$^\ast$ & A0-D0 & 32.011 m &  71.014$^\circ$ & 64 mas\\
                C & E0-H0 & 48.004 m &  71.005$^\circ$ & 42 mas\\
                D & D0-H0 & 64.005 m &  71.012$^\circ$ & 32 mas\\
                E & D0-G1 & 71.557 m & 134.443$^\circ$ & 28 mas\\
                F & H0-G1 & 71.555 m &   7.576$^\circ$ & 28 mas\\
            \noalign{\smallskip}
            \hline
    \end{tabular}
\end{table*}
%----------------

%###########################################################################################
%###########################################################################################

\begin{center}
\footnotesize 
\begin{longtable}{ccccccccccc}
\caption{Detailed observation logs of R~Aql, R~Aqr, R~Hya, and V~Hya.} \label{TableApp_log}\\
\toprule
   \noalign{\smallskip}        
     \multicolumn{11}{l}{\textbf{Observation log of R~Aql:}} \\
   \noalign{\smallskip}        
     Date                    & UT time & P$^{\mathrm{a}}$ & AT$^{\mathrm{b}}$ & Disp$^{\mathrm{c}}$ & Cal$^{\mathrm{d}}$ & B$^{\mathrm{e}}$ (m) & PA$^{\mathrm{f}}$ ($^\circ$) & Phase$^{\mathrm{g}}$ & H$^{\mathrm{h}}$ ($^\circ$) & QF$^{\mathrm{i}}$ \\
%   \noalign{\smallskip}        
\midrule
\endfirsthead
\toprule
     Date                    & UT time & P$^{\mathrm{a}}$ & AT$^{\mathrm{b}}$ & Disp$^{\mathrm{c}}$ & Cal$^{\mathrm{d}}$ & B$^{\mathrm{e}}$ (m) & PA$^{\mathrm{f}}$ ($^\circ$) & Phase$^{\mathrm{g}}$ & H$^{\mathrm{h}}$ ($^\circ$) & QF$^{\mathrm{i}}$ \\
\midrule
\endhead
\multicolumn{3}{r}{{...}continues on next page} \\
\endfoot
\endlastfoot
  2007-04-17  &  09:06:23.779  &  P79  &  B  &  GRISM  &  2 of 3  &  29.05  &  74.77  &  0.78  &  53.87  &  used  \\
  2007-04-25  &  08:09:12.338  &  P79  &  D  &  GRISM  &  2 of 2  &  54.90  &  74.83  &  0.81  &  50.84  &  used  \\
  2007-05-09  &  09:51:15.000  &  P79  &  D  &  GRISM  &  0 of 1  &  63.39  &  71.22  &  0.85  &  52.97  &  not used  \\
  2007-06-19  &  08:25:08.000  &  P79  &  B  &  GRISM  &  0 of 2  &  28.98  &  66.54  &  0.02  &  41.70  &  not used  \\
  2007-06-20  &  03:28:40.000  &  P79  &  C  &  GRISM  &  4 of 6  &  34.14  &  73.90  &  0.02  &  41.23  &  used  \\
  2007-06-20  &  04:42:25.000  &  P79  &  B  &  GRISM  &  0 of 1  &  28.64  &  74.81  &  0.02  &  52.51  &  not used  \\
  2007-06-21  &  08:30:51.000  &  P79  &  A  &  GRISM  &  0 of 2  &  14.04  &  65.19  &  0.02  &  39.22  &  not used  \\
  2007-06-22  &  08:05:24.000  &  P79  &  C  &  GRISM  &  1 of 1  &  44.27  &  67.37  &  0.02  &  43.09  &  used  \\
  2007-10-02  &  23:37:29.000  &  P80  &  A  &  PRISM  &  1 of 1  &  16.01  &  72.65  &  0.40  &  56.10  &  used  \\
  2007-10-05  &  00:24:43.000  &  P79  &  B  &  GRISM  &  1 of 2  &  31.24  &  70.22  &  0.41  &  50.57  &  used  \\
  2007-10-05  &  23:33:31.000  &  P80  &  B  &  PRISM  &  1 of 1  &  32.00  &  72.56  &  0.41  &  55.85  &  used  \\
  2007-10-06  &  23:43:58.000  &  P80  &  C  &  PRISM  &  2 of 2  &  47.87  &  71.90  &  0.42  &  54.73  &  used  \\
  2007-10-07  &  00:11:23.000  &  P80  &  C  &  PRISM  &  2 of 2  &  47.08  &  70.51  &  0.42  &  51.29  &  used  \\
  2008-03-06  &  09:16:57.330  &  P80  &  D  &  PRISM  &  1 of 1  &  31.21  &  69.28  &  0.98  &  28.48  &  used  \\
  2008-03-06  &  09:26:46.124  &  P80  &  D  &  PRISM  &  1 of 1  &  33.42  &  70.36  &  0.98  &  30.49  &  used  \\
  2008-03-25  &  09:31:47.779  &  P79  &  A  &  GRISM  &  0 of 1  &  12.33  &  74.45  &  0.05  &  45.38  &  not used  \\
  2008-04-02  &  09:00:39.945  &  P81  &  A  &  PRISM  &  3 of 3  &  12.34  &  74.46  &  0.07  &  45.44  &  used  \\
  2008-04-02  &  09:39:51.946  &  P81  &  C  &  PRISM  &  3 of 4  &  41.50  &  74.83  &  0.07  &  51.25  &  used  \\
  2008-04-04  &  09:30:40.946  &  P81  &  B  &  PRISM  &  1 of 1  &  27.57  &  74.82  &  0.08  &  51.08  &  used  \\
  2008-04-04  &  09:39:46.960  &  P81  &  B  &  PRISM  &  1 of 1  &  28.17  &  74.82  &  0.08  &  52.22  &  used  \\
  2008-04-27  &  08:24:58.000  &  P81  &  D  &  PRISM  &  1 of 1  &  58.66  &  74.76  &  0.17  &  53.96  &  used  \\
  2008-04-27  &  08:36:46.960  &  P81  &  D  &  PRISM  &  1 of 1  &  59.45  &  74.69  &  0.17  &  55.04  &  used  \\
  2008-05-23  &  05:54:09.000  &  P79  &  D  &  GRISM  &  1 of 1  &  52.03  &  74.69  &  0.26  &  47.67  &  used  \\
  2008-05-26  &  05:17:58.000  &  P81  &  D  &  PRISM  &  1 of 1  &  49.44  &  74.46  &  0.27  &  43.68  &  used  \\
  2008-06-01  &  06:08:05.138  &  P81  &  A  &  PRISM  &  4 of 4  &  70.28  &  128.8  &  0.30  &  54.02  &  used  \\
  2008-06-01  &  06:16:17.960  &  P81  &  A  &  PRISM  &  4 of 4  &  69.88  &  128.8  &  0.30  &  54.79  &  used  \\
  2008-06-01  &  06:56:41.946  &  P81  &  A  &  PRISM  &  4 of 4  &  67.14  &  129.2  &  0.30  &  57.02  &  used  \\
  2008-06-01  &  07:04:33.959  &  P81  &  A  &  PRISM  &  0 of 4  &  66.46  &  129.4  &  0.30  &  57.13  &  not used  \\
  2008-06-06  &  07:46:19.000  &  P81  &  A  &  PRISM  &  0 of 2  &  15.96  &  71.90  &  0.32  &  54.21  &  not used  \\
  2008-06-06  &  09:40:04.000  &  P81  &  C  &  PRISM  &  0 of 1  &  41.28  &  64.31  &  0.32  &  36.72  &  not used  \\
  2008-06-07  &  06:16:17.945  &  P81  &  C  &  PRISM  &  1 of 1  &  45.96  &  74.42  &  0.32  &  56.43  &  used  \\
  2008-06-07  &  06:39:05.000  &  P81  &  B  &  PRISM  &  3 of 3  &  31.67  &  73.66  &  0.32  &  57.11  &  used  \\
  2009-04-20  &  08:01:43.946  &  P83  &  B  &  PRISM  &  3 of 3  &  25.57  &  74.61  &  0.49  &  47.18  &  used  \\
  2009-04-23  &  06:40:13.946  &  P83  &  C  &  PRISM  &  2 of 6  &  28.53  &  72.10  &  0.50  &  34.71  &  used  \\
  2009-04-23  &  07:53:25.946  &  P83  &  C  &  PRISM  &  6 of 6  &  38.78  &  74.66  &  0.50  &  47.72  &  used  \\
  2009-05-02  &  09:38:45.000  &  P83  &  D  &  PRISM  &  3 of 3  &  63.97  &  72.88  &  0.53  &  56.23  &  used  \\
  2009-06-03  &  05:35:12.000  &  P83  &  A  &  PRISM  &  1 of 3  &  13.88  &  74.85  &  0.65  &  51.00  &  used  \\
  2009-06-03  &  06:52:23.000  &  P83  &  A  &  PRISM  &  1 of 3  &  15.69  &  74.01  &  0.65  &  57.07  &  used  \\
  2009-06-26  &  04:31:29.000  &  P83  &  D  &  PRISM  &  1 of 2  &  58.70  &  74.75  &  0.73  &  54.09  &  used  \\
  2009-07-03  &  05:21:54.000  &  P83  &  B  &  PRISM  &  1 of 2  &  31.88  &  73.27  &  0.76  &  56.82  &  used  \\
   \noalign{\smallskip}        
   \hline        
   \noalign{\smallskip}
     \multicolumn{11}{l}{\textbf{Observation log of R~Aqr:}} \\
   \noalign{\smallskip}        
     Date                    & UT time & P$^{\mathrm{a}}$ & AT$^{\mathrm{b}}$ & Disp$^{\mathrm{c}}$ & Cal$^{\mathrm{d}}$ & B$^{\mathrm{e}}$ (m) & PA$^{\mathrm{f}}$ ($^\circ$) & Phase$^{\mathrm{g}}$ & H$^{\mathrm{h}}$ ($^\circ$) & QF$^{\mathrm{i}}$ \\
   \noalign{\smallskip}        
   \hline        
   \noalign{\smallskip}        
  2007-05-16  &  10:06:28.000  &  P79  &  D  &  GRISM  &  1 of 1  &  51.09  &  53.81  &  0.50  &  50.38  &  used  \\
  2007-06-18  &  09:46:44.000  &  P79  &  A  &  GRISM  &  2 of 4  &  15.53  &  67.42  &  0.59  &  74.20  &  used  \\
  2007-06-19  &  09:34:00.000  &  P79  &  B  &  GRISM  &  0 of 2  &  30.79  &  66.67  &  0.59  &  72.48  &  not used  \\
  2007-06-20  &  08:25:38.000  &  P79  &  C  &  GRISM  &  4 of 6  &  42.81  &  61.15  &  0.59  &  58.66  &  used  \\
  2007-06-21  &  06:59:42.000  &  P79  &  A  &  GRISM  &  2 of 2  &  11.32  &  45.06  &  0.59  &  40.09  &  used  \\
  2007-07-02  &  07:18:59.000  &  P79  &  B  &  GRISM  &  2 of 3  &  26.64  &  56.61  &  0.62  &  54.29  &  used  \\
  2007-07-27  &  10:07:03.000  &  P79  &  D  &  GRISM  &  1 of 1  &  57.40  &  78.05  &  0.68  &  60.39  &  used  \\
  2007-10-04  &  06:26:11.000  &  P79  &  A  &  GRISM  &  0 of 1  &  12.59  &  79.91  &  0.86  &  49.04  &  not used  \\
  2007-10-04  &  02:12:15.000  &  P80  &  B  &  PRISM  &  2 of 2  &  30.08  &  64.81  &  0.86  &  68.14  &  used  \\
  2007-10-04  &  04:10:49.000  &  P80  &  B  &  PRISM  &  2 of 2  &  31.80  &  73.85  &  0.86  &  77.57  &  used  \\
  2007-10-04  &  05:28:51.000  &  P80  &  A  &  PRISM  &  1 of 1  &  14.50  &  77.84  &  0.86  &  61.91  &  used  \\
  2007-10-05  &  05:16:14.000  &  P79  &  B  &  GRISM  &  0 of 2  &  29.57  &  77.35  &  0.86  &  63.83  &  not used  \\
  2007-10-06  &  01:57:28.000  &  P80  &  C  &  PRISM  &  2 of 2  &  44.52  &  63.84  &  0.87  &  66.67  &  used  \\
  2007-10-07  &  05:54:46.000  &  P79  &  C  &  GRISM  &  1 of 1  &  40.31  &  79.13  &  0.87  &  53.48  &  used  \\
  2007-10-07  &  01:16:29.000  &  P80  &  C  &  PRISM  &  2 of 2  &  41.29  &  58.76  &  0.87  &  58.52  &  used  \\
  2007-10-07  &  02:00:53.165  &  P80  &  A  &  PRISM  &  1 of 1  &  14.88  &  64.02  &  0.87  &  68.23  &  used  \\
  2007-11-30  &  00:31:14.000  &  P80  &  D  &  PRISM  &  0 of 1  &  47.51  &  74.32  &  0.01  &  76.85  &  not used  \\
  2007-12-02  &  00:23:50.000  &  P80  &  D  &  PRISM  &  0 of 3  &  63.42  &  74.20  &  0.01  &  76.78  &  not used  \\
  2008-05-22  &  09:16:47.000  &  P81  &  D  &  PRISM  &  0 of 3  &  48.10  &  49.62  &  0.43  &  45.12  &  not used  \\
  2008-05-30  &  07:15:31.945  &  P80  &  D  &  PRISM  &  0 of 5  &  36.48  &  21.26  &  0.47  &  24.69  &  not used  \\
  2008-05-30  &  09:49:08.137  &  P80  &  D  &  PRISM  &  2 of 5  &  55.36  &  59.13  &  0.47  &  59.52  &  used  \\
  2008-06-06  &  08:15:59.945  &  P81  &  A  &  PRISM  &  0 of 2  &  11.72  &  47.73  &  0.49  &  44.70  &  not used  \\
  2008-06-06  &  09:53:36.946  &  P81  &  C  &  PRISM  &  1 of 1  &  44.07  &  63.13  &  0.49  &  66.54  &  used  \\
  2008-06-07  &  09:21:11.000  &  P81  &  B  &  PRISM  &  3 of 3  &  28.21  &  60.38  &  0.49  &  60.30  &  used  \\
  2008-06-07  &  09:54:00.945  &  P81  &  B  &  PRISM  &  3 of 3  &  29.58  &  63.60  &  0.49  &  67.47  &  used  \\
  2008-09-25  &  05:26:02.000  &  P81  &  D  &  PRISM  &  1 of 1  &  61.26  &  76.25  &  0.78  &  69.53  &  used  \\
  2008-09-28  &  05:26:37.330  &  P81  &  A  &  PRISM  &  1 of 1  &  15.23  &  76.49  &  0.78  &  66.90  &  used  \\
  2008-12-28  &  01:14:16.000  &  P82  &  A  &  PRISM  &  0 of 1  &  12.25  &  80.18  &  0.01  &  43.58  &  not used  \\
  2008-12-31  &  00:43:23.000  &  P82  &  B  &  PRISM  &  1 of 2  &  26.09  &  79.50  &  0.02  &  47.72  &  used  \\
  2008-12-31  &  01:14:25.000  &  P82  &  B  &  PRISM  &  1 of 2  &  23.39  &  80.52  &  0.02  &  40.95  &  used  \\
  2009-01-01  &  00:12:38.000  &  P82  &  C  &  PRISM  &  1 of 1  &  39.63  &  79.35  &  0.03  &  53.72  &  used  \\
  2009-05-25  &  09:28:26.000  &  P83  &  D  &  PRISM  &  0 of 1  &  50.56  &  53.10  &  0.39  &  50.22  &  not used  \\
  2009-06-03  &  09:04:03.000  &  P83  &  A  &  PRISM  &  2 of 3  &  13.01  &  55.02  &  0.42  &  52.70  &  used  \\
  2009-06-03  &  09:40:43.000  &  P83  &  C  &  PRISM  &  1 of 1  &  42.32  &  60.38  &  0.42  &  60.93  &  used  \\
  2009-07-03  &  06:48:28.000  &  P83  &  B  &  PRISM  &  1 of 2  &  24.83  &  51.87  &  0.50  &  48.71  &  used  \\
  2009-07-30  &  08:40:36.000  &  P83  &  D  &  PRISM  &  1 of 1  &  63.29  &  74.41  &  0.57  &  75.51  &  used  \\
  2009-08-14  &  08:16:39.000  &  P83  &  C  &  PRISM  &  0 of 1  &  45.81  &  76.36  &  0.62  &  68.58  &  not used  \\
  2009-08-14  &  08:38:16.000  &  P83  &  A  &  PRISM  &  0 of 1  &  14.85  &  77.26  &  0.62  &  63.93  &  not used  \\
  2009-08-14  &  09:30:22.000  &  P83  &  B  &  PRISM  &  0 of 1  &  26.11  &  79.49  &  0.62  &  52.30  &  not used  \\
   \noalign{\smallskip}        
   \hline        
   \noalign{\smallskip}
     \multicolumn{11}{l}{\textbf{Observation log of R~Hya:}} \\
   \noalign{\smallskip}        
     Date                    & UT time & P$^{\mathrm{a}}$ & AT$^{\mathrm{b}}$ & Disp$^{\mathrm{c}}$ & Cal$^{\mathrm{d}}$ & B$^{\mathrm{e}}$ (m) & PA$^{\mathrm{f}}$ ($^\circ$) & Phase$^{\mathrm{g}}$ & H$^{\mathrm{h}}$ ($^\circ$) & QF$^{\mathrm{i}}$ \\
   \noalign{\smallskip}        
   \hline        
   \noalign{\smallskip}        
  2007-04-12  &  02:08:09.000  &  P79  &  A  &  GRISM  &  2 of 3  &  13.48  &  48.88  &  0.60  &  52.64  &  used  \\
  2007-04-17  &  07:28:33.338  &  P79  &  B  &  GRISM  &  2 of 3  &  24.73  &  86.35  &  0.61  &  49.77  &  used  \\
  2007-04-18  &  07:10:38.780  &  P79  &  A  &  GRISM  &  3 of 3  &  12.92  &  85.18  &  0.61  &  52.93  &  used  \\
  2007-04-22  &  01:35:49.071  &  P79  &  D  &  GRISM  &  8 of 8  &  54.75  &  50.43  &  0.62  &  55.11  &  used  \\
  2007-04-22  &  07:13:32.070  &  P79  &  D  &  GRISM  &  8 of 8  &  48.03  &  87.13  &  0.62  &  47.83  &  used  \\
  2007-04-25  &  01:23:03.779  &  P79  &  E  &  GRISM  &  0 of 2  &  63.50  &  123.9  &  0.63  &  54.00  &  not used  \\
  2007-04-25  &  04:23:39.780  &  P79  &  F  &  GRISM  &  2 of 2  &  71.55  &  9.98   &  0.63  &  84.59  &  used  \\
  2007-05-16  &  23:52:59.000  &  P79  &  B  &  GRISM  &  0 of 1  &  27.96  &  52.53  &  0.69  &  52.30  &  not used  \\
  2007-06-20  &  01:20:22.000  &  P79  &  C  &  GRISM  &  6 of 6  &  46.25  &  77.60  &  0.78  &  76.25  &  used  \\
  2007-06-21  &  22:54:49.000  &  P79  &  C  &  GRISM  &  1 of 1  &  46.13  &  62.66  &  0.78  &  71.24  &  used  \\
  2007-07-02  &  23:17:47.000  &  P79  &  C  &  GRISM  &  5 of 5  &  47.95  &  69.99  &  0.81  &  86.20  &  used  \\
  2007-07-02  &  01:42:07.000  &  P79  &  C  &  GRISM  &  0 of 5  &  41.92  &  82.60  &  0.81  &  60.55  &  not used  \\
  2007-07-03  &  22:54:48.000  &  P79  &  A  &  GRISM  &  0 of 2  &  15.91  &  68.30  &  0.82  &  81.96  &  not used  \\
  2007-07-03  &  00:28:49.000  &  P79  &  A  &  GRISM  &  2 of 2  &  15.57  &  76.81  &  0.82  &  76.34  &  used  \\
  2007-07-04  &  23:58:37.000  &  P79  &  B  &  GRISM  &  0 of 4  &  31.72  &  74.45  &  0.82  &  82.30  &  not used  \\
  2007-07-04  &  02:19:46.000  &  P79  &  B  &  GRISM  &  0 of 4  &  24.68  &  86.41  &  0.82  &  50.24  &  not used  \\
  2007-12-26  &  08:39:14.000  &  P79  &  D  &  GRISM  &  1 of 1  &  51.86  &  44.87  &  0.29  &  45.73  &  used  \\
  2008-02-20  &  07:53:42.000  &  P80  &  D  &  PRISM  &  4 of 4  &  63.89  &  69.61  &  0.44  &  85.31  &  used  \\
  2008-02-20  &  08:42:11.000  &  P80  &  D  &  PRISM  &  6 of 4  &  63.50  &  74.33  &  0.44  &  83.27  &  used  \\
  2008-02-21  &  09:02:26.000  &  P79  &  D  &  GRISM  &  0 of 2  &  62.50  &  76.42  &  0.44  &  77.80  &  used  \\
  2008-02-22  &  07:02:15.000  &  P80  &  B* &  PRISM  &  0 of 6  &  31.22  &  64.71  &  0.45  &  75.47  &  not used  \\
  2008-02-22  &  07:54:30.156  &  P80  &  B* &  PRISM  &  0 of 6  &  32.00  &  70.49  &  0.45  &  87.16  &  not used  \\
  2008-03-02  &  06:36:33.000  &  P80  &  E  &  PRISM  &  2 of 2  &  70.92  &  129.9  &  0.47  &  77.67  &  used  \\
  2008-03-02  &  07:52:58.000  &  P80  &  F  &  PRISM  &  1 of 2  &  71.55  &  10.27  &  0.47  &  84.61  &  used  \\
  2008-03-02  &  08:28:33.000  &  P80  &  D  &  PRISM  &  2 of 2  &  62.29  &  76.75  &  0.47  &  76.56  &  used  \\
  2008-03-13  &  04:43:47.000  &  P80  &  B  &  PRISM  &  7 of 8  &  29.03  &  56.32  &  0.50  &  61.84  &  used  \\
  2008-03-13  &  07:00:13.945  &  P80  &  B  &  PRISM  &  7 of 8  &  31.97  &  72.36  &  0.50  &  86.68  &  used  \\
  2008-03-13  &  09:09:41.945  &  P80  &  B  &  PRISM  &  7 of 8  &  27.22  &  83.52  &  0.50  &  57.36  &  used  \\
  2008-03-14  &  05:39:07.000  &  P80  &  A  &  PRISM  &  4 of 6  &  15.57  &  64.36  &  0.50  &  75.34  &  used  \\
  2008-03-14  &  07:02:11.000  &  P80  &  A  &  PRISM  &  4 of 6  &  15.96  &  73.21  &  0.50  &  85.40  &  used  \\
  2008-03-14  &  08:51:47.000  &  P80  &  A  &  PRISM  &  4 of 6  &  13.98  &  82.60  &  0.50  &  60.53  &  used  \\
  2008-03-25  &  06:08:58.000  &  P80  &  C  &  PRISM  &  6 of 6  &  47.97  &  72.15  &  0.53  &  87.51  &  used  \\
  2008-03-25  &  07:59:34.945  &  P80  &  C  &  PRISM  &  6 of 6  &  42.97  &  81.63  &  0.53  &  62.57  &  used  \\
  2008-03-31  &  04:48:31.000  &  P80  &  C  &  PRISM  &  1 of 1  &  47.29  &  66.29  &  0.54  &  79.04  &  used  \\
  2008-04-01  &  03:50:22.946  &  P81  &  B  &  PRISM  &  2 of 2  &  29.75  &  58.88  &  0.55  &  66.69  &  used  \\
  2008-04-02  &  06:42:37.945  &  P81  &  C  &  PRISM  &  4 of 4  &  46.11  &  77.84  &  0.55  &  72.93  &  used  \\
  2008-04-02  &  08:01:14.946  &  P81  &  A  &  PRISM  &  3 of 3  &  13.26  &  84.39  &  0.55  &  55.05  &  used  \\
  2008-04-03  &  07:04:02.946  &  P81  &  B  &  PRISM  &  2 of 2  &  29.69  &  79.96  &  0.55  &  67.16  &  used  \\
  2008-04-28  &  01:55:35.946  &  P81  &  D  &  PRISM  &  6 of 6  &  58.79  &  57.60  &  0.62  &  64.72  &  used  \\
  2008-04-28  &  02:57:39.946  &  P81  &  E  &  PRISM  &  2 of 2  &  70.99  &  130.1  &  0.62  &  78.85  &  used  \\
  2008-04-28  &  04:06:01.945  &  P81  &  F  &  PRISM  &  1 of 2  &  71.55  &  9.69   &  0.62  &  85.24  &  used  \\
  2008-04-28  &  07:13:08.945  &  P81  &  D  &  PRISM  &  6 of 6  &  44.07  &  89.17  &  0.62  &  42.84  &  used  \\
  2008-05-25  &  00:16:55.000  &  P81  &  D  &  PRISM  &  2 of 2  &  59.74  &  59.31  &  0.69  &  66.43  &  used  \\
  2008-07-03  &  01:31:24.000  &  P81  &  A  &  PRISM  &  1 of 2  &  14.10  &  82.30  &  0.79  &  61.43  &  used  \\
  2008-07-03  &  02:43:24.000  &  P81  &  C  &  PRISM  &  2 of 2  &  33.75  &  88.69  &  0.79  &  45.14  &  used  \\
  2008-07-06  &  03:20:45.000  &  P81  &  D  &  PRISM  &  2 of 2  &  35.68  &  93.62  &  0.80  &  34.14  &  used  \\
  2009-01-16  &  08:03:46.946  &  P82  &  F  &  PRISM  &  4 of 5  &  71.54  &  -7.84  &  0.32  &  57.04  &  used  \\
  2009-01-16  &  08:34:11.000  &  P82  &  D  &  PRISM  &  1 of 2  &  58.75  &  57.54  &  0.32  &  63.95  &  used  \\
  2009-01-17  &  07:50:45.946  &  P82  &  E  &  PRISM  &  2 of 2  &  63.99  &  124.0  &  0.32  &  54.99  &  used  \\
  2009-01-21  &  07:17:41.000  &  P81  &  B  &  PRISM  &  5 of 5  &  26.59  &  47.49  &  0.33  &  51.06  &  used  \\
  2009-01-21  &  09:02:59.000  &  P82  &  B  &  PRISM  &  5 of 5  &  31.10  &  64.15  &  0.33  &  74.98  &  used  \\
  2009-01-22  &  07:23:42.945  &  P81  &  C  &  PRISM  &  4 of 5  &  40.45  &  48.91  &  0.34  &  53.31  &  used  \\
  2009-01-22  &  07:36:00.946  &  P81  &  C  &  PRISM  &  4 of 5  &  41.40  &  51.25  &  0.34  &  56.10  &  used  \\
  2009-01-25  &  08:13:46.945  &  P82  &  A  &  PRISM  &  8 of 8  &  14.94  &  59.33  &  0.34  &  67.36  &  used  \\
  2009-01-25  &  08:50:11.945  &  P82  &  A  &  PRISM  &  8 of 8  &  15.56  &  64.18  &  0.34  &  75.65  &  used  \\
  2009-01-26  &  09:11:32.000  &  P82  &  B  &  PRISM  &  1 of 1  &  31.71  &  67.44  &  0.35  &  81.40  &  used  \\
  2009-01-27  &  08:13:44.945  &  P82  &  C  &  PRISM  &  4 of 4  &  45.26  &  60.42  &  0.35  &  69.14  &  used  \\
  2009-01-27  &  08:39:07.946  &  P82  &  C  &  PRISM  &  4 of 4  &  46.52  &  63.77  &  0.35  &  74.92  &  used  \\
  2009-01-27  &  08:46:50.123  &  P82  &  C  &  PRISM  &  4 of 4  &  46.83  &  64.72  &  0.35  &  76.68  &  used  \\
  2009-02-16  &  07:39:19.000  &  P82  &  D  &  PRISM  &  2 of 2  &  63.27  &  66.95  &  0.40  &  79.20  &  used  \\
  2009-03-16  &  08:08:24.000  &  P82  &  D  &  PRISM  &  2 of 3  &  59.80  &  79.60  &  0.48  &  68.84  &  used  \\
  2009-03-16  &  08:36:32.000  &  P81  &  D  &  PRISM  &  2 of 3  &  56.78  &  82.00  &  0.48  &  62.43  &  used  \\
  2009-04-20  &  07:34:42.945  &  P83  &  B  &  PRISM  &  2 of 3  &  23.02  &  88.14  &  0.57  &  45.29  &  used  \\
  2009-04-23  &  05:28:10.000  &  P83  &  C  &  PRISM  &  6 of 6  &  45.43  &  78.85  &  0.58  &  71.30  &  used  \\
  2009-04-23  &  07:05:14.946  &  P83  &  C  &  PRISM  &  6 of 6  &  36.81  &  86.57  &  0.58  &  49.27  &  used  \\
  2009-04-24  &  06:05:35.000  &  P83  &  B  &  PRISM  &  2 of 2  &  28.16  &  82.31  &  0.58  &  61.90  &  used  \\
  2009-05-02  &  01:44:52.000  &  P83  &  D  &  PRISM  &  3 of 3  &  59.37  &  58.63  &  0.60  &  65.64  &  used  \\
  2009-05-03  &  01:50:41.946  &  P83  &  E  &  PRISM  &  2 of 2  &  68.81  &  126.7  &  0.61  &  67.85  &  used  \\
  2009-05-03  &  06:19:20.000  &  P83  &  F  &  PRISM  &  1 of 3  &  70.55  &  22.78  &  0.61  &  50.75  &  used  \\
  2009-05-24  &  23:13:50.000  &  P83  &  D  &  PRISM  &  1 of 1  &  53.58  &  48.26  &  0.66  &  51.89  &  used  \\
  2009-06-04  &  23:30:42.000  &  P83  &  C  &  PRISM  &  3 of 3  &  44.27  &  58.02  &  0.69  &  64.64  &  used  \\
  2009-06-04  &  03:46:00.000  &  P83  &  A  &  PRISM  &  1 of 2  &  13.39  &  84.09  &  0.69  &  56.99  &  used  \\
   \noalign{\smallskip}        
   \hline        
   \noalign{\smallskip}
     \multicolumn{11}{l}{\textbf{Observation log of V~Hya:}} \\
   \noalign{\smallskip}        
     Date                    & UT time & P$^{\mathrm{a}}$ & AT$^{\mathrm{b}}$ & Disp$^{\mathrm{c}}$ & Cal$^{\mathrm{d}}$ & B$^{\mathrm{e}}$ (m) & PA$^{\mathrm{f}}$ ($^\circ$) & Phase$^{\mathrm{g}}$ & H$^{\mathrm{h}}$ ($^\circ$) & QF$^{\mathrm{i}}$ \\
   \noalign{\smallskip}        
   \hline        
   \noalign{\smallskip}        
  2007-04-12  &  03:45:42.000  &  P79  &  A  &  GRISM  &  3 of 3  &  15.01  &  78.79  &  0.68  &  68.63  &  used  \\
  2007-04-14  &  04:01:23.388  &  P79  &  C  &  GRISM  &  1 of 1  &  43.49  &  80.29  &  0.68  &  63.90  &  used  \\
  2007-04-14  &  04:56:33.000  &  P79  &  B  &  GRISM  &  0 of 1  &  24.46  &  84.98  &  0.68  &  50.71  &  not used  \\
  2007-04-22  &  23:36:20.000  &  P79  &  D  &  GRISM  &  1 of 8  &  58.41  &  58.54  &  0.70  &  63.41  &  used  \\
  2007-04-22  &  03:48:12.779  &  P79  &  D  &  GRISM  &  2 of 8  &  55.34  &  81.89  &  0.70  &  58.22  &  used  \\
  2007-12-02  &  07:58:13.000  &  P80  &  D  &  PRISM  &  0 of 3  &  52.05  &  48.04  &  0.12  &  50.30  &  not used  \\
  2007-12-02  &  08:58:00.000  &  P80  &  D  &  PRISM  &  1 of 3  &  58.65  &  58.93  &  0.12  &  63.90  &  used  \\
  2007-12-12  &  06:16:34.000  &  P80  &  F  &  PRISM  &  1 of 1  &  71.54  &  -16.21 &  0.14  &  36.30  &  used  \\
  2007-12-12  &  07:28:52.000  &  P80  &  D  &  PRISM  &  2 of 3  &  53.33  &  50.29  &  0.14  &  52.60  &  used  \\
  2007-12-12  &  08:19:04.000  &  P80  &  E  &  PRISM  &  1 of 1  &  68.45  &  126.5  &  0.14  &  64.00  &  used  \\
  2007-12-12  &  09:05:20.166  &  P80  &  D  &  PRISM  &  2 of 3  &  61.77  &  64.26  &  0.14  &  74.50  &  used  \\
  2008-01-10  &  07:13:33.000  &  P80  &  B  &  PRISM  &  2 of 2  &  31.14  &  65.29  &  0.20  &  76.50  &  used  \\
  2008-01-10  &  05:03:32.000  &  P79  &  B  &  GRISM  &  1 of 1  &  25.01  &  44.22  &  0.20  &  45.49  &  used  \\
  2008-01-11  &  04:54:29.000  &  P80  &  B  &  PRISM  &  1 of 1  &  24.73  &  43.11  &  0.20  &  45.60  &  used  \\
  2008-01-12  &  05:27:07.000  &  P80  &  A  &  PRISM  &  0 of 2  &  13.39  &  50.69  &  0.20  &  52.60  &  not used  \\
  2008-01-12  &  06:18:56.000  &  P80  &  A  &  PRISM  &  0 of 2  &  14.65  &  58.83  &  0.20  &  64.40  &  not used  \\
  2008-01-12  &  04:37:00.000  &  P79  &  A  &  GRISM  &  0 of 1  &  11.96  &  39.65  &  0.20  &  41.26  &  not used  \\
  2008-02-20  &  04:42:41.000  &  P80  &  D  &  PRISM  &  0 of 4  &  62.75  &  66.32  &  0.28  &  78.50  &  not used  \\
  2008-02-21  &  03:27:22.000  &  P80  &  D  &  PRISM  &  0 of 3  &  57.55  &  57.17  &  0.28  &  61.20  &  not used  \\
  2008-02-22  &  03:37:17.000  &  P80  &  B* &  PRISM  &  0 of 6  &  29.39  &  59.09  &  0.28  &  64.30  &  not used  \\
  2008-02-22  &  04:25:40.000  &  P80  &  B* &  PRISM  &  0 of 6  &  31.15  &  65.28  &  0.28  &  75.30  &  not used  \\
  2008-03-11  &  02:14:28.945  &  P80  &  A  &  PRISM  &  3 of 3  &  14.29  &  56.53  &  0.31  &  61.60  &  used  \\
  2008-03-11  &  02:48:44.946  &  P80  &  A  &  PRISM  &  3 of 3  &  15.05  &  61.44  &  0.31  &  69.40  &  used  \\
  2008-03-13  &  01:21:28.946  &  P80  &  B  &  PRISM  &  7 of 8  &  26.17  &  48.54  &  0.32  &  51.30  &  used  \\
  2008-03-13  &  01:53:55.946  &  P80  &  B  &  PRISM  &  7 of 8  &  27.93  &  54.46  &  0.32  &  58.70  &  used  \\
  2008-03-13  &  03:34:35.946  &  P80  &  B  &  PRISM  &  7 of 8  &  31.64  &  67.68  &  0.32  &  81.40  &  used  \\
  2008-03-25  &  02:22:58.000  &  P80  &  C  &  PRISM  &  6 of 6  &  46.74  &  65.36  &  0.34  &  76.70  &  used  \\
  2008-03-25  &  02:02:04.000  &  P79  &  C  &  GRISM  &  1 of 1  &  45.74  &  62.80  &  0.34  &  71.28  &  used  \\
  2008-03-27  &  00:03:49.946  &  P80  &  A  &  PRISM  &  1 of 1  &  12.56  &  44.63  &  0.34  &  47.10  &  used  \\
  2008-03-27  &  00:54:34.945  &  P80  &  C  &  PRISM  &  1 of 2  &  41.87  &  54.41  &  0.34  &  58.60  &  used  \\
  2008-04-02  &  05:13:23.946  &  P81  &  C  &  PRISM  &  2 of 4  &  40.95  &  82.29  &  0.35  &  56.96  &  used  \\
  2008-04-07  &  00:24:01.946  &  P81  &  A  &  PRISM  &  1 of 1  &  14.18  &  55.86  &  0.36  &  60.61  &  used  \\
  2008-04-07  &  01:01:29.000  &  P81  &  C  &  PRISM  &  1 of 1  &  45.24  &  61.65  &  0.36  &  69.13  &  used  \\
  2008-04-07  &  01:47:45.945  &  P81  &  B  &  PRISM  &  2 of 2  &  31.47  &  66.78  &  0.36  &  79.52  &  used  \\
  2008-04-08  &  00:39:00.946  &  P81  &  A  &  PRISM  &  1 of 1  &  14.64  &  58.73  &  0.36  &  64.91  &  used  \\
  2008-04-08  &  01:37:05.946  &  P81  &  B  &  PRISM  &  2 of 2  &  31.32  &  66.04  &  0.36  &  78.03  &  used  \\
  2008-04-28  &  01:24:40.000  &  P81  &  D  &  PRISM  &  6 of 6  &  63.91  &  72.77  &  0.40  &  84.97  &  used  \\
  2008-11-30  &  07:49:13.000  &  P81  &  D  &  PRISM  &  1 of 1  &  51.02  &  46.15  &  0.81  &  47.18  &  used  \\
  2008-12-27  &  07:24:08.000  &  P82  &  A  &  PRISM  &  1 of 1  &  14.84  &  60.02  &  0.86  &  65.61  &  used  \\
  2008-12-27  &  08:14:23.000  &  P80  &  C  &  PRISM  &  0 of 1  &  47.04  &  66.21  &  0.86  &  76.98  &  not used  \\
  2009-01-16  &  05:01:58.946  &  P82  &  E  &  PRISM  &  0 of 1  &  62.98  &  124.5  &  0.90  &  57.84  &  not used  \\
  2009-01-16  &  06:54:20.000  &  P82  &  F  &  PRISM  &  4 of 5  &  71.35  &   1.97  &  0.90  &  76.67  &  used  \\
  2009-01-19  &  06:37:49.000  &  P82  &  D  &  PRISM  &  2 of 2  &  62.41  &  65.54  &  0.91  &  75.61  &  used  \\
  2009-01-19  &  06:51:54.959  &  P82  &  D  &  PRISM  &  2 of 2  &  62.80  &  66.42  &  0.91  &  78.75  &  used  \\
  2009-01-20  &  04:05:35.000  &  P80  &  C  &  PRISM  &  2 of 4  &  36.00  &  39.97  &  0.91  &  41.95  &  used  \\
  2009-01-20  &  04:50:06.946  &  P82  &  C  &  PRISM  &  1 of 4  &  39.53  &  49.19  &  0.91  &  52.04  &  used  \\
  2009-01-20  &  06:41:18.946  &  P82  &  B  &  PRISM  &  1 of 1  &  31.23  &  65.66  &  0.91  &  77.27  &  used  \\
  2009-01-22  &  05:59:08.945  &  P82  &  B  &  PRISM  &  0 of 1  &  30.12  &  61.51  &  0.91  &  69.54  &  not used  \\
  2009-01-22  &  06:07:04.959  &  P82  &  B  &  PRISM  &  0 of 1  &  30.41  &  62.53  &  0.91  &  71.34  &  not used  \\
  2009-01-22  &  06:42:31.946  &  P82  &  C  &  PRISM  &  3 of 5  &  47.18  &  66.67  &  0.91  &  79.28  &  used  \\
  2009-01-25  &  03:35:25.945  &  P82  &  A  &  PRISM  &  2 of 8  &  11.64  &  36.53  &  0.92  &  39.57  &  used  \\
  2009-01-25  &  07:17:58.945  &  P82  &  A  &  PRISM  &  8 of 8  &  16.01  &  71.37  &  0.92  &  86.60  &  used  \\
  2009-02-19  &  01:53:34.000  &  P82  &  D  &  PRISM  &  0 of 1  &  46.81  &  37.21  &  0.97  &  38.77  &  not used  \\
  2009-02-27  &  06:54:56.945  &  P82  &  B  &  PRISM  &  1 of 1  &  29.27  &  79.89  &  0.98  &  64.50  &  used  \\
  2009-02-28  &  03:28:46.946  &  P82  &  B  &  PRISM  &  2 of 2  &  29.92  &  60.86  &  0.98  &  68.43  &  used  \\
  2009-02-28  &  04:04:29.946  &  P82  &  B  &  PRISM  &  2 of 2  &  31.14  &  65.27  &  0.98  &  76.49  &  used  \\
  2009-03-08  &  00:42:31.015  &  P82  &  C  &  PRISM  &  3 of 3  &  34.29  &  34.38  &  1.00  &  37.82  &  used  \\
  2009-03-08  &  02:14:47.000  &  P80  &  C  &  PRISM  &  3 of 3  &  42.15  &  55.01  &  1.00  &  58.75  &  used  \\
  2009-03-08  &  02:52:36.000  &  P80  &  C  &  PRISM  &  3 of 3  &  44.78  &  60.63  &  1.00  &  67.36  &  used  \\
  2009-03-08  &  03:50:00.000  &  P82  &  A  &  PRISM  &  1 of 1  &  15.81  &  67.46  &  1.00  &  80.24  &  used  \\
  2009-03-17  &  03:16:00.000  &  P82  &  D  &  PRISM  &  2 of 2  &  63.26  &  67.61  &  0.01  &  80.54  &  used  \\
  2009-03-17  &  03:48:12.000  &  P82  &  D  &  PRISM  &  2 of 2  &  63.98  &  70.78  &  0.01  &  86.38  &  used  \\
  2009-05-03  &  23:43:46.000  &  P83  &  D  &  PRISM  &  0 of 2  &  62.16  &  65.02  &  0.09  &  75.35  &  not used  \\
\noalign{\smallskip}
\hline
\end{longtable}
    \begin{flushleft}
       \textbf{Notes. }
       $^{\mathrm{a}}$~Semester of proposal. 
       $^{\mathrm{b}}$~AT station: A~=~E0$-$G0, B~=~G0$-$H0, B*~=~A0$-$D0, C~=~E0$-$H0, D~=~D0$-$H0, E~=~D0$-$G1 or F~=~H0$-$G1 (cf.~Table~\ref{App_ObsLog}). 
       $^{\mathrm{c}}$~Dispersive element. 
       $^{\mathrm{d}}$~Calibrators used out of those available. 
       $^{\mathrm{e}}$~Projected baseline length. 
       $^{\mathrm{f}}$~Position angle of the projected baseline on the sky. 
       $^{\mathrm{g}}$~Visual light phase. 
       $^{\mathrm{h}}$~Elevation. 
       $^{\mathrm{i}}$~Quality flag showing if that observation is used for the model fitting or not.
     \end{flushleft}
\end{center}

%###########################################################################################
%###########################################################################################

\end{document}